%% file: main.tex
\documentclass{ieeetmlcn}
\usepackage{amsmath,amssymb,amsfonts}
\usepackage{algorithmic}
\usepackage{textcomp}
\usepackage{array}       
\usepackage[caption=false]{subfig}
\usepackage{tikz}
\usepackage{booktabs}    
\usepackage{array}       
\usepackage{color}
\usepackage{tikz}
\usepackage[normalem]{ulem}
\useunder{\uline}{\ul}{}
\usepackage{pgfplots}
\usepgfplotslibrary{groupplots}
\usepackage{times}
\usepackage{algorithmic}
\usepackage{amssymb}
\usepackage{amsxtra}
\usepackage{multirow}
\usepackage{mathtools}
\usepackage{algorithm}
\usepackage[normalem]{ulem}
\usepackage{tabu}

\usepackage{amsmath}
\usepackage{comment}
\usepackage{makecell} 
\usepackage{hyperref}
\hypersetup{hidelinks=true}
\def\BibTeX{{\rm B\kern-.05em{\sc i\kern-.025em b}\kern-.08em
    T\kern-.1667em\lower.7ex\hbox{E}\kern-.125emX}}
\AtBeginDocument{\definecolor{tmlcncolor}{cmyk}{0.93,0.59,0.15,0.02}\definecolor{NavyBlue}{RGB}{0,86,125}}
\usepackage{cite}
\usepackage{flushend}
\usepackage{hyperref}
\usepackage{url}
\usepackage{faktor}

\usepackage{graphicx}

\usepackage{cite}
\usepackage{amsmath,amssymb,amsfonts}
\usepackage{graphicx,color}
\usepackage{textcomp}
\usepackage{xcolor}
\usepackage{hyperref}
\hypersetup{hidelinks=true}
\usepackage{algorithm,algorithmic}

\def\BibTeX{{\rm B\kern-.05em{\sc i\kern-.025em b}\kern-.08em
    T\kern-.1667em\lower.7ex\hbox{E}\kern-.125emX}}

\def\OJlogo{\vspace{-4pt}\includegraphics[height=18pt]{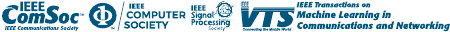}}
\def\seclogo{\vspace{10pt}\includegraphics[height=18pt]{new_logo}}

\def\authorrefmark#1{\ensuremath{^{\textbf{#1}}}}

\definecolor{green}{RGB}{0,128,0}
\definecolor{orange}{RGB}{255,165,0}
\definecolor{dimgray85}{RGB}{85,85,85}
\definecolor{gainsboro229}{RGB}{229,229,229}
\definecolor{lightgray204}{RGB}{204,204,204}
\definecolor{darkgray176}{RGB}{176,176,176}
\definecolor{gray128}{RGB}{100,100,100}    

\def\OJlogo{\vspace{-4pt}\includegraphics[height=18pt]{new_logo}}
\def\seclogo{\vspace{10pt}\includegraphics[height=18pt]{new_logo}}

\def\authorrefmark#1{\ensuremath{^{\textbf{#1}}}}
\AtBeginDocument{\definecolor{tmlcncolor}{cmyk}{0.93,0.59,0.15,0.02}\definecolor{NavyBlue}{RGB}{0,86,125}}
\begin{document}
\receiveddate{XX Month, XXXX}
\reviseddate{XX Month, XXXX}
\accepteddate{XX Month, XXXX}
\publisheddate{XX Month, XXXX}
\currentdate{XX Month, XXXX}
\doiinfo{TMLCN.2022.1234567}

\markboth{}{Hamrouni {et al.}}

\title{Resource Allocation in Hybrid Radio-Optical IoT
Networks using GNN with Multi-task Learning}

\author{Aymen Hamrouni\authorrefmark{1}, Student Member, IEEE, Sofie Pollin\authorrefmark{1}, Senior Member, IEEE,\\ and Hazem Sallouha\authorrefmark{1}, Senior Member, IEEE}

\affil{WaveCore at the Department of Electrical Engineering (ESAT), KU Leuven, Belgium. \\
E-mails: \{aymen.hamrouni, sofie.pollin, hazem.sallouha\}@kuleuven.be.}
\corresp{Corresponding author: Aymen Hamrouni (email: aymen.hamrouni@kuleuven.be).}
\authornote{This work was supported in part by the SUPERIOT project (SNS JU under the European Union's Horizon Europe research and innovation program, Grant Agreement N°101096021) and the 6thSense project (HORIZON-MSCA-DN-2022, Grant Agreement N°101119652).\vspace{-0.7cm}}

\begin{abstract}
\textcolor{black}{This paper addresses the problem of dual-technology scheduling in hybrid Internet-of-things (IoT) networks that integrate Optical Wireless Communication (OWC) alongside Radio Frequency (RF). We begin by presenting an optimization formulation that jointly considers throughput maximization and delivery-based Age of Information (AoI) minimization between access points and IoT nodes under energy and link availability constraints. However, given the intractability of solving such NP-hard problems at scale and the impractical assumption of full channel observability, we propose the Dual-Graph Embedding with Transformer (DGET) framework, a supervised multi-task learning architecture combining a two-stage Graph Neural Networks (GNNs) with a Transformer-based encoder. The first stage employs a transductive GNN that encodes the known graph topology and initial node and link states (e.g., energy levels, available links, and queued transmissions). The second stage introduces an inductive GNN for temporal refinement, which learns to generalize these embeddings to the evolved states of the same network, capturing changes in energy and queue dynamics over time, by aligning them with ground-truth scheduling decisions through a consistency loss. These enriched embeddings are then processed by a classifier for the communication links with a Transformer encoder that captures cross-link dependencies through multi-head self-attention via classification loss.  Simulation results show that hybrid RF-OWC networks outperform standalone RF systems by handling higher traffic loads more efficiently and reducing AoI by up to $20\%$, all while maintaining comparable energy consumption. The proposed DGET framework, compared to traditional optimization-based methods, achieves near-optimal scheduling with over $90\%$ classification accuracy, reduces computational complexity, and demonstrates higher robustness under partial channel observability.}
\end{abstract}


\maketitle
\section{Introduction}
\subsection{Motivation}
\IEEEPARstart{I}{n} recent years, the proliferation of Internet-of-things (IoT) devices has revolutionized the way we live and work. From wearable gadgets that monitor our health to smart home systems that enhance our living environment, the ubiquity of IoT technologies is undeniable. One key challenge in IoT networks lies in the efficient collection and transmission of data in real-time~\cite{Hamr2412:AoI,10568668}. \textcolor{black}{For instance, in healthcare, IoT enables continuous remote monitoring of critical patient vitals, such as heart rate and blood pressure, requiring prompt transmission for timely medical interventions~\cite{10433153,9709545}. Industrial IoT (IIoT) requires real-time data transmission for monitoring machinery and production lines, enabling preemptive maintenance~\cite{10638762}. Another key challenge in IoT networks is
energy efficiency. Many devices are low-powered or even batteryless~\cite{10089846}, relying on energy harvesting.} While prompt data transmission is crucial, the energy consumption involved in this process must also be taken into account as it significantly impacts the long-term operation of IoT devices. 
\textcolor{black}{Conventional IoT networks mainly rely on single-band Radio Frequency (RF) technologies, such as Bluetooth Low Energy (BLE) at 2.4 GHz, which are designed for low-power operation~\cite{11079647} but operate in a heavily shared spectrum, suffering from congestion, interference, and limited bandwidth~\cite{10136454}. These effects become particularly severe in dense deployments~\cite{7786107}. }

\begin{figure}[t]
\begin{center}
\hspace{-0.3cm}
\includegraphics[width=0.5\textwidth]{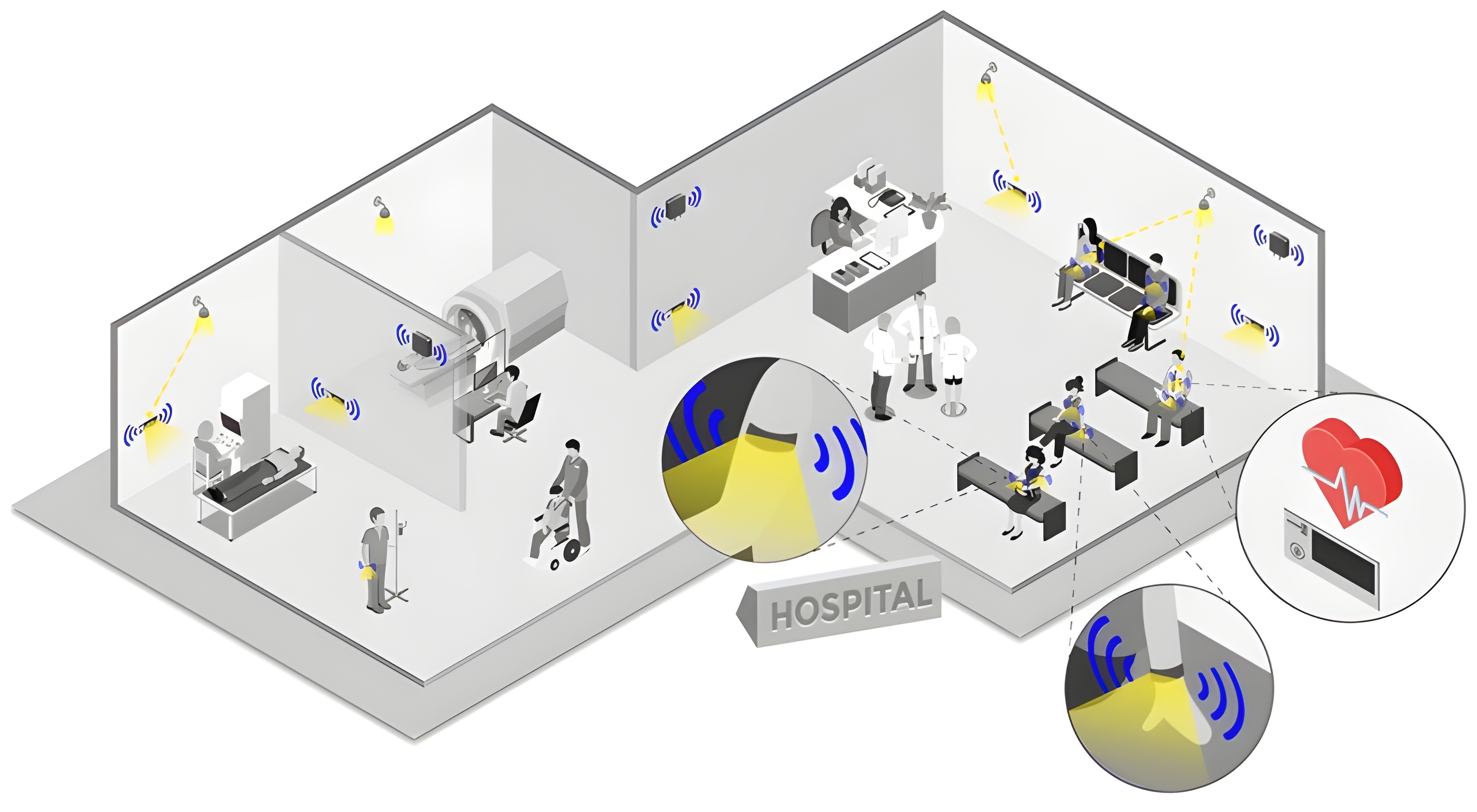}
\end{center}
\caption{\textcolor{black}{Illustration of a hybrid RF and OWC deployment in a hospital environment. RF ensures broad coverage and mobility for staff and patient monitoring, while OWC provides secure, interference-free communication for critical areas such as operating rooms and diagnostic imaging\cite{Katz2024}.\vspace{-0.5cm}}}
\label{example}
\end{figure}

Multi-band communication is a promising solution\cite{10179155,10539139,10034753} for next-generation IoT networks. By leveraging the use of multiple communication bands, such as a combination of RF bands and Optical Wireless Communication (OWC) technologies (e.g., visible light, infrared), multi-band networks can allow IoT devices to dynamically select the most suitable band based on real-time factors such as the IoT application's data rate and latency requirements, environment, or the devices' condition (e.g., energy budget).  
The intelligent integration of RF and OWC in a hybrid system has the potential to exploit their complementary strengths. RF remains essential for providing long-range connectivity and non-line-of-sight (NLoS) communication, whereas OWC, operating in visible light, infrared, or terahertz (THz) bands, offers high-capacity, low-latency, and interference-free communication~\cite{Katz_2024}.
\textcolor{black}{Such integration not only mitigates congestion in heavily utilized bands but also improves resilience by providing an alternative communication medium in case of network failures\cite{8960379}.  Optical components such as infrared LEDs and photodetectors are low-cost, energy-efficient, and widely available,  making hybrid RF-OWC architectures feasible with limited additional hardware complexity\cite{10008721,10539390}. Fig.~\ref{example} illustrates a smart hospital scenario with multiple access points (APs) and IoT nodes equipped with RF and OWC, where RF supports general wireless connectivity and OWC is used in sensitive areas such as MRI rooms to avoid RF interference. Recent works~\cite{Katz_2024,10539390}  have demonstrated the practical feasibility of RF-OWC IoT networks that combine RF and Visible Light Communications (VLC) for data transfer and energy harvesting, enabling battery-less or ultra-low-power operation.}

Despite the enormous advantages, the integration of RF and OWC technologies into a unified hybrid communication system presents several challenges in optimizing the RF-OWC resources allocated to serve IoT nodes. First, OWC and RF have different propagation characteristics, processing, and energy profiles, increasing the optimization problem complexity. This challenge becomes even more pronounced when considering practical IoT implementations that often involve dual and sometimes conflicting objectives, such as maximizing network throughput while minimizing latency.
Second, the scalability of the hybrid RF-OWC resource allocation solution is particularly critical in IoT networks due to networks typically consisting of multiple APs, with the number of IoT nodes being at least an order of magnitude higher~\cite{iotanalytics2024}.
Moreover, seamless integration of RF and OWC technologies requires adaptive and context-aware resource management that can respond to varying link quality, dynamic availability, and the constrained energy budgets of IoT devices. This is especially difficult in low-power IoT networks where decisions must be made promptly and often under uncertainty due to incomplete or changing channel conditions. Despite the growing body of research on hybrid RF-OWC systems, to the best of our knowledge, addressing these resource allocation challenges holistically in hybrid RF-OWC IoT networks remains an open research question.

\vspace{-0.1cm}
\subsection{Contribution}
\textcolor{black}{In this work, we aim to investigate the resource allocation challenges in a hybrid dual-band IoT network that utilizes RF and OWC technologies\cite{es1252}. We first formulate the optimization problem as a Mixed-Integer Non-Linear Programming (MINLP) which captures the full nonlinear interactions among energy consumption, link availability, and scheduling decisions. The MINLP is then approximated as a MILP using standard linearization techniques. The optimization problem has  two-fold objectives: 1) maximize the throughput across the network with respect to energy consumption and 2) minimize the average delivery-based AoI in the network. As direct minimization of the time-average AoI is known to be intractable in general multi-source, multi-destination networks\cite{10145068} due to the coupling of scheduling decisions over time and across links, we adopt a tractable surrogate that minimizes the age of each update at its delivery instant. }
\textcolor{black}{Moreover, the proposed design considers the resource allocation problem over coherence time windows, in which the channel is assumed quasi-static. This is particularly valid in low-mobility IoT indoor environments where devices and surroundings exhibit minimal variation. In such settings, the quasi-static nature of the environment allows for reliable short-term predictions of Signal-to-noise Ratio (SNR) over short durations  (e.g., $500$ ms)\cite{8395053,8805349,9044427,9860403}. This formulation enables dynamic, context-aware band selection, allowing IoT devices to autonomously select between RF and OWC based on slowly varying channel-state measurements and network conditions within each coherence-time window.}

\textcolor{black}{This paper extends our previous work~\cite{Hamr2412:AoI} by reformulating the optimization problem to include more realistic RF and optical channel models, device-level energy and communication constraints, and comprehensive performance evaluations to validate the scalability, generalizability, and practical feasibility for hybrid RF-OWC IoT networks.
Moreover, deterministic optimization methods are highly sensitive to link SNR estimates because they require precise, real-time knowledge of link states to compute feasible and optimal allocations. Therefore, given the NP-hardness of hybrid RF-OWC resource allocation~\cite{articlesssss} and the unrealistic assumption of full system knowledge, this work proposes an alternative data-driven Dual-Graph Embedding with Transformer (DGET) architecture utilizing Graph Neural Networks (GNNs)~\cite{9944643} and Transformers\cite{vaswani2023attentionneed}. }We first model the IoT network as a time-series graph, where each snapshot represents the network's state at a given time. The \textit{input snapshots} consist of known network states, including initial devices' energy levels, transmission schedules, and available communication links over coherence time. The \textit{recorded snapshots} represent the same network as it evolves, reflecting updates in energy, link status, and message queues due to ongoing communications. They are constructed by solving the scheduling problem using optimization techniques, and they contain ground truth values for the network changes.

\textcolor{black}{The design of DGET is driven by the intrinsic characteristics of hybrid RF-OWC wireless networks, in which link relevance is governed by physical-layer factors such as visibility, blockage, interference, and device energy states, and where reliable observability is typically limited to a subset of known network snapshots. DGET explicitly exploits this asymmetry by first processing input snapshots through a transductive GNN~\cite{10191134}, which learns fine-grained graph embeddings that capture both local and global structural patterns from the known network topology. These transductive embeddings are tailored to the hybrid RF-OWC IoT setting by jointly encoding device-level attributes (e.g., initial energy states) and edge-level features (e.g., RF-OWC link availability and scheduled transmissions), thereby enforcing stable structural priors and enabling attention mechanisms to identify physically meaningful neighbor-importance patterns across heterogeneous links, as supported by prior studies on fully observed graphs~\cite{tang2023understandinggeneralizationgraphneural}. While this transductive stage yields optimized and consistent representations for observed network states, it is inherently limited in capturing instantaneous and evolving dynamics induced by time-varying conditions such as energy depletion, queue evolution, and adaptive scheduling decisions. This limitation is particularly critical in dynamic IoT environments, where device states, channel conditions, and scheduling queues change continuously as communication unfolds. To address this limitation, DGET incorporates an inductive GNN~\cite{li2024combininginductiontransductionabstract} that operates on top of the transductive embeddings, refining them by integrating decision-level information from recorded snapshots, including selected links, scheduling outcomes, and energy updates. It learns an aggregation function that can generalize across graphs and also preserve the structure and reinforce the learned relationships obtained from the transductive embeddings through attention on the recorded snapshots. It is designed to promote knowledge transfer from the known snapshots seen during training through a domain adaptation mechanism, guided by a consistency loss aligning reconstructed embeddings with recorded graphs.}

\textcolor{black}{The proposed DGET model is trained using a multi-task learning framework.
As the first task aligns reconstructed inductive embeddings with the ground truth recorded graphs, the second task performs communication link type classification using a Transformer-based architecture with a supervised learning setup. The Transformer’s attention mechanism on the resultant inductive embeddings enables global aggregation of these embeddings and reasoning over spatially distant devices and over multiple network snapshots, allowing the model to identify cross-device and temporal correlations in the IoT network to predict the optimal communication technology for each link.  The supervised learning setup is motivated by the availability of high-quality optimal scheduling labels obtained from solving the MILP formulation across multiple network realizations, which further guide the DGET model towards learning physically consistent mappings between network states and scheduling decisions. This multi-task learning approach, designed for the hybrid RF-OWC IoT environment, allows DGET to simultaneously optimize two coupled tasks;  aligning the embeddings reconstruction to the ground truth recorded graph and also improving the accuracy of link prediction.}

The main contributions \footnote{The code for this paper has been made open source and can be accessed here: https://github.com/aymenhamrouni/DGET-Dual-Graph-Embedding-Transformer-for-Hybrid-RF-OWC-IoT-Networks}
in this paper can be summarized as follows: 
\begin{itemize}
    \item  We present a system formulation for a hybrid RF-OWC network and formulate a constrained MINLP optimization problem, approximated as a MILP, that seeks to maximize the overall throughput of the network while taking into account the energy consumption and the delivery-based AoI. 
    \item We introduce the DGET framework that combines two GNN components with a Transformer encoder to solve the resource allocation problem with lower overhead compared to solving the MILP problem. \textcolor{black}{The transductive GNN learns fixed structural patterns from the known IoT network topology, while the inductive GNN generalizes these embeddings to temporally evolved configurations of the IoT network, capturing changes in energy, queue states, and link dynamics.} A Transformer-based encoder applies self-attention to model complex dependencies and global relationships on the inductive GNN embeddings to predict active links and their corresponding transmission technologies. The entire model is trained end-to-end using a multi-task learning framework with two coupled loss functions.
    \item  We evaluate the performance of the hybrid RF-OWC network under varying conditions, including different network loads and sizes. Results demonstrate that the hybrid network can achieve up to $15\%$ higher successful transmission rate and up to $20\%$ reduction in Age-of-Information (AoI) compared to a standalone RF network. Under high network load scenarios, the RF-OWC configuration demonstrates improved energy efficiency. Furthermore, the proposed DGET framework achieves over $90\%$ classification accuracy with up to $8$x lower inference complexity compared to the optimization model and demonstrates robustness under partial system observability.
\end{itemize}
The remainder of this paper is organized as follows: Section~\ref{sec2} reviews the related work in the areas of hybrid RF-OWC communication and scheduling. In Section~\ref{sec3}, we describe the system model, detailing the channel characteristics, network architecture, and communication setup. Section~\ref{sec4} formulates the optimal RF-OWC scheduling problem through a mathematical optimization framework. The proposed DGET framework is introduced in Section~\ref{sec5}, outlining its architecture and key components. Section~\ref{sec6} presents the experimental evaluation used to assess the performance of our approach. Finally, Section~\ref{sec7} concludes the paper and outlines directions for future research.

\section{Related Work}
\label{sec2}
\textcolor{black}{Given the growing interest in exploiting the complementary advantages of RF and OWC technologies, numerous studies have investigated dual-technology integration to enhance communication performance in various IoT scenarios.} Tang et al.~\cite{10502287} investigated the resource allocation optimization for indoor hybrid RF-OWC systems. They formulated an optimization problem to maximize the minimum data collection rate across all IoT devices within the given time period. Fakirah et al.~\cite{1} investigated the integration of Visible Light Communication (VLC) with RF systems by using hybrid access points to facilitate seamless information exchange between vehicles and roadside units in a vehicle-to-infrastructure mode. Obeed et al.~\cite{3} proposed an iterative optimization-based algorithm for load balancing and power allocation in a hybrid RF-OWC system, consisting of a single RF access point and multiple OWC access points. Xiao et al.~\cite{5} examined a hybrid downlink system combining VLC and RF with a cognitive-based resource allocation policy to optimize network performance. Guo et al.~\cite{9493176} explored a relay-assisted network designed for simultaneous wireless information and power transfer, leveraging hybrid VLC-RF communication where a light-emitting diode access point serves multiple users engaged in both information exchange and energy harvesting. Alenezi et al.~\cite{9128892} proposed a Reinforcement Learning (RL) algorithm for the hybrid WiFi-VLC scheduling problem, designed to operate on the WiFi side with the goal of maximizing throughput. \textcolor{black}{Khalili et al. \cite{10978689} presented a multi-attribute decision-making framework to manage handovers in RF-OWC heterogeneous networks, systematically evaluating APs under criteria including handover cost, data-rate demand vs. achievable rate, and capacity utilization.  Collectively, these works underscore a crucial insight: RF alone faces spectrum congestion, interference, and coverage limitations, whereas OWC (or VLC) offers high capacity, unlicensed spectrum access, and improved security but suffers from line-of-sight and mobility constraints. Thus, dual‐technology access becomes necessary to exploit their complementary strengths and meet the heterogeneous demands of dynamic IoT networks.}

GNNs have been adopted in IoT networks \cite{9446513,9912428,9859333}, and recently in wireless communications networks\cite{10183787}, as they can effectively exploit domain knowledge\cite{10183787,10593270}, i.e., the graph topology in wireless communications.  Suarez-Varela et al.~\cite{9846958} highlighted the use of GNNs for the modeling, control, and management of communication networks. Shen et al.~\cite{9944643} demonstrated that GNNs can achieve near-optimal performance in wireless networks with fewer training samples compared to traditional neural networks. Yang et al.~\cite{10183787} proposed a GNN-based framework for optimizing wireless network performance, utilizing wireless interference graphs.  Cheng et al.\cite{10.1145/3570236.3570293} proposed a model for resource allocation in heterogeneous D2D networks, addressing the limitations of traditional GNNs that often ignore graph devices' features and edges' features.  Eisen et al.\cite{Eisen_2020} addressed the resource allocation in wireless networks through a model-free approach using Random Edge Graph Neural Networks (REGNNs) that operate over interference-induced random graphs. 
Chen et al.\cite{9462385} proposed a supervised GNN framework for device-to-device (D2D) resource allocation in wireless IoT networks, where communication links are represented as graph nodes and interference as edges. Zhao et al.\cite{9414098} addressed distributed link scheduling in wireless networks by formulating it as a Maximum Weighted Independent Set (MWIS) problem, which is NP-hard. They proposed a graph convolutional network (GCN)-based approach that learns topology-aware graph node embeddings to improve greedy scheduling methods.

Existing studies on hybrid RF-OWC networks primarily use optimization methods that are computationally intensive. Moreover, they largely neglect information freshness metrics like AoI. While Graph Neural Networks have demonstrated efficacy in wireless resource management, their application to hybrid RF-OWC resource allocation, particularly when formulated as a classification problem incorporating device energy, packet requirements, and link quality, remains unexplored. In this work, we aim to address this research gap.

\section{System Model}
\label{sec3}

In this section, we introduce the system model, represented by the network setup, channel model, and data communication. For clarity, we provide the main notation used throughout the paper in Table~\ref{tableN}.

\begin{table}[t]
\begin{center}
\caption{\label{tableN} \textcolor{black}{Main notations and their descriptions. }}
\addtolength{\tabcolsep}{-5pt}
\begin{tabular}{|c || c|}
  \hline
 \textbf{Notations} & \textbf{Descriptions} \\
  \hline \hline
$\mathcal{N}_d$ & Set of IoT nodes \\\hline
$\mathcal{N}_{\text{AP}}$ & Set of APs \\\hline
$\mathcal{N}$ &  Total set of devices, $\mathcal{N}_d \cup \mathcal{N}_{\text{AP}}$ \\\hline
$\mathcal{T}$ & Set of discrete time steps \\\hline
$\mathcal{M}$ & Set of communication technologies \\\hline
$\xi_i$ & Energy level of device $i$ \\\hline
$\xi_s^m$, $\xi_r^m$ & \makecell[c]{Energy to transmit/receive a message \\ using technology $m$} \\\hline
$\Phi_{i,j}$ & \makecell[c]{Collection of valid time steps of \\ messages from $i$ to $j$} \\\hline
$\phi_{i,j}^{f,l}$ & \makecell[c]{Valid time steps of message $f$ of type $l$ \\ from $i$ to $j$} \\\hline
$\left (\phi_{i,j}^{f,l} \right )_s, \left (\phi_{i,j}^{f,l} \right )_e$ & \makecell[c]{Generation and discard time steps \\ of message $f$ with type $l$ from $i$ to $j$} \\\hline
$\Gamma_{i,j}^f$ & \makecell[c]{Number of packets in message $f$ \\ from $i$ to $j$} \\\hline
$\bold{V}$ & \makecell[c]{Visibility matrix for communication technologies} \\\hline
$x_{i,j,m,k}$ & \makecell[c]{Binary value indicating transmitter, receiver, \\ technology, and time step $k$} \\\hline
$\delta_{i,j,f}$ & \makecell[c]{Time between generation and \\ receipt of message $f$} \\\hline
$\tilde{y}_{i,j,m,f,k}$ & \makecell[c]{Number of transmitted packets of message $f$ \\ using technology $m$ at time $k$} \\\hline
$\bold{v}^\text{I}_i$, $\bold{v}^\text{R}_i$ & Input and Recorded feature vector of device $i$ \\\hline
$\bold{e}^\text{I}_{ij}$, $\bold{e}^\text{R}_{ij}$& Input and Recorded edge feature from $i$ to $j$ \\\hline
$\mathcal{E}_t$ & Set of allowed communication technology\\\hline
$\mathcal{E}_c$, $\mathcal{E}'_c$ & Set of initial and augmented communication status \\\hline
$\mathcal{G}^{I}_k$, $\mathcal{G}^{R}_k$ & Input and recorded graphs at time $k$ \\\hline
\end{tabular}
\vspace{-0.7cm}
\end{center}
\end{table}

\subsection{Network Setup}
 
\begin{figure}[t]
\begin{center}
\includegraphics[width=0.4\textwidth]{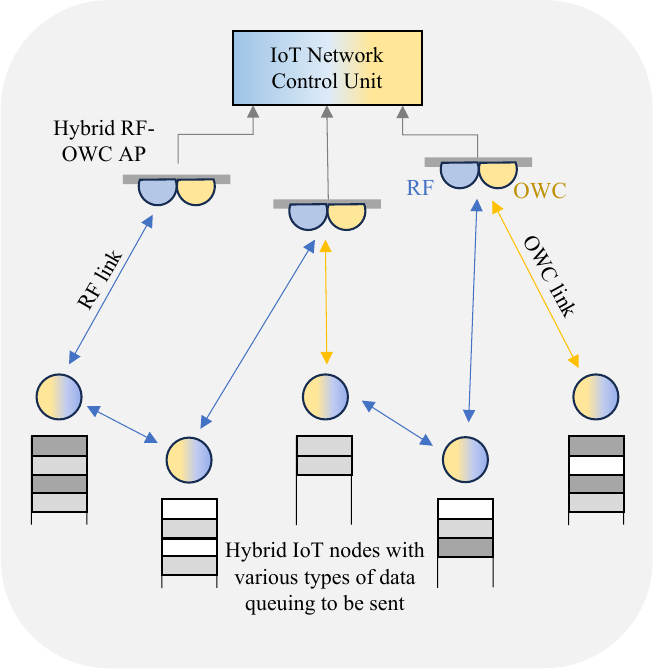}
\end{center}
\caption{A visualization of the RF-OWC model showing a hybrid RF-OWC network exchanging different types of applications' data.\vspace{-0.5cm}}
\label{fig0}
\end{figure}

We consider a hybrid IoT network consisting of access points (APs) and nodes that can communicate through RF and OWC links.
We assume that nodes communicate with APs using either OWC or RF, whereas only RF is used for inter-node communication. RF transmissions are assumed to happen in pre-assigned time slots using Time Division Multiple Access (TDMA), ensuring no two devices transmit simultaneously over the same channel. Moreover, we suppose that OWC uses tightly focused beams of visible or infrared light, which are inherently directional. Therefore, each device pair can establish either a LoS RF-OWC link or rely on a quasi-LoS RF link, minimizing inter-device interference. The system model with an RF-OWC hybrid configuration is illustrated in Fig.~\ref{fig0} with different IoT nodes transmitting multiple types of application data, such as sensor readings, status updates, and alarm messages.  

Let $\mathcal{N}_d$ be the index set of IoT nodes and $\mathcal{N}_{AP}$ be the index set of APs in the system. The index set of total devices\footnote{Throughout this work, the term device collectively refers to both IoT nodes and APs.} in the network is denoted by $\mathcal{N}=\mathcal{N}_{d}  \cup \mathcal{N}_{AP}$, with $\mathcal{N}_{d}  \cap \mathcal{N}_{AP} = \emptyset$. Let $\mathcal{M}$ be the set of possible communication technologies. In the network configuration considered, $\mathcal{M}=\{0,1\}$ with $m=0,\, m \in \mathcal{M}$ representing RF and $m=1,\, m \in \mathcal{M}$ denoting OWC. Let $\mathcal{L}$ be the set of different data types in the system. For example, in the case of a healthcare scenario, where IoT sensors exchange two types of messages, e.g., heart rates and oxygen levels; we have $|\mathcal{L}|=2$.  
We model the communication in the network over time by an equal-length discrete time horizon $\mathcal{T}$, which reflects the channel coherence time. This horizon is segmented into time steps $[1, 2, \ldots, |\mathcal{T}|]$, where each time step is a single time unit denoted by the scalar $\tau$ (in seconds).

\vspace{-0.1cm}
\subsection{Channel Model}
The path loss, $PL$, in dB considered for an indoor RF channel\cite{Goldsmith_2005} is given as:

\begin{align}
&P L[\mathrm{dB}]=P L_{0}[\mathrm{dB}]+10 \, \alpha \,\log _{10} \frac{d}{d_0}+\psi[\mathrm{dB}],
\end{align}
where $P L_{0}$ is the path loss at a reference distance $d_0=1 \mathrm{~m}$ from transmitter, $ \alpha$ is the path loss exponent, $d$ is the distance between transmitter and receiver, and $\psi$ is a log-normal shadow fading with zero mean and variance $\sigma^2_\psi$.
Consequently, the received power accounting for both large-scale and small-scale fading effects is expressed as:
\begin{align}
   P_{r}^{\text{RF}} = P_t^{\text{RF}} \, G_t \, G_r \, 10^{-\frac{PL\,[\mathrm{dB}]}{10}} \, |h|^2,
    \label{eq:received_power}
\end{align}
where  \(P_t^{\text{RF}}\)  is the transmitted power, \(G_t, G_r\) are the gains of the transmitter and receiver antennas. Here \( h \) is the small-scale Rician fading coefficient, given by\cite{Goldsmith_2005}:
\[
h = \underbrace{\sqrt{\frac{K}{1 + K}} \, e^{j \gamma}}_{\text{LoS component}} + \underbrace{\sqrt{\frac{1}{1 + K}} \, z}_{\text{NLoS component}},
\]
where \( K \) is the Rician factor, representing the ratio of the power in the LoS component to the power in the NLoS components, $\gamma \sim \mathcal{U}(0,2 \pi)$ being a uniformly distributed random variable for the LoS phase, and \( z \sim \mathcal{CN}(0, 1) \) is a complex Gaussian random variable representing the NLoS components. 
To characterize the quality of the RF communication signals, we rely on the SNR\cite{Goldsmith_2005}, which is given as:
\begin{equation}
SNR^{\text{RF}}=\frac{P_r^{\text{RF}}}{N_{\text{RF}}}, 
\end{equation}
where $N_{\text{RF}}=k_{B} \, T \, B_\text{RF}$ represents the noise power. Here, $B_{\mathrm{RF}}$ is the RF bandwidth, $k_{B}$ is the Boltzmann constant $\left(1.38 \, 10^{-23} \mathrm{~J} / \mathrm{K}\right)$  %
and $T$ is the room temperature (in Kelvin). Given the time-varying nature of the wireless channel, we define $SNR^{\text{RF}}_{{i,j,k}}$ as the SNR for the RF signals from device $i$ and device $j$ at time $k$.
The achievable rate between any device pairs at time $k$  is computed as $C^{\text{RF}}_{i,j,k} = B_{\mathrm{RF}} \, \log_2\left(1 + SNR^{\text{RF}}_{{i,j,k}}\right)$.

In LED-based OWC, the optical power received $P_{\text{r, OWC}}$ is determined by the Lambertian radiation pattern of the LED, geometric attenuation, atmospheric loss, and receiver characteristics\cite{10283503}. It is computed as follows:
\begin{equation}
P_r^{\text{OWC}}= \kappa \,  P_t^{\text{OWC}} \, T_{\text {Opt }} \, \frac{(u+1) \, A_{\operatorname{Rec}} \, \cos ^u(\theta) \, \cos (\phi)}{2 \pi \, d^2},
\end{equation}
with \textcolor{black}{$\kappa$ is a coefficient $\in (0,1]$ } that models power splitting between actual data modulation and lighting,  $P_t^{\text{OWC}}$ is the transmitted optical power in Watts (W), $T_{\text {Opt }}$ is the total optical efficiency (lens/filter losses), $u$ is the Lambertian order,  $A_{\text {Rec }}$ is the photodiode effective area (in meters$^2$), $\theta$ is the LED irradiance angle at the transmitter, $\phi$ is the photodiode incidence angle at the receiver, and $d$ is the link distance (in meters). 
The photodiode converts $P_r^{\text{OWC}}$ to electrical current via responsivity $\eta_r$ (Ampere/Watt): 
\begin{equation}  
I_r=\eta_r \, P_r^{\text{OWC}}.
\end{equation}
The SNR formula can then be expressed as follows~\cite{7003937}:
\begin{align}
\centering
   SNR^{\text {OWC}}&=\frac{I_r^2}{N_{\mathrm{shot}}^2+N_{\text {thermal }}^2}
    \nonumber\\& =\frac{\left(\eta_r  \, P_r^{\text{OWC}}\right)^2}{2 q\left(\eta_r \, P_r^{\text{OWC}}+\eta_r \, P_{n}\right) B_{\mathrm{OWC}}+\frac{4 \, k_B \, T}{R_L} \, B_{\mathrm{OWC}}},
\end{align}
where $B_{\mathrm{OWC}}$ is the OWC bandwidth and the shot noise is expressed as $N_{\mathrm{shot}}=\sqrt{2 q \, \left(I_r+I_{\mathrm{n}}\right) B_{\mathrm{OWC}}}$ with $q$ the Electron charge $\left(1.6 \, 10^{-19} \mathrm{C}\right)$ and 
$I_{n}=\eta_r \, P_{n}$ is the background light-induced current from the ambient light’s noise power $P_{n}$. The thermal noise is also expressed as $N_{\text {thermal }}=\sqrt{\frac{4 \, k_B \, T}{R_L} B_{\mathrm{OWC}}}
$ with $R_L$ being the load resistance ~$(\Omega)$. Similarly as RF, we define $SNR^{\text{OWC}}_{{i,j,k}}$ from device $i$ and device $j$  at time $k$.
The achievable rate for OWC between a transmitter $i$ and a receiver $j$ at time $k$ is $C^{\text{OWC}}_{i,j,k} = B_{\mathrm{OWC}} \, \log_2\left(1 + SNR^{\text{OWC}}_{{i,j,k}}\right)$. 

\begin{figure}[t]
\begin{center}
\includegraphics[width=0.4\textwidth]{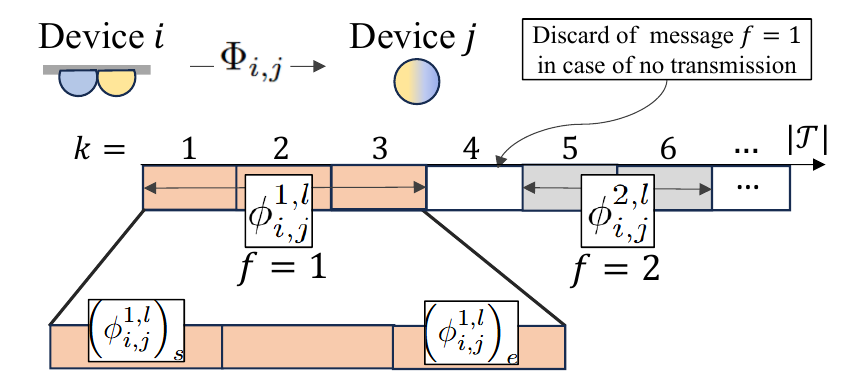}
\end{center}
\caption{Illustration of the communication model and the lifespan of messages.\vspace{-0.3cm}}
\label{fig:comm}
\end{figure}
\subsection{Communication Setup}

In order to capture the quasi-stationary indoor communication channel, we define a 4-D visibility matrix $\bold{V}$, with dimension $|\mathcal{N}| \times |\mathcal{N}| \times |\mathcal{T}| \times |\mathcal{M}|$, and where its elements are the scalar $SNR^m_{i,j,k}$. This visibility matrix governs the communication flow between any two devices in the network. Given that IoT nodes cannot communicate with each other using OWC, we represent this by setting $SNR^{1}_{i,j,k}=-\infty, \forall (i,j) \in \mathcal{N}_{d} \times \mathcal{N}_{d}, \forall k \in \mathcal{T}$. Similarly, as communications between APs are assumed to be done on a separate dedicated medium, $SNR^m_{i,j,k}=-\infty,  \forall (i,j) \in \mathcal{N}_{AP} \times \mathcal{N}_{AP}, \forall m \in \mathcal{M}, \forall k \in \mathcal{T}$. To model the communication over time, we introduce the matrix $\bold{P}$, which defines the communication schedule between any two devices at any given time. The elements of this matrix, $\rho^k_{i,j}$, are binary values, where $\rho^k_{i,j} = 1$ indicates that communication will occur between devices $i$ and $j$ at time step $k$, and $\rho^k_{i,j} = 0$ otherwise.

In our communication model, data are transmitted between devices in the form of messages, where each message is composed of a series of independent packets. A packet represents a smaller unit of data within a message, and the size of each packet in bits is denoted by $s_p$.

To define the lifespan for generated messages, we denote $\Phi_{i,j}=\{\phi_{i,j}^{1,l},\phi_{i,j}^{2,l},\cdots,\phi_{i,j}^{|\Phi_{i,j}|,l}\}$ as the collection of valid time steps for the messages to be sent from the device $i$ to the device $j$. In this sense, ${\phi_{i,j}^{1,l}}$ represents the set of valid time steps corresponding to the first message in the queue that needs to be sent from device $i$ to device $j$. For instance, assume we have two messages to be sent from device $i$ to device $j$ with $\Phi_{i,j}=\{\{1,2,3\},\{5,6\}\}$. This implies that the first and the second messages can only be communicated during one of the time steps $\{1,2,3\}$ and $\{5 ,6\}$, respectively, before they get discarded. An illustrative example is shown in Fig.~\ref{fig:comm}. Packets belonging to a message in $\Phi_{i,j}$ must be sent at most once during their validity window represented by the message generation time and discard time (e.g., $1$ is the generation time of the first message in the previous example, while $4$ is its discard time). The discard time represents the maximum time at which the packets of the message could be sent before they are considered aged. We denote $\left ({\phi_{i,j}^{{1,l}}} \right )_s$ as the generation time of message $f$ and $\left ({\phi_{i,j}^{{1,l}}}\right )_e$  as its discard time.  Intuitively,  there is no overlap between the time steps of any two distinct messages for a given device. In other words, $\phi_{i,j}^{f,l} \cap \phi_{i,j}^{f',l}=\emptyset, \forall f \neq f'$.  In addition, we assume that each generated message can have a single type of data. Hence, $\phi_{i,j}^{f,l} \cap \phi_{i,j}^{f,l'}= \emptyset, \forall l \neq l'$.
Let $\Gamma_{i,j}^f \in \mathbb{Z}_{>0}, f \in [1,\dots,|\bold{\Phi}_{i,j}|]$ represent the amount of data included in message $f$ in terms of number of unit packet size $s_p$.

Each IoT device $i \in \mathcal{N}$ is assigned a total energy budget $\xi_i$. The energy (joules/bit) for transmitting or receiving between device $i$ and device $j$ using RF is computed as follows\cite{Goldsmith_2005}:
\begin{align}
&\xi_s^{\text {RF}}=  \frac{P_t^{\text{RF}}}{C^{\text {RF}} },\,\, \xi_r^{\text {RF}}=  \frac{P_r^{\text{\text {RF}}}}{C^{\text {RF}} }.
\end{align}

The energy (joules/bit) for transmitting or receiving between device $i$ and device $j$ using OWC is computed as follows\cite{Miller:12}: 
\begin{align}
& \xi^{\text {OWC}}_s=  \frac{\frac{P_t^{\text{OWC}}}{ \eta_{c}}}{C^{\text {OWC}} },\,\, \xi^{\text {OWC}}_r=  \frac{I_r \,  V_{\text{bias}}}{ C^{\text {OWC}} },
\end{align}
where $\eta_{c}$ is the electrical-to-optical conversion efficiency and $V_{\text{bias}}$ is the bias voltage of the photodetector during reception. 

\vspace{-0.1cm}
\section{Optimal RF-OWC Scheduling}
\label{sec4}

This section defines the optimization problem in RF-OWC IoT settings by outlining the decision variables, the constraints, and the overall objective function. 
\vspace{-0.1cm}
\subsection{\textcolor{black}{Decision Variables}}

In order to indicate the chosen 2-tuples (transmitter, receiver) and to keep track of assigned communication technology at each time step, we introduce the \textit{main} binary decision variable $x_{i,j,m,k}$ defined as follows:\begin{align} \label{x}
& x_{i,j,m,k}= 
     \begin{cases}
       \text{1,} &\,\text{if device $i$ transmits a message to device $j$} \\  & \text{using technology $m$ at time step $k$,} \\
       \text{0,} &\,\text{otherwise,} \\ 
     \end{cases}\notag\\
 &\hspace{2cm}\forall \, (i,j)\in \mathcal{N} \times \mathcal{N}, \forall \, m \in \mathcal{M}, \forall k \in \mathcal{T}.
     \end{align}

Unless otherwise stated, indices $i,j$ denote devices, $k$ denotes the time step, and $m$ denotes the communication technology.
To keep track of which technology device $i$ is using at time step $k$, we introduce a continuous \textit{endogenous} binary variable $s_{i,k}$, which is 0 if device $i$ is using RF, and 1 if it is using OWC.  Throughout this paper, we assume that the receiver's queuing delay, propagation delay, and processing delay are negligible and focus solely on the transmitter's queuing delay and transmission delay. The transmission delay is treated as the unit of time for both RF and OWC. 


\vspace{-0.1cm}
\subsection{\textcolor{black}{Constraints}}

\subsubsection{Scheduling Constraints}

To ensure that each IoT device within the network can communicate with at most one other device and use either RF or OWC at each time step, we include the following constraints:
\begin{equation}
\begin{split}
    \sum_{j \in \mathcal{N}} \sum_{m \in \mathcal{M}} x_{i,j,m,k} \leq 1, \quad \forall i , \forall k 
       \\
        \sum_{i \in \mathcal{N}}\sum_{m \in \mathcal{M}} x_{i,j,m,k} \leq 1, \quad \forall j , \forall k 
    \end{split}   
    \label{const1}
    \end{equation}

Any device $j$ engaged in communicating at a given time step with another device $i$ cannot simultaneously communicate with another device $j'$. This is set by the following constraints:
\begin{align}
    x_{i,j,m,k} + x_{j,j',m',k} \leq 1, \quad &\forall (i, j, j') \in \mathcal{N} \times \mathcal{N} \times \mathcal{N}, \nonumber\\ &\forall (m, m') \in   \mathcal{M} \times \mathcal{M}, \forall k \in \mathcal{T}
\label{const5} 
\end{align} 

 In the event of no communication during the time step $k$, the value $s_{i,k}$ is equal to $s_{i,k-1}$. Initially, every device starts with $s_{i,-1}=0$ (i.e., RF initialization).  The constraints are then written as:
\begin{align}
\label{const11}
	s_{j,k} =
	&\begin{cases}
		m, & \hspace{-0.3cm} \displaystyle\sum_{i \in \mathcal{N}} (x_{i,j,m,k} + x_{j,i,m,k}) = 1\\
		1,& \hspace{-0.3cm}\displaystyle\sum_{m \in \mathcal{M}}\sum_{i \in \mathcal{N}} (x_{i,j,m,k} + x_{j,i,m,k}) = 0 \land \, s_{j,k-1} = 1 \\
        0,& \hspace{-0.3cm}\displaystyle\sum_{m \in \mathcal{M}} \sum_{i \in \mathcal{N}} (x_{i,j,m,k} + x_{j,i,m,k}) = 0 \land \, s_{j,k-1} = 0
	\end{cases} \\ 
    &\forall j , \forall m, \forall k, \nonumber
\end{align}
with $\land$ being the logical AND operation. To avoid frequent technology switching between RF and OWC, we introduce the following constraints:
\begin{equation}
         \sum_{k \in \mathcal{T}} \left| s_{j,k} - s_{j,k-1} \right| \leq \Omega, \forall j, 
         \label{const14}
\end{equation}
with $\Omega$ is a custom threshold that defines the maximum switching rate for each device in the network. The highest value of $\Omega$ is $\frac{|\mathcal{T}|}{2}$, which in this case means that each device in the network is allowed to consecutively alternate between RF and OWC. We should note that these constraints can be integrated with the previous constraints to incur an energy loss or a delay to the network in case of frequent switching between technologies. 
\vspace{-0.5cm}
\subsubsection{Traffic and Message Validity}

To ensure that communication between devices can happen only when there is an actual message to be transmitted, we add the following constraints:
\begin{equation}
\begin{split}
   \sum_{m  \in \mathcal{M}} x_{i,j,m,k}  \leq \rho^k_{i,j}, \quad \forall i, j, \forall k.
    \end{split}
 \label{const2} \end{equation}

The following constraints in~\eqref{const10} extract the AoI at the delivery instant and are computed using the relative time offset within the valid window of a message sent from device $i$ to device $j$. 
If the message was sent successfully at the $k$-th time step, this duration of elapsed time is $k -  \left ({\phi_{i,j}^{f,l}} \right)_s+1$ between the current time and the generation time step of message $f$ and counting the unit transmission delay. In the absence of transmission, $\delta_{i,j,f}$ is set to an upper bound $\chi$ that represents a strict bound on the time duration of the longest message in the system (i.e., max($|\phi_{i,j}^{f,l}|$), $f \in [1,\dots,|\Phi_{i,j}|]$, $l \in \mathcal{L}$). These constraints are written as follows:
\begin{align}
\label{const10}
	\delta_{i,j,f} =
	&\begin{cases}
		k -  \left ({\phi_{i,j}^{f,l}} \right)_s+1, & \, \displaystyle\sum_{m \in \mathcal{M}} x_{i,j,m,k} = 1\\
		\chi, & \, \displaystyle\sum_{m \in \mathcal{M}} x_{i,j,m,k} = 0
	\end{cases} \\ 
    &\forall k \in \phi_{i,j}^{f,l},  \forall l, \forall f \in \{1, \ldots, |\Phi_{i,j}|\}, \forall i,j. \nonumber
\end{align}

Any device in the system can transmit multiple messages to a single receiver, and each of these messages can have different types of data. As each message has a duration of validity over several time steps, as discussed in the previous section, we impose the following constraints so that packets within the message, regardless of the latter's data type, can be sent at most once within their valid time window:
    \begin{equation}
    \begin{split}
        \sum_{m \in \mathcal{M}} \sum_{k \in \phi_{i,j}^{f,l}} x_{i,j,m,k} \leq 1, \quad \forall i, j, \forall f \in [1,\dots,|\Phi_{i,j}|], \forall l \in \mathcal{L}
    \end{split} 
     \label{const6} \end{equation}
where $\phi_{i,j}^{f,l}$ represents the valid time slots of message $f$ with data type $l$ that should be sent from device $i$ to device $j$.

In our network setup, message size $\Gamma_{i,j}^f$ is defined as a positive integer, representing the number of packets $s_p$ in message $f$. If device $i$ has more than one unit of a packet to transmit to device $j$ (i.e., $\Gamma_{i,j}^f>1$ for $f \in [1,\dots,|\Phi_{i,j}|]$), it may either transmit all the packets within that message, none, or a part of that message.  In this design, all packets can be transmitted within their corresponding message’s validity window. However, not all packets need to be sent; a subset of the packets can be transmitted at any given time. For example, in large-scale IoT networks where an IoT node collects air quality data from different nearby locations, an IoT device has a limited time window to transmit the data. During this time window, the device may transmit all $20$ packets, only $3$ packets, or none at all if the device is unable to transmit due to network congestion. To model such behavior, we introduce the following constraints: 
\begin{equation}
\begin{split}
& 0 \leq x_{i,j,m,k} \, \left ( \Gamma_{i,j}^f - y_{i,j,f} \right ) \leq \frac{C^{m}_{i,j,k}}{s_p} \, \tau 
 , \\ & \forall k \in \phi_{i,j}^{f,l},  \forall l, \forall m, \forall f \in \{1, \ldots, |\Phi_{i,j}|\}, \forall i,j, 
 \end{split}
  \label{eq:star}
  \end{equation}
where $y_{i,j,f}$ is an \textit{endogenous} integer variable representing the non-communicated number of packets. It is greater than or equal to zero and less than the maximum number of packets per message $\overline{\sigma}$ (e.g., $\overline{\sigma}=max(\Gamma_{i,j}^f)$). We note here that the non-communicated packets during a time step $k$ can still be communicated at $k'>k$ using either available communication technology as long as the packets did not reach their discard time.
We have a product of two variables, the \textit{main} decision variables $x_{i,j,m,k}$ and the \textit{endogenous} variable $y_{i,j,f}$. \textcolor{black}{ We linearize the product of binary and continuous variables in ~\eqref{eq:star} using the McCormick~\cite{McCormick1976} linearization by introducing a new \textit{auxiliary}  non-negative integer variable denoted by $\tilde{y}_{i,j,m,f,k}$ such that $\tilde{y}_{i,j,m,f,k} = x_{i,j,m,k} \, \left( \Gamma_{i,j}^f - y_{i,j,f} \right)$}. 
\vspace{-0.5cm}
 \subsubsection{Physical and Link Feasibility}
Despite the inherent assumption that can be made in the communication matrix $\bold{P}$ (or in $\bold{V}^m$) that denies self-communication, the following constraint explicitly enforces that rule within the optimization model.
\begin{equation}
\begin{split}
    x_{i,i,m,k} = 0, \quad \forall i , \forall m, 
 \forall k 
 \end{split}
  \label{const4} \end{equation}
  
If a device opts to communicate using a communication technology, the communication must be possible. To this end, we set a pre-defined threshold $SNR^m_{min}, m \in \mathcal{M} $ that determines the minimum SNR value that must be satisfied by the visibility matrix $\bold{V}^{m}$ for the communication to be considered successful, which is represented by:
\begin{equation}
\begin{split}
  SNR^m_{min} - U
 (1 - x_{i,j,m,k} ) \leq SNR^m_{i,j,k}, \\ \quad \forall i, j, \forall m, \forall k 
 \end{split}
  \label{const3} \end{equation}
with $U$ a large number that accounts for the SNR values less than $SNR^m_{min}$.

In order to prevent devices from exceeding the energy budget allocated to their communication technology, we introduce the following constraint:
    \begin{equation}
    \begin{split}
     \!\! \!\sum_{m \in \mathcal{M}}  \sum_{j \in \mathcal{N}} \!\! \!  \!\! \!\!\!\!\sum_{\substack{k \in \phi_{i,j}^{f,l} \\ f \in \{1, \ldots, |\Phi_{i,j}|\}}} 
  \!\!\!\!\!\!\!\!\Bigg[ \tilde{y}_{i,j,m,f,k} \, \xi^m_s  + \tilde{y}_{j,i,m,f,k} \, \xi^m_r  \Bigg ]
       \leq \frac{\xi_i}{s_p},  \forall i.
    \end{split}  \label{const9} \end{equation}

\textcolor{black}{The previous conditional constraints \eqref{const11}, \eqref{const14}, and \eqref{const10} are inherently nonlinear, but we can approximate and enforce these conditions in linear programming by using the Big-M method~\cite{articless}.}

\begin{figure*}[t]
\vspace{-0.1cm}
\begin{center}
\includegraphics[width=1\textwidth]{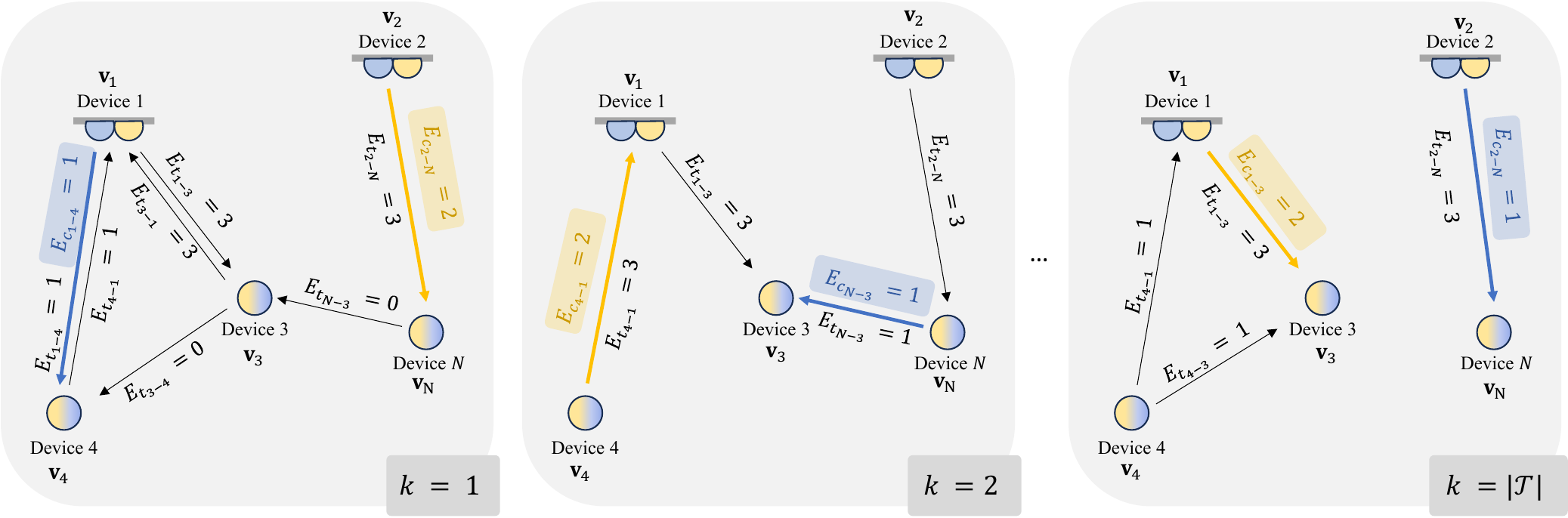}
\end{center}
\caption{An example of a graph modeling for a network with $N$ devices over a period of time $\mathcal{T}$. $E_{t_{i-j}}$ represents the allowed communication links from device $i$ to device $j$. $E_{c_{i-j}}$, on the other hand, represents the chosen communication technology for that link. For illustration, we have omitted some of the communication links that cannot be established from a device $i$ to a device $j$ to showcase the environment visibility  (i.e., $E_{t_{i-j}}=0$ ). The edges in yellow/blue represent the chosen links and their value (e.g., $E_{c_{2-3}}=2$ refers to OWC) at each time $k \in \mathcal{T}$.\vspace{-0.4cm} }
\label{illustrativeexmaple}
\end{figure*}

\subsection{Problem Formulation}
Given a hybrid RF-OWC IoT network with devices engaged in exchanging messages over discrete time steps, the objective function is designed to maximize the overall throughput in the network with respect to energy consumption while minimizing the average packet delivery.  
\textcolor{black}{Rather than optimizing the time-average AoI state process, which is intractable in multi-source networks, we minimize the age of each update at its delivery time $\delta_{i,j,f}$. This surrogate eliminates temporal state coupling and yields an additive, tractable objective.
} This bi-objective optimization can be formulated as follows:
\begin{align}
\centering
     \text{P1:} \quad & \underset{x,\delta, \tilde{y}}{\min}
    \hspace{0.1cm}\alpha_1 
    \frac{\sum_{i,j,f} \delta_{i,j,f}}{S_1}
      - \alpha_2 
    \frac{\sum_{i,j,m,f,k} \frac{ \tilde{y}_{i,j,m,f,k} } {( \xi_r^m + 
        \xi_s^m )} }{S_2}, \label{P1} \notag \\ 
    &   \text{subject to: } \notag 
\eqref{const1}-\eqref{const9}. \notag
\end{align}

In (P1), the terms $S_1$ and $S_2$ are normalization coefficients that take the maximum possible value for each sub-objective. However, such normalization alone often proves insufficient as certain terms can disproportionately dominate the optimization process or cause solution saturation. This issue arises because each term has a different sensitivity to the input change, causing some to inherently carry more weight even when normalized. Therefore, to mitigate the scaling issue between the terms, we introduce the regularization factors $\alpha_1$ and $\alpha_2$, where $\sum_{i=1}^{2} \alpha_i =1$.

\textcolor{black}{The initial formulation has nonlinear constraints (e.g., Eq.~\eqref{const10} \& Eq.~\eqref{const11}) and a non-convex term (e.g., Eq.~\eqref{eq:star}) defining a conceptual MINLP.  Applying the McCormick and Big-M reformulations yields an equivalent MILP, where the objective and all constraints are linear. Although the continuous relaxation is convex, the presence of binary variables makes the feasible region non-convex and the problem combinatorial. Such problems are known to be NP-hard~\cite{articlesssss}. 
State-of-the-art solvers, such as branch-and-bound or branch-and-cut~\cite{MORRISON201679}, have a worst-case runtime that grows exponentially with a time complexity of $\approx O\left[2^{\left(|\mathcal{N}|^2 \times |\mathcal{M}| \times |\mathcal{T}|\right)}\right]$, which is computationally expensive for large-scale IoT networks. Nevertheless, the MILP formulation serves as a performance oracle to generate globally optimal scheduling labels and to benchmark the learning-based approach. In the next section, we introduce our model DGET that relies on graph embedding techniques along with the Transformer-based architectures to learn the devices' scheduling timetable and technology usage at each time with lower overhead. }

\vspace{-0.2cm}
\section{Proposed Graph Transformers Network for RF-OWC IoT Networks}
\label{sec5}
In this section, we present the designed DGET framework for resource allocation in hybrid RF-OWC IoT networks. 
We begin the section with the graph modeling approach and explain how the hybrid RF-OWC IoT network is modeled into temporal snapshots. We then introduce the dual embedding framework via GNNs\footnote{Readers that are not familiar with GNNs are encouraged to check~\cite{
9335498,8995280,hamilton2018inductiverepresentationlearninglarge}.} and present the Transformer-based classifier that operates on the generated embeddings to enable resource allocation decisions. Finally, we detail the customized learning strategies that enhance model learning and prediction robustness.

\subsection{Graph Modeling}

As an alternative to solving the computationally expensive problem (P1), the RF-OWC hybrid IoT network can be modeled as a temporal directed graph $\mathcal G_{k}(\mathcal V,\mathcal E), k \in \mathcal{T}$, where the vertices represent the total IoT devices (ie, $|\mathcal V|=|\mathcal{N}|$, including IoT nodes and APs) and the edges denote the hybrid communication links at any time step $k \in \mathcal{T}$.  An example illustrating the concept is provided in Fig.~\ref{illustrativeexmaple}. Since the graph is directed and self-loops are disallowed (i.e., a device cannot communicate with itself), the total number of potential edges is $|\mathcal{E}| = |\mathcal{N}| \, (|\mathcal{N}| - 1)$.

To capture the time-varying nature of the network, we define two types of time-series graph snapshots:  $\mathcal G^{I}_{k}(\mathcal V,\mathcal E)$ and $\mathcal G^{R}_{k}(\mathcal V,\mathcal E)$. The first graph, $\mathcal G^{I}_{k}(\mathcal V, \mathcal E)$, which will be referred to as \textit{input graph}, captures the state of the environment at time $k$ and includes prior knowledge of the network including allowable communication technologies between devices, initial energy levels, and feasibility and quality of links.
In contrast, the second graph, $\mathcal G^{R}_{k}(\mathcal V, \mathcal E)$, which will be referred to as \textit{recorded graph}, represents the resulting state of the system after communication activities have taken place. This graph reflects communication updates such as exchanged messages, link selections, and energy depletion resulting from actual communication activities. These states depend on the scheduling decisions made by solving the optimization problem formulated in Section~\ref{sec4}.
\begin{figure*}[t]
\begin{center}
\includegraphics[width=1\textwidth]{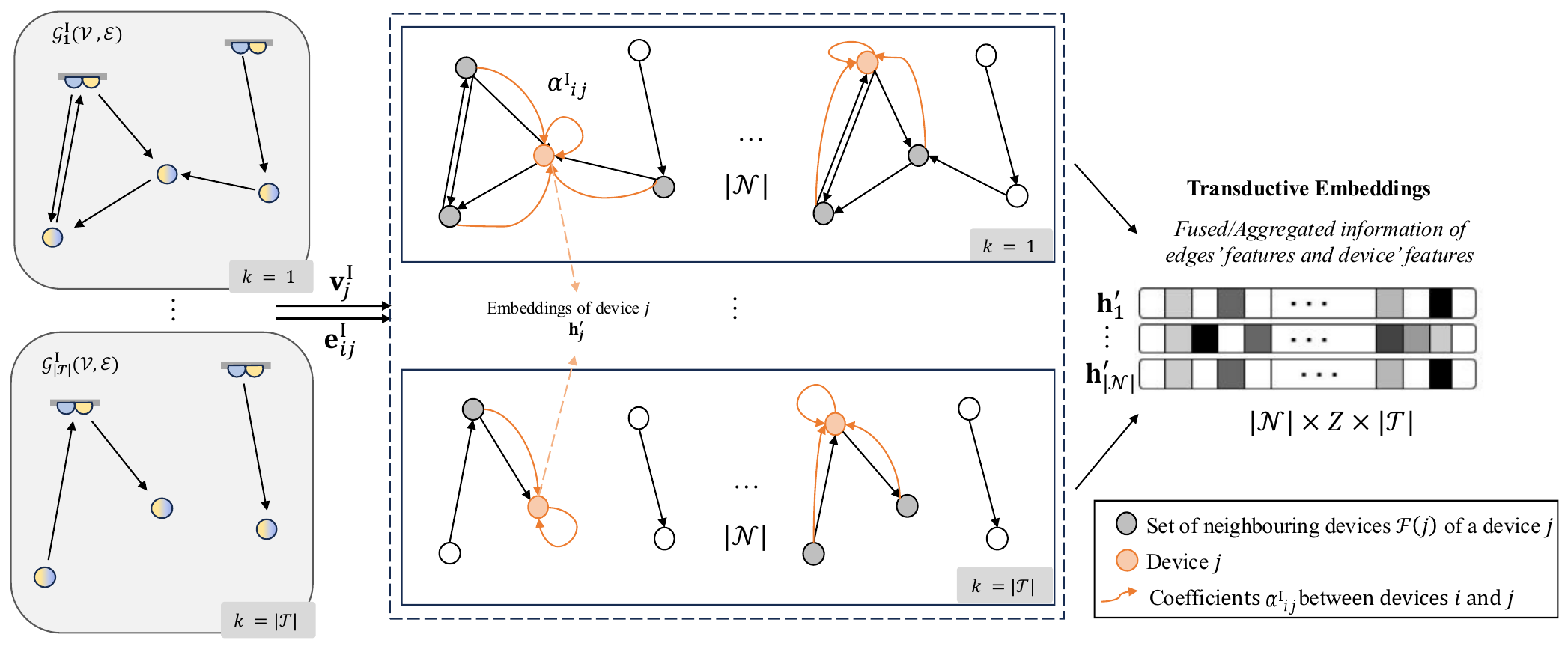}
\end{center}
\caption{Visualization of device-link transductive embeddings from the network graph, resulting in a numerical representation for each device in the graph. Each temporal input graph $\mathcal{G}^{\text{I}}_{k}(\mathcal V,\mathcal E)$ is transformed into embeddings with a size $ |\mathcal{N}| \times Z$, with $Z$ the embedding dimension.\vspace{-0.3cm}}
\label{gnnworkflow}
\end{figure*}

For the input graph, each device $i \in \mathcal{V}$ is associated with a feature vector $\bold{v}^{\text{I}}_i$, which encodes its type and local state information (e.g., available packet for transmission, initial energy level, energy per bit consumption, total outgoing packets, etc.). In the recorded graph, the devices' features are referred to as $\mathbf{v}^{\text{R}}_{i}$, which contain $\mathbf{v}^{\text{I}}_i$ but also the parameters evolution over time to reflect communication decisions made at each step (e.g., energy reduction due to data transmission).  Similarly, the input graph includes a link feature vector from device $i$ to device $j$, denoted as $\mathbf{e}^{\text{I}}_{ij}$. The latter includes the prior knowledge link-specific information, such as the per-edge queue of messages, their types, and the link quality. The edges feature vector $\mathbf{e}^{\text{I}}_{ij}$ contains the possible communication links $\mathcal{E}_t = \{0, 1, 2, 3\}$, which encodes the allowed communication technologies between two devices: 0 means no communication is possible, 1 allows only RF, 2 allows only OWC, and 3 allows either RF or OWC. This information is prior knowledge of both the input and the recorded graphs.   In contrast, the recorded graph includes a link feature vector from device $i$ to device $j$, denoted as $\mathbf{e}^{\text{R}}_{ij}$, which mirrors the input graph link feature and also includes the ground truth link status $\mathcal{E}_c = \{0, 1, 2\}$. The latter indicates the chosen communication links at a given time, with  0 denoting no communication occurred, 1 denoting RF communication was selected, and 2 denoting communication was via OWC. In this context, we note $\mathcal{E}_{t_{i,j}}$ as the permissible communication technologies from device $i$ to device $j$, while $\mathcal{E}_{c_{i,j}}$ indicates the actual communication technology selected for transmission from device $i$ to device $j$. Both the recorded device feature $\mathbf{v}^{\text{R}}_{i}$ and the link's status $\mathbf{e}^{\text{R}}_{ij}$ are obtained as a result of solving (P1) in Section~\ref{sec4} and recording the network parameter evolution.  By definition, the selected communication status $c \in \mathcal{E}_c$ must satisfy $c \leq f$ for any edge, with $f \in \mathcal{E}_t$, ensuring the chosen technology complies with the available link constraints.

This modeling allows us to shift the resource allocation problem in the IoT network to predicting the value of $c \in \mathcal{E}_c$ for each pair $i,i'$ in the graph $\mathcal G^{I}_{k}(\mathcal V,\mathcal E)$ at any given time step $k \in \mathcal{T}$. We reformulate the task as a classification problem acting on the input graph. Specifically, we aim to predict the appropriate communication status $c \in \mathcal{E}_c$ for each edge in the input graph $\mathcal{G}^{I}_k(\mathcal{V}, \mathcal{E})$ at each time step $k$ by establishing a learning process from the information in the recorded graph.    We encode topology, traffic availability, link feasibility, and pre-decision link quality on edges, with static device priors on nodes. Known graphs contain only pre-decision information, while recorded graphs mirror those features and add post-decision information, such as energy changes and chosen technology.

\vspace{-0.1cm}
\subsection{Device-Link Embedding using GNN}

Once we build the time-series graphs $\mathcal G^{\text{I}}_{k}(\mathcal V,\mathcal E)$ and $\mathcal G^{\text{R}}_{k}(\mathcal V,\mathcal E)$,  we proceed with embedding the device and edges into a latent space using a two-stage GNN pipeline that combines transductive and inductive learning. The first stage applies a transductive GNN on the input graph $\mathcal G^{\text{I}}_{k}(\mathcal V,\mathcal E)$ to generate an initial set of devices' embeddings.  The transductive GNN operates over fully observed training graphs and has access to the entire graph topology and feature distribution. This allows it to generate globally consistent and context-aware embeddings with local and global interactions present in the input graph.  In the second stage, we employ an inductive GNN to enrich the transductive representation by learning decision patterns from the recorded graph $\mathcal G^{\text{R}}_{k}(\mathcal V,\mathcal E)$, with the latter acting as a supervised representation. In other words, the inductive GNN maps the transductive embeddings into richer and semantically aligned embeddings that resemble those obtained from the  recorded graph snapshots given by $\mathcal G^{R}_{k}(\mathcal V,\mathcal E)$. The inductive GNN acts as a semantic enhancer to refine and enrich transductive embeddings  by learning how to incorporate the information seen only during training, such as the actual communication decisions and energy updates.

The key goal for transductive GNN is to produce rich embeddings of the graph while incorporating the features of the devices and their relationships at specific points in time. Furthermore, we are interested in extracting network information from the vertices, i.e., devices, and also from the edges, i.e., links. Therefore, we propose to leverage a mixture of vertex-edge embedding for each step. The mapping function from the graph to vectors is represented by a component that captures the vertices' features and the respective links with their features. Graph Attention Networks (GATs)~\cite{veličković2018graphattentionnetworks}~ are particularly well-suited for this task, as their attention mechanism adaptively weighs the influence of neighboring IoT devices based on both their features and the type or quality of the connecting link~\cite{Chen2023}. 
\textcolor{black}{In the transductive setting, where the full network snapshot is observable, attention enables the model to learn discriminative structural priors by emphasizing technologically relevant RF-OWC interactions, such as feasible OWC links or energy-efficient RF paths, while down-weighting less informative or infeasible connections.  This is critical at the transductive stage, where the goal is to extract stable structural priors that reflect the topology of the network rather than treating all neighbors uniformly. Building on this, we extend the attention formulation to also incorporate edge attributes so that the model can differentiate between heterogeneous RF-OWC connections. We define the following aggregation score \( A_{ij}^{\text{I}} \) between two devices \( i \) and \( j \):
\begin{align}
A_{ij}^{\text{I}} = \sigma \left( \beta (\mathbf{e}^{\text{I}}_{ij}) \, \mathbf{a}^\top \,  \textit{Conc}   \left( \bold{W} \,  \mathbf{v}^{\text{I}}_{i} , \bold{W} \, \mathbf{v}^{\text{I}}_{j}  \right)  \right), \\ \forall \, (i,j)\in \mathcal{N} \times \mathcal{N}, \notag
\end{align}
where $ \mathbf{v}^{\text{I}}_{i}$ and $\mathbf{v}^{\text{I}}_{j}$ are the feature vectors of devices $i$ and $j$ from the input graph $\mathcal G^{\text{I}}_{k}(\mathcal V,\mathcal E)$, $\bold{W}$ is a learnable weight matrix, \textit{Conc}$(,)$ denotes the concatenation function, $\mathbf{a}$  is also a learnable attention vector, $\sigma$ is a nonlinear activation function, and $\beta$ the Min-Max normalization. For ease of explanation, we have included a high-level visualization in Fig.~\ref{gnnworkflow} highlighting the transductive embedding process. The inclusion of the normalized $E_{t{_{ij}}}$ ensures that the importance of the connection between devices $ i$ and $j$, described by the possible communication technology in this case, is adjusted according to the weight of the edge.  Then, the normalized  coefficient \( \alpha_{ij}^\text{I} \) for the input graph is computed via a softmax function:
\begin{align} 
&\alpha_{ij}^\text{I}  = \frac{\exp(A_{ij}^{\text{I}} )}{\sum_{n \in \mathcal V} \exp(A_{in}^{\text{I}} )},  \forall \, (i,j) \in \mathcal{N} \times \mathcal{N}. 
\end{align}
Finally, the updated feature vector for device \( i \) is computed as:
\begin{align}
&\mathbf{h}_i' = \sigma \left( \sum_{j \in \mathcal{F}(i) } \alpha_{ij}^\text{I}  \, \mathbf{W'} \, \mathbf{v}^{\text{I}}_{j} \right),  \forall \, i\in \mathcal{N},
\end{align}
where $\mathcal{F}(i)$ the set of neighbor devices to $i$ and $\bold{W'} $ is a learnable weight matrix. }

\textcolor{black}{
While GATs employ parameter sharing, allowing their attention weights to be reused across compatible graphs, this property alone does not ensure meaningful generalization in dynamic environments. In a purely transductive setting, the learned embeddings are optimized for a fixed and fully observed graph, capturing specific node features, link states, and connectivity patterns. As a result, when the network topology or device attributes (e.g., energy levels, queue lengths, or hybrid RF-OWC link conditions) evolve over time, these embeddings must be recomputed and cannot directly transfer to unseen configurations. The inductive GNN in DGET addresses this limitation by learning a generalizable aggregation function that adapts the transductive embeddings to dynamically changing graphs.} The inductive GNN first acts on the recorded graph $\mathcal G^{\text{R}}_{k}(\mathcal V,\mathcal E)$, where the goal is to embed the recorded graph into a supervised representation. In other words, inductive embeddings learn to enrich and refine the transductive embeddings given by the input snapshots $\mathcal G^{\text{I}}_{k}(\mathcal V,\mathcal E)$ while relying on the recorded graph embedding $\mathcal G^{\text{R}}_{k}(\mathcal V,\mathcal E)$ to generalize beyond the transductive learned patterns.
To achieve this, we draw inspiration from the GraphSAGE framework\cite{hamilton2018inductiverepresentationlearninglarge} and incorporate attention-based neighbor weighting with explicit edge features into the aggregation process. \textcolor{black}{Unlike vanilla GraphSAGE, which aggregates only node embeddings, our inductive GNN design concatenates edge attributes through a skip connection and modulates neighbor contributions based on the known input states.} The output of the inductive embedding at the layer $p$  is denoted by  $\mathbf{o}_i^{p}$ and expressed as follows: 
\begin{align} 
& \mathbf{o}_i^{p} = \sigma \left( \sum_{j \in \mathcal{F}(i)} \alpha_{ij}^\text{I}  \, \mathbf{W}^{p}  \, \mathbf{o}_j^{p'} \right), \forall \, i\in  \mathcal{N}, \notag  \\ 
 &\mathbf{o}_j^{p'} = \sigma \left( \mathbf{W}^{p'} \,  \textit{Agr} \left(  \textit{Conc}(\mathbf{o}^{p-1}_i, \mathbf{{e}}^{\text{I}}_{ji}) \right) \right), \forall i \in \mathcal{F}(j)\,, 
 \label{eqInductive}
\end{align}
\textcolor{black}{where $\mathbf{o}^{p-1}_j$ is the embedding of the device $j$ in the $p-1$ layer obtained by following the chain equation in \eqref{eqInductive}, $ \textit{Agr}(,)$ is a general neighborhood aggregation function (e.g., mean, pooling, or LSTM-based), and  both $ \bold{W}^{p}$ and $ \bold{W}^{p'} $ are learnable weight matrices for layer $p$.} The initialization value of $\mathbf{o}^{0}_j$ is obtained from the transductive embedding $h'_i$. The inductive GNN is also provided with the edges' features $\mathbf{{e}}^{I}_{ji}$ through a skip connection, which acts as an informational anchor that ensures the updated devices' embeddings remain consistent with the link-level constraints defined in the input graph.
The obtained final embedding size $ \mathbf{o}_i^{p}$ is $|\mathcal T| \times Z$, where the rows represent the time dimension and $Z$ is the embedding dimension.



\begin{figure*}[t]
\includegraphics[width=1\textwidth]{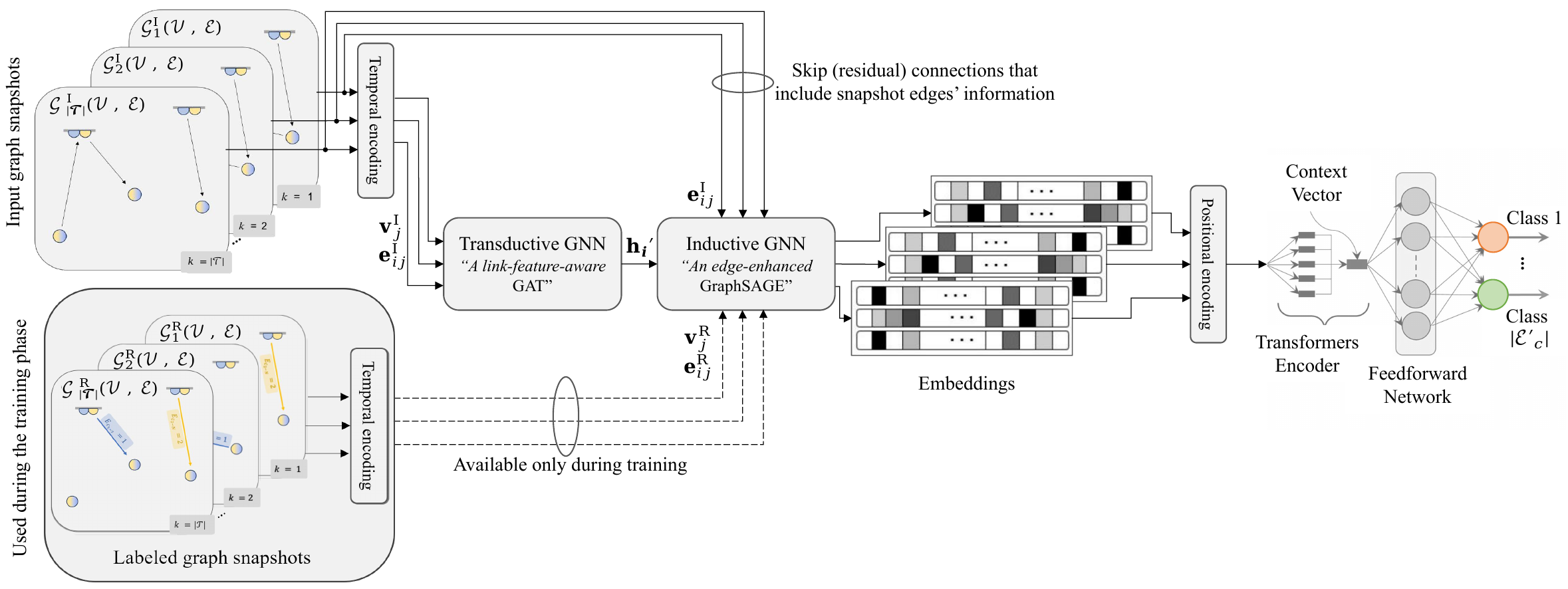}
\caption{\textcolor{black}{High-level architecture illustrating the different components of the DGET model. \vspace{-0.4cm}}}
\label{GTNdetails}
\end{figure*}

\vspace{-0.2cm}
\subsection{Consistency Embeddings}

In the previous section, we discussed how the input time-series graph $\mathcal G^{I}_{k}(\mathcal V,\mathcal E)$ is transformed to embeddings by encoding vertex features and edge attributes through a series of transductive and inductive GNNs. In this section, we explain the overall flow of the DGET framework. For clarity, we have included an illustration of the high-level architecture in Fig.~\ref{GTNdetails}.

The transductive GNN operates over fully observed graphs and has access to the entire graph topology and feature distribution from the input graph $\mathcal G^{I}_{k}(\mathcal V,\mathcal E)$. These embeddings are rich in structural priors and capture local and global interactions present in the graph. Once the transductive GNN has generated embeddings from the input graph $\mathcal G^{I}_{k}(\mathcal V,\mathcal E)$, we pass them to an inductive GNN in the subsequent layer. The inductive GNN focuses on generalizing from the learned embeddings by using local neighborhood aggregation, which allows it to capture temporal variations in device and link features during inference.  The inductive model uses these initial embeddings from input snapshots as starting points, and through neighborhood-based computations, as shown in the previous subsection, it attempts to reconstruct or refine the embeddings by learning patterns from recorded snapshots $\mathcal G^{R}_{k}(\mathcal V,\mathcal E)$.

To ensure these generated embeddings faithfully represent the ground-truth embeddings obtained from the recorded graph, we use a consistency loss function\cite{kim2021structuredconsistencylosssemisupervised} that optimizes embedding accuracy by measuring the similarity between predicted embeddings from the input graph $\mathcal G^{I}_{k}(\mathcal V,\mathcal E)$ and the true embeddings $\mathcal G^{R}_{k}(\mathcal V,\mathcal E)$ from the recorded graph, which are only observed during training. The loss aims to align reconstructed embeddings with actual graph $\mathcal G^{R}_{k}$ embeddings, minimizing error across time steps. This loss is represented as follows: 

\begin{align}
\bold{L}_{\text{consistency}} = \frac{1}{|\mathcal N|} \sum_{k \in \mathcal{T}, j \in \mathcal{V}} \| \bold{o}_{j}^{p}(k) - \bold{g}_{j}^{p}(k) \|^2_2 \,,
\end{align}
where $\bold{o}_{j}^{p}(k)$ is the final layer $p$ predicted embedding for device $j$ at time $k$ and  $\bold{g}_{j}^{p}(k)$ is the final layer $p$ true embedding for device $j$ at time $k$. \textcolor{black}{The true embeddings  $\bold{g}_{j}^{p}(k)$ are obtained using \eqref{eqInductive} acting on the recorded graph  $\mathcal G^{R}_{k}(\mathcal V,\mathcal E)$, therefore $\mathbf{v}^{R}_i$ and $\mathbf{e}^{R}_{ij}$, and while recomputing the attention $\alpha_{ij}^\text{R}$ to capture not only the initial state but also evolving node and link states. Attention is applied in both GNN stages but serves different purposes. The transductive GNN uses attention to encode stable structural importance from fully observed network snapshots. The inductive GNN re-applies attention to dynamically re-weight neighbors to adapt to dynamic changes in node energy, queue occupancy, link availability, and evolving connectivity patterns.} 

The consistency loss aims to minimize the discrepancy between the true observation, as a reference, on the recorded graph and the initial learned representation from the input graph by applying the mean squared error over all time steps.  To capture temporal information, we include a temporal encoding to each time snapshot graph, distinguishing each of their order in time. Temporal encoding enables the model to learn the evolution or time-based interactions between devices or edges.  We should note that the inductive GNN operates in two modes.  During training, the inductive GNN receives  embeddings from the transductive GNN and the recorded graph with ground-truth outcomes. It learns to map the transductive embeddings to richer, semantically aligned embeddings using the consistency loss that penalizes discrepancies between reconstructed embeddings, from the input graph, and true embeddings, from the recorded graph. At inference, the inductive GNN uses only transductive embeddings and transforms these embeddings into enriched representations that capture the learned decision patterns and semantic structures.

 \vspace{-0.1cm}
\subsection{Transformer-based Classification}

To classify the communication links over time, we employ a Transformer-based classifier.  This classifier utilizes an attention mechanism that operates on the sequence of devices' embeddings generated by the inductive GNN, allowing the model to weigh and select relevant information from each time step's embedding based on the relationships between embeddings across time. Positional encoding is applied to the generated graph embeddings to allow the Transformer-based classifier to be aware of the structure or relative position of embeddings input\cite{vaswani2023attentionneed}.  The Transformer block consists of multiple layers of multi-head self-attention and feedforward neural networks.  
The final layers are a feedforward neural network that yields, through a custom weighted loss function, the predicted class per link at time step $k$. In other words, the classifier outputs which devices will communicate with which devices and using which technology and at which time step $k \in \mathcal{T}$.  The proposed DGET architecture leverages the strengths of both graph-based and sequence-based learning to provide a solution for predicting the scheduling of communication between the devices, along with the type of medium used. The classification  loss function is a negative log-likelihood loss represented as follows:
\begin{equation}
\begin{aligned}
\bold{L}_{\text{classification}} =  - \frac{1}{|\mathcal N|} \sum_{i \in \mathcal{N}}^{} \log \left [ p(y_i | x_i) \right],
\label{clasificationloss1}
\end{aligned}
\end{equation}
where  \( y_i \) is the true label for the \( i \)-th sample, \( x_i \) is the input for the \( i \)-th sample, and
 \( p(y_i | x_i) \) is the predicted probability of the true class for the \( i \)-th sample.

The final loss function for the overall architecture is represented as follows: 
\begin{align}
 \bold{L} =   \bold{L}_{\text{classification}} + \lambda  \bold{L}_{\text{consistency}},   
\end{align}
with  \( \lambda \) is a hyperparameter that controls the balance between the classification and consistency losses.
 \vspace{-0.1cm}
\subsection{Enhanced Learning Strategies and Post-processing}
The graph $\mathcal G^{R}_{k}(\mathcal V,\mathcal E)$ has, at each time step, different $|\mathcal E|=|\mathcal{N}| \, (|\mathcal{N}|-1)$ edges that reflect the scheduling between the devices at each time. Therefore, at any time $k \in \mathcal{T}$, there can be at most  $\left\lfloor \frac{|\mathcal{N}|}{2} \right\rfloor$ non-zero class edges. This is because of the constraints that each time step, a device can communicate with only another device.  Since $ \left\lfloor \frac{|\mathcal{N}|}{2} \right\rfloor \ll |\mathcal{N}| \, (|\mathcal{N}|-1) $, the vast majority of edge labels correspond to the no-communication class (i.e., class zero), resulting in a highly imbalanced class distribution. To address this challenge and improve classification performance, we propose a three-fold refinement and stability strategy.

The first fold involves augmenting the edge class space by correlating the communication class $ \mathcal{E}_c $ with the residual communication features $\mathcal{E}_t$, as shown in Fig.~\ref{GTNdetails}. Originally, the classifier outputs $3$ classes (i.e., $ \mathcal{E}_c = \{0,1,2\} $), representing the communication type (or no communication). However, by combining these with edge features $ \mathcal{E}_t = \{0,1,2,3\} $, we expand the space to $ 8 $ augmented classes $ \mathcal{E}_c' = \{0,1,2,3,4,5,6,7\} $. Each class in $ \mathcal{E}_c' $ reflects a joint event: the communication feasibility and the classifier's prediction. For example, $ \mathcal{E}_c' = 0 $ means no communication was possible and correctly predicted as such; $ \mathcal{E}_c' = 1 $ means communication via RF was possible but not predicted; and $ \mathcal{E}_c' = 2 $ means communication via RF was both possible and predicted. This conditional labeling improves representation learning and balances the dataset by leveraging hidden semantic patterns in the edge features.
\begin{table}[t]
\centering
\caption{Simulation parameters for RF and OWC Systems.}
\label{tab:sim-setup}
\begin{tabular}{|l||l|}
\hline
\textbf{Parameter} & \textbf{Value} \\
\hline
\hline
\multicolumn{2}{|c|}{\textbf{RF System}} \\
\hline
Bandwidth ($B_{\text{RF}}$) & 2 MHz \\
Transmit Power ($P_t^{\text{RF}}$) & 50 mW \\
Path Loss Exponent $\alpha$ & 3 \\
$PL_0$ & 40 dB \\
$G_t$ = $G_r$ & 5 dBi \\
RF SNR Threshold ($\text{SNR}^0_{\text{min}}$) & 15 dB \\
\hline
\multicolumn{2}{|c|}{\textbf{OWC System}} \\
\hline
Bandwidth ($B_{\text{OWC}}$) & 1 GHz \\
Transmit Power ($P_t^{\text{OWC}}$) & 5 W \\
Optical Transmission Coefficient ($T_{\text{Opt.}}$) & 0.99 \\
Lambertian order ($u$) & 1 \\
Illumination Coefficient ($\kappa$) & 0.3 \\
Incidence and Irradiance angles ( $\phi$, $\theta$) & $60^{\circ}$ \\
Photodiode Effective Area ($A_{Rec}$)  & $0.0025$ m$^2$ \\
 Photodiode Responsivity ($\eta_r$) & 0.6 A/W \\
  Background Light power ($P_{n}$) & $10^{-9}$ W \\
 OWC SNR Threshold 
 ($\text{SNR}^1_{\text{min}}$) & 10 dB \\
\hline
\multicolumn{2}{|c|}{\textbf{DGET Parameters}} \\
\hline
Number of GATs layers  & $64$  \\
Number of GATs Attention heads & $4$  \\
Number of GraphSAGE layers  & $64$  \\
Embeddings Dimension of GraphSAGE  & $256$  \\
GraphSAGE Aggregation  $(\textit{Agr})$ & \textit{mean()}  \\
Transformer encoder layers & $6$  \\
Number of Transformer attention heads & $8$  \\
Transformer Embedding size  & $256$  \\
Dataset Split Train/Val-Test & $80\%$:$20\%$ \\
Cross-Validation & Stratified $10$-fold  \\
Fold Train-Val & $7$:$3$ \\
Epochs per Fold & $70$ \\
Regularization & $L_2$  \\
Dropout Rate & $0.3$ \\
Weight Initialization & Xavier initialization \\
Activation Function ($\sigma$) & LeakyReLU \\
\hline
\multicolumn{2}{|c|}{\textbf{System Parameters}} \\
\hline
Devices' Distance Range & $\mathcal{U}(3, 20)$ m  \\
Sensitivity Coefficients $\alpha_1$, $\alpha_2$ & $0.35$, $0.65$  \\
\hline
\end{tabular}
\vspace{-0.3cm}
\end{table}

The second fold is to leverage a weighted loss function that gives more focus to under-represented classes (i.e., those that appear less frequently) and less importance to over-represented classes. The classification loss presented in~\eqref{clasificationloss1} is therefore updated to a negative log-likelihood loss with weighted classes, and it is represented as follows:
\begin{equation}
\begin{aligned}
\bold{L}_{\text{classification}} =  - \frac{1}{|\mathcal N|} \sum_{i \in \mathcal{N}}^{} \left( \sum_{c=0}^{C-1} w_c \, y_{ic} \, \log(\hat{y}_{ic}) \,  W_{c, \hat{y}_i} \right), \label{Lclassification}
\end{aligned}
\end{equation}
where $w_c$  is the inverse-frequency weight for class $c$, $y_{ic}$ is the true label indicator, $\hat{y}_{ic}$  is the predicted probability for class $c$, and $ W_{c,\hat{y}_{i}}$ is the penalty weight that applies when the true class is $c$ and the model predicts class $\hat{y}_{i}$. The matrix $ \bold{W}_{c, c'}$, where $c,c' \in \mathbb{R}^{C \times C}$ represents a penalty factor that adjusts the loss based on the predicted class to allow penalizing certain misclassifications. The weights $w_c$ for each class are generally calculated based on the inverse of their frequency in the dataset.

The third fold incorporates a post-processing correction phase that ensures the feasibility of the predicted scheduling. After prediction, we perform a constraint check to detect violations (e.g., a device communicating with multiple others at the same time step \( k \in \mathcal{T} \)). If infeasible predictions are found, we perform a top-2 accuracy search: the second-highest class from the classifier's output distribution is selected, and the feasibility is re-evaluated. The rationale is that most errors occur between semantically adjacent classes (e.g., \( \mathcal{E}_c' = 1 \) vs. \( \mathcal{E}_c' = 2 \)), due to the structured residual encoding introduced by the augmented labels and class penalty design $\bold{W}_{c, c'}$.  Indeed, this post-processing approach is more effective when the initial classifier solution results in more communication between devices than is permissible. Therefore, we adapt the weighted loss function in~\eqref{Lclassification} to attribute high penalization to the nearby classes except for the classes that result in over-predicting. This leads to a powerful elimination of false positives and reduces the number of false negatives. After that, and to backtrack and eliminate non-allowed communication, we create a mapping matrix with the indices of the feasible set in the solution vector. Let this mapping be the matrix $\bold{M}_p$ with size $\frac{|\mathcal{N}| \,  (|\mathcal{N}|-1) }{2}$ and $(|\mathcal{N}| \,  (|\mathcal{N}|-1))-((|\mathcal{N}|-2)\,(|\mathcal{N}|-3)) $. For clarity, we can take the example of a network of $|\mathcal{N}|=4$ devices. We have then $|\mathcal{N}| \, (|\mathcal{N}|-1)=12$ potential links between these devices. If we enumerate these links, we have the following arrangements: $\{1-2,1-3,1-4,2-1,2-3,2-4,3-1,3-2,3-4,4-1,4-2,4-3\,\}$. The feasibility matrix has a size of $6 \times 10$ with the rows representing the combination of three devices as $\{1-2,1-3,1-4,2-3,2-4,3-4\}$ and the columns representing the set for each pair of these devices that overlap (i.e., only one of these sets can communicate). In this case, the first column has all the pairs of devices that include device 1 and device 2 which is $\{1-2,1-3,1-4,2-1,2-3,2-4,3-1,3-2,4-1,4-2\}$, whereas the third column has all the pairs of devices that include device 2 and device 3, which is $\{1-2,1-3,2-1,2-3,2-4,3-1,3-2,3-4,4-2,4-3\}$, and so on.

\vspace{-0.15cm}
\section{Experimental Results}
This section outlines the simulation settings and the evaluation metrics used to assess the resource allocation performance across various network scenarios. We compare the performance of hybrid RF-OWC networks against RF networks under diverse conditions.  Furthermore, we benchmark the proposed DGET framework against optimal solutions obtained via the optimization model in (P1) to demonstrate its effectiveness, scalability, and robustness towards partial system parameter knowledge. 

\label{sec6}
\subsection{Simulation Setup and Evaluation Metrics}

\subsubsection{System Parameters and Experimental Setup}

The simulation parameters for RF, OWC, the proposed DGET, and the overall system are included in Table \ref{tab:sim-setup}. All the simulations in this section are achieved using Monte-Carlo with $10,000$ iterations. All algorithms are implemented in a Python $3.10$ environment and run on a 72-socket Intel(R) Xeon(R) Gold 5220 CPU @ 2.20GHz  with 256GB of RAM. For solving the MILP problem, we use the academic CPLEX, an off-the-shelf optimization software package.

\begin{figure}[t]
\centering  \input{figures/EnergyPerbit_TTexchanged1}   \caption{\textcolor{black}{Energy per bit in logarithmic scale (marker) and avg. exchanged packets (dashed) vs. avg. number of packets per device for RF, OWC, and hybrid RF-OWC systems.  \vspace{-0.3cm}}}
        \label{fig12-a}  
 \end{figure}
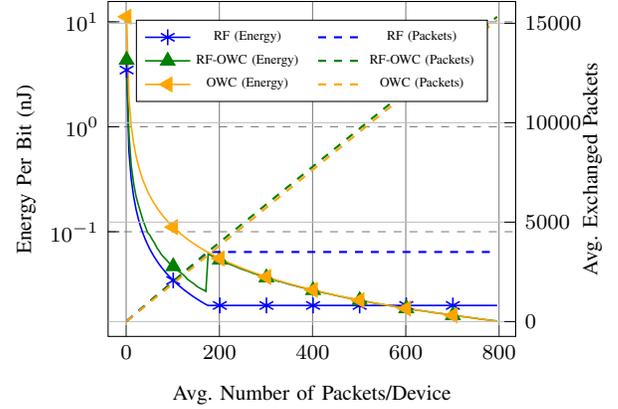
\begin{figure}[t]
\centering   
\hspace{0.28cm}\input{figures/Switching_TTexchanged} 
       \caption{ Avg. technology switching/energy consumption rate per device for hybrid RF-OWC.}\vspace{-0.3cm}%
       \label{fig12-b}
\end{figure}
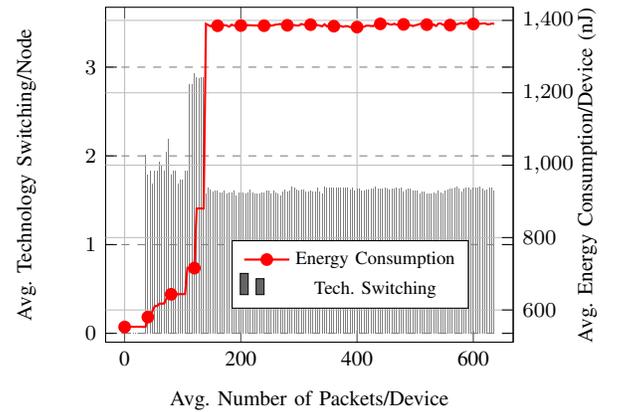

We consider a time step of $\tau = 50$ms and a total simulation time of $10$ time steps, hence $|\mathcal T|=500$ms. The distance between the different IoT devices is randomly generated from $\mathcal{U}(3, 20)$ meters.
The total available energy of all nodes also follows the uniform distribution $\mathcal{U}(0.05, 0.1)$ joules for both RF and OWC technologies. The data arrival process from each device $i$ is modeled as a Bernoulli process, as in \cite{10015118}. Accordingly, we generate the communication matrix $\bold{P}$ such that each entry $\bold{P}_{a,b}$ represents a potential message transmission from device $a$ to device $b$ at a given time step. Each APs-to-node pair is independently sampled from a Bernoulli distribution with probability $0.7$. Similarly, node-to-node communications are generated with a Bernoulli probability of $0.3$. 
Each time slot includes multiple independent packet generation events per device, each following a Bernoulli process.
Messages are constrained to have a maximum duration of 4 time steps, i.e., $\left(\phi_{i,j}^{f,l}\right )_e - \left (\phi_{i,j}^{f,l} \right )_s\leq 4$. 

We model link availability as a two-state Markov process to capture temporal correlation in outages. Each RF and OWC link alternates between an available state and a blocked state. For OWC links, the blocked state represents object-shadowing events with an average duration between 3 and 8 time steps, while for RF links, the blocked state represents congestion-induced interruptions with an average duration between 2 and 6 time steps. The number of available types of data is set to $|\mathcal L|=2$. Relying on the same sensitivity analysis approach from our previous work \cite{Hamr2412:AoI}, we choose the value of $\alpha_1=0.35$ and hence, $\alpha_2=0.65$.

\begin{figure}[t]
\centering
 \subfloat[]{%
       \resizebox{0.235\textwidth}{!}{\input{figures/AoIvs.Nodes}}  
      \label{fig15} }  \hspace{-1cm}
 \qquad
 \subfloat[]{ %
        \resizebox{0.235\textwidth}{!}{\input{figures/AoIvs.Aps}}  
      \label{fig16} }
       \caption{ (a) M-AoI and P-AoI vs. number of IoT nodes when $|\mathcal{N}_{AP}|=3$ and (b) M-AoI and P-AoI vs. number of APs when $|\mathcal{N}_{d}|=5$, for both hybrid RF-OWC and RF systems.\vspace{-0.3cm}}%
        \label{fig:extra}
\end{figure}
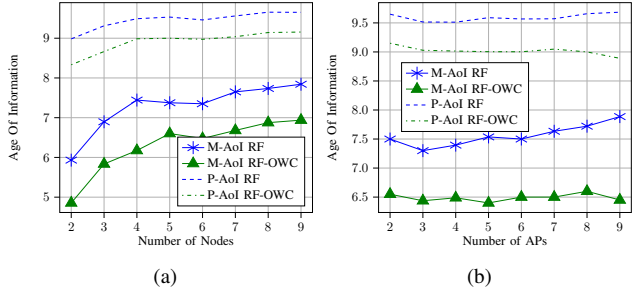

\vspace{-0.1cm}
\subsubsection{Evaluation Metrics}

We evaluate our system using a range of performance metrics that capture both network-level efficiency and model-level learning behavior.

First, we consider the linear AoI~\cite{10000990}, the most widely used and recognized metric to measure the freshness of information. AoI is defined as the difference between the current time and the time at which the last successfully received update was generated at the source~\cite{10032496,10000990}. At time $k \in \mathcal{T}$, if the freshest update received by device $j \in \mathcal{N}$ from device $i \in \mathcal{N}$ was generated at $u(k) = \left (\phi_{i,j}^{f,l} \right )_s$, then the AoI is defined as:
\begin{align}
\Delta(k) = k - u(k)\,.
\end{align}
AoI increases linearly with time and resets upon the reception of a newer update. To evaluate AoI performance, we consider two metrics: the Mean AoI (M-AoI) and the Peak AoI (P-AoI). Given a set of update reception times \( k_1, k_2, \ldots, k_n \), the M-AoI and P-AoI are defined as:
\begin{align}
\text{M-AoI} &= \frac{1}{|\mathcal{T}|} \int_0^{|\mathcal{T}|} \Delta(k) \, dk \,,\\
\text{P-AoI} &= \frac{1}{|\mathcal{N}|} \sum_{i \in \mathcal{N}} \Delta(k_i^-)\,.
\end{align}
M-AoI reflects the average age of information across the network, indicating the system's overall freshness. P-AoI captures the maximum age just before an update is received, offering insight into the worst-case data staleness.

Furthermore, we evaluate the network's reliability using energy-centric indicators, such as total energy consumption and energy per bit, and communication-centric indicators, such as the successful transmission rate and overall network exchanged packets.  For all simulations with OWC technology, the communication-related energy is considered while excluding illumination energy, as reflected by the scaling factor $\kappa = 0.3$.

For evaluating the performance of the DGET model, we monitor the consistency loss, classification loss, and total model loss in training and validation. We also evaluate the GATs embeddings and the reconstructed GraphSAGE embeddings and present the classification accuracy and AoI of the resultant DGET resource allocation. Moreover, we assess the confusion matrix to showcase the reliability of class-wise predictions and perform a complexity analysis for running time to analyze the computational complexity of the different approaches. Finally, we investigate the DGET robustness towards channel imperfections.

\subsection{Hybrid RF-OWC Regime Analysis}

We solve (P1) for a network setup of $|\mathcal{N}_{AP}|=2$ and $|\mathcal{N}_{d}|=3$  under three configurations: RF only, OWC only, and hybrid RF-OWC, and present a comparative analysis in terms of energy efficiency and avg. successful packet exchange, as shown in Fig.~\ref{fig12-a}.  At low traffic levels, RF achieves the best energy efficiency. However, as the number of packets per device increases, RF performance saturates. The RF-OWC system initially offers better energy efficiency than OWC and continues to perform reliably beyond the RF saturation point (around $180$ packets). Although RF-OWC starts with higher energy consumption than RF, it improves steadily and surpasses RF at approximately $600$ packets per device. Overall, RF-OWC demonstrates superior scalability and supports higher traffic volumes while maintaining energy-efficient communication.

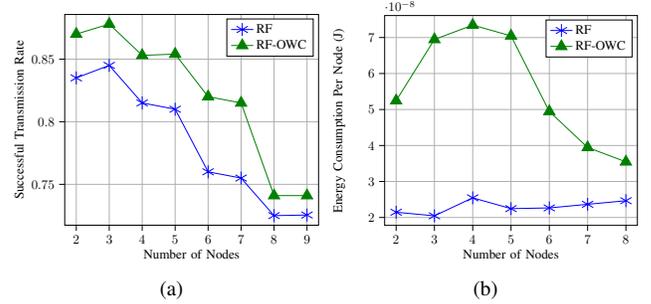
\begin{figure}[t]
\vspace{-0.3cm} \centering
 \subfloat[]{%
       \resizebox{0.235\textwidth}{!}{\input{figures/PDRvs.Nodes_NOWOC}}  
      \label{fig1} }  \hspace{-1cm}
 \qquad
 \subfloat[]{ %
        \resizebox{0.235\textwidth}{!}{\input{figures/Energyvs.Nodes_NOWOC}}  
      \label{fig2} }
       \caption{(a) Successful transmission rate and (b) energy consumption rate per device vs. number of IoT nodes for RF and RF-OWC with $|\mathcal{N}_{AP}|=2$.\vspace{-0.3cm}}%
        \label{fig:images1}
\end{figure}

To further understand the behavior of the hybrid system under varying traffic loads, Fig.~\ref{fig12-b} shows the relationship between the number of technology switches per device and the corresponding energy consumption as the average packet load increases for the same network size setup. At low traffic levels, RF is predominantly used due to its energy efficiency. As the load increases, the system frequently switches between RF and OWC to manage the growing demand. However, beyond approximately ~$170$ packets per device, the number of switches stabilizes, indicating that the system consistently relies on the same technology for communication. This behavior aligns with the observation in Fig.~\ref{fig12-a}, where OWC becomes more favorable under high traffic conditions. In all the subsequent simulations, a default network generation rate of $100$ packets per device is adopted to ensure that both RF and OWC communication modes are actively utilized.

\subsection{Optimization Results}

In Fig.~\ref{fig:extra}, we analyze the Peak Age of Information (P-AoI) and Mean Age of Information (M-AoI) metrics for both RF and RF-OWC systems. In Fig.~\ref{fig:extra}-(a), we fix the number of APs to $|\mathcal{N}_{AP}|=3$ and vary the number of nodes. Both metrics increase as the number of nodes grows, indicating a decline in information timeliness with network scale. Notably, the RF-OWC system achieves approximately $14\%$ lower AoI values compared to the RF-only system. In Fig.~\ref{fig:extra}-(b), we fix the number of nodes to \textcolor{black}{$|\mathcal{N}_{d}|=5$} and vary the number of APs. The M-AoI and P-AoI slightly increase in the RF system with the increase of APs but remain somehow stable in the RF-OWC system. This is attributed to the predominance of node-to-APs traffic and the ability of newly introduced APs in the RF-OWC system to leverage OWC, thereby alleviating RF congestion and preserving overall information freshness.

\begin{figure}[t]
\centering
 \subfloat[]{%
       \resizebox{0.235\textwidth}{!}{\input{figures/PDRvs.APs_NOWOC}}  
      \label{fig1} }  \hspace{-1cm}
 \qquad
 \subfloat[]{ %
        \resizebox{0.24\textwidth}{!}{\input{figures/Energyvs.APs_NOWOC}}  
      \label{fig2} }
       \caption{(a) Successful transmission rate and (b)  energy consumption rate per device vs. number of APs for RF and hybrid RF-OWC systems with $|\mathcal{N}_{d}|=5$.}%
        \label{fig:images3} \vspace{-0.5cm}
\end{figure}
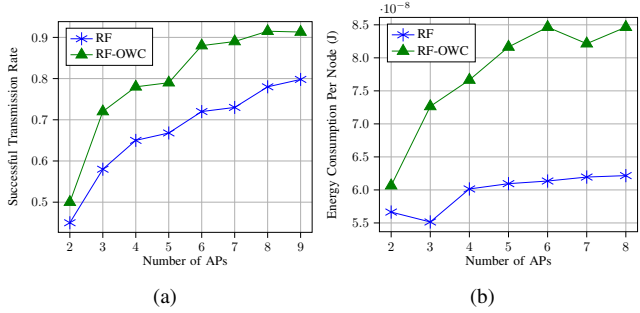

In Fig.~\ref{fig:images1}, we compare the performance of RF and hybrid RF-OWC systems with the number of APs fixed at $|\mathcal{N}_{AP}|=2$. Both systems show a general decrease in successful transmission rate as the number of nodes increases; however, the hybrid RF-OWC system consistently achieves higher success rates. The performance gap between the two systems narrows as the network size grows. Regarding energy consumption, the RF-OWC system initially experiences an increase followed by a decline, indicating that newly added nodes may no longer access the OWC channel, likely due to channel saturation by existing nodes and the limited number of APs. As a result, energy consumption in the RF-OWC system trends closer to that of the RF-only system. Meanwhile, the RF system’s energy consumption remains relatively stable across different node counts.

Fig.~\ref{fig:images3} explores a scenario with a fixed number of $|\mathcal{N}_{d}|=5$ IoT nodes while varying the number of APs. The hybrid RF-OWC system again shows the highest successful transmission rate, which improves as the number of APs increases, reflecting more communication opportunities. Energy consumption rises for both RF-OWC and RF systems, with a notably sharper increase observed in the hybrid RF-OWC configuration. This sharper increase is attributed to the added APs being OWC-enabled, leading to more frequent reliance on the OWC interface for communication.

\subsection{Training of DGET}

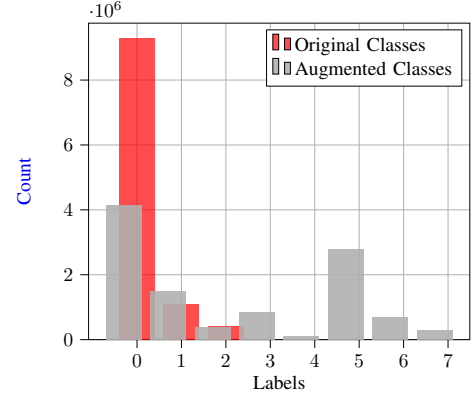
\begin{figure}[t]
\centering
       \resizebox{0.35\textwidth}{!}{\input{figures/LabelOccurencesfigure}}  
  \caption{\textcolor{black}{Histogram of labels before and after label augmentation for a dataset of $|\mathcal{N}_{d}|=5$ and $|\mathcal{N}_{AP}|=2$. \vspace{-0.5cm}}}%
        \label{fig:16-1}
\end{figure}

In order to train the proposed DGET model, we generate approximately $500,000
$ solution instances by solving (P1) and recording the communication evolution over the time horizon $\mathcal{T}$. Each solution yields a sequence of $|\mathcal{T}|$ known-labeled graph pairs. As shown in the histogram in Fig.~\ref{fig:16-1}, the dataset exhibits a significant class imbalance, predominantly in favor of class 0. This skew arises naturally from the data generation process and the system constraints, where most time steps involve no active communication. Such an imbalance can severely bias a machine learning model, degrading its performance in minority classes. To mitigate this, we augment the dataset by leveraging available communication links, thereby rebalancing the label distribution and expanding it to $8$ classes (cf. Section V-E). 
The DGET training parameters are provided in Table~\ref{tab:sim-setup}.
Notably, we apply the post-processing strategy during validation, as in testing, to ensure consistency and fair evaluation of generalization performance. We also employ the OneCycle learning rate scheduler. The learning rate starts at $2 \times 10^{-4}$, increases to a peak of $5 \times 10^{-3}$, and then decays to a final value of $5 \times 10^{-7}$. This schedule enables rapid initial exploration followed by granular convergence in the later training phases.  Additionally, we introduce a time-varying consistency loss coefficient $\lambda$ that decays linearly over the training epochs with $\lambda = 0.5 \times \left(1 - \frac{\text{epoch}}{\text{epochs}}\right)$. This shifts the model's focus to optimizing stable representations early on, then prioritizes classification performance in later epochs. The DGET model is trained on 8 NVIDIA GeForce RTX 2080 Ti GPUs, each with 11 GB of GDDR6 memory. For positional encoding, we adopt a learnable approach by initializing random vectors for each device's embeddings and allowing the model to optimize them during training. Temporal encoding is implemented using sinusoidal functions to capture periodic patterns, as described in \cite{bansal2025temporalencodingstrategiesenergy}.

 \begin{figure}[t]
\centering
    \subfloat[]{%
        \resizebox{0.24\textwidth}{!}{\input{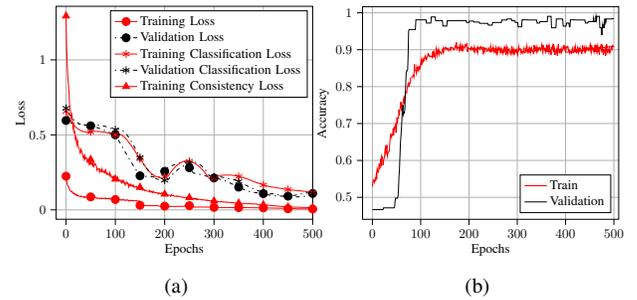}} 
    }  \hspace{-1.2cm}
 \centering 
    \qquad
    \subfloat[]{%
        \resizebox{0.24\textwidth}{!}{\input{figures/accuracy}} %
    }
       \caption{(a) Overall training and validation losses, training and validation losses for classification, and training loss for consistency vs. epochs, and (b) overall training and validation accuracy vs. epochs.\vspace{-0.3cm}}%
        \label{fig:16}
\end{figure}

\begin{figure}[t]
    \centering
        \resizebox{0.35\textwidth}{!}{\input{figures/ConfusionMatrix}}  
      \label{confusionmatrix} 
    \caption{Classification confusion matrix on test data after post-processing highlighted by green color if the change is positive with post-processing or red otherwise.\vspace{-0.5cm}}%
        \label{fig:confusionmatrix}
\end{figure}
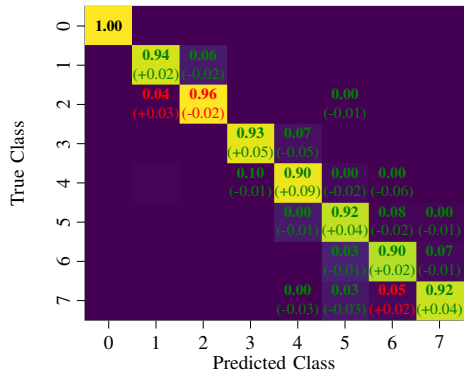

Fig.~\ref{fig:16}-(a) shows the DGET's training and validation loss curves, along with the breakdown of individual components: the consistency and classification losses. The overall training loss exhibits a sharp decline during the initial training phase, primarily driven by the consistency loss, which carries greater weight early in training. This phase is followed by stabilization after approximately $250$ epochs. The classification loss, though more variable throughout the training process, steadily decreases and stabilizes at a low value. The validation loss, in contrast, reflects only the classification component, as the consistency loss is not computed during validation (i.e., validation uses only known data and does not involve unlabeled or predicted graph embeddings). Such separation ensures that validation directly assesses the model’s classification ability without the influence of consistency regularization.

Fig.~\ref{fig:16}-(b) depicts the training and validation classification accuracy curves. Initially, the validation accuracy lags behind the training accuracy. This discrepancy arises not only from the model still learning but also from the inclusion of post-processing during validation. In the early epochs, the post-processing logic often fails to improve prediction accuracy because it relies on structural coherence (cf. Section V-E), which is not yet established in the model's output. Consequently, post-processing can even slightly degrade accuracy in the initial stages. However, around epoch 70, validation accuracy with post-processing begins to surpass raw training accuracy. As training progresses and the model starts generating more consistent and meaningful predictions, the post-processing becomes increasingly effective, ultimately yielding nearly a $7\%$ improvement in overall accuracy (while also accounting for the dropout effect).

Fig.~\ref{fig:confusionmatrix} presents the confusion matrices for the DGET model before and after post-processing on hold-out $20\%$ testing set. The green and red annotations in the matrix respectively highlight positive and negative changes, offering intuitive insights into where the model improved or regressed. The post-processing significantly improved classification performance, especially in reducing inter-class confusion for adjacent and similar classes. The diagonal elements, representing correct classifications, improved across nearly all classes, especially for Class 3 (0.87 to 0.96), Class 4 (0.52 to 0.97), and Class 7 (0.77 to 0.92). Off-diagonal confusions were also significantly reduced, such as class 3, which was misclassified as Class 4 in $7\%$ of cases without post-processing.

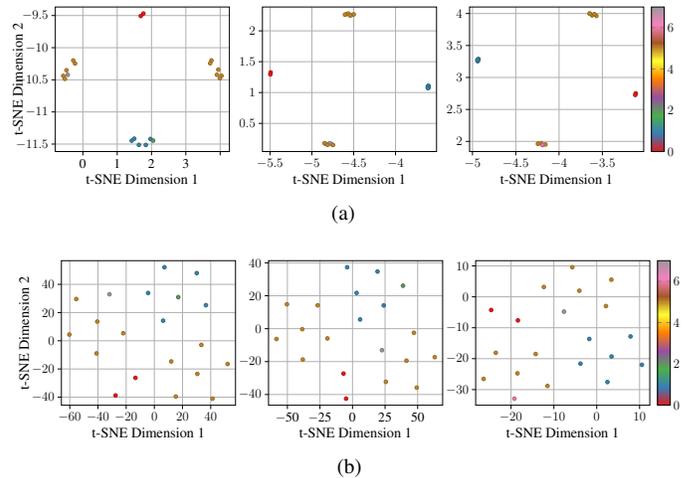
\begin{figure}[t]
  \hspace{-0.25cm}\subfloat[]{%
   \resizebox{0.505\textwidth}{!}  {  \input{figures/Trans_Embeddings}}}
     \label{fig1}
 \qquad

 \hspace{-0.25cm} \subfloat []{%
   \resizebox{0.505\textwidth}{!}  {  \input{figures/Indu_Embeddings}}}
      \label{fig2}
       \caption{\textcolor{black}{t-SNE visualization of transductive (top) and inductive (bottom) embeddings across three snapshots ($|\mathcal{T}|=3$), with each column representing one snapshot in time.}}%
        \label{roccurves} 
\end{figure}

To better illustrate the advantages of incorporating inductive learning on top of transductive embeddings, we visualize the learned representations using t-distributed Stochastic Neighbor Embedding (t-SNE)~\cite{7966046} projected into 2D space. This visualization offers a tangible view of how well the model captures class separability over time. As shown in the top row of Fig.~\ref{roccurves}, transductive embeddings exhibit low-variance features and tend to cluster tightly, making it harder to define a decision boundary distinguishing between communication classes. In contrast, the inductive embeddings in the bottom row are more dispersed, revealing a clearer and more structured separation between ground truth classes (e.g., no communication, RF, or OWC). This increased spread indicates a more generalizable feature space.

\subsection{Inference with DGET}

In Table~\ref{AoItable}, we evaluate the performance of three communication configurations using $|\mathcal{N}_{d}|=9$ devices and $|\mathcal{N}_{AP}|=2$ access points, following the training and simulation setup previously described. The results compare M-AoI and P-AoI across: (1) an RF-only system optimized via MILP, (2) a hybrid RF-OWC system also optimized via MILP, and (3) a hybrid RF-OWC system where scheduling is handled by the DGET model.

The optimized RF-OWC configuration yields the best performance, with a M-AoI of $6.1$ and a P-AoI of $7.8$, showing clear improvements over the RF-only baseline (M-AoI of $7.3$, P-AoI of $8.9$). These gains are consistent across both data categories (i.e., Data Type 1 and Data Type 2).  The DGET model exhibits slightly higher AoI values (e.g., M-AoI of $6.4$ vs. $6.1$ for all data), but the differences are minor. This indicates that despite not being globally optimal, the learned DGET policy can closely approximate the performance of the optimization-based solution in terms of AoI.

In Fig.\ref{runningtime}, we evaluate the running time to highlight the inference computational efficiency of the DGET compared to the optimization solver. We maintain the ratio of 1 APs per 3 IoT nodes, and increase the overall number of IoT devices, respectively. The RF-OWC system, which incorporates both RF and optical wireless communication capabilities, shows the highest computational cost due to the added complexity of managing multi-band interfaces and link selection processes. In contrast, the traditional RF system presents a more moderate increase in execution time, reflecting its simpler communication model. The DGET framework consistently outperforms both baseline methods in terms of execution efficiency. Despite incorporating advanced graph-based reasoning and sequence modeling, it maintains the lowest running time across all network sizes, demonstrating a polynomial trend that is estimated $~O\left( |\mathcal{N}|^2 \, \varphi \right)$ with $\varphi$ is the aggregated hidden dimensionality across model components.

\begin{table}[t]
\caption{M-AoI and P-AoI for both RF (MILP), RF-OWC (MILP), and RF-OWC (DGET) configurations computed for the overall system, data type 1, and data type 2.}
\label{AoItable}
\resizebox{8.7cm}{!}{
\begin{tabular}{c|c|c|c|}
\cline{2-4}
 & RF  (MILP)& RF-OWC  (MILP)&  RF-OWC (DGET) \\ \hline
\multicolumn{1}{|c|}{All Data} &
  \begin{tabular}[c]{@{}c@{}}M-AoI: 7.3\\  P-AoI: 8.9\end{tabular} &
  \begin{tabular}[c]{@{}c@{}}M-AoI: 6.1\\  P-AoI: 7.8\end{tabular} &
  \begin{tabular}[c]{@{}c@{}}M-AoI: 6.4\\  P-AoI: 8\end{tabular} \\ \hline
\multicolumn{1}{|c|}{Data Type 1} &
  \begin{tabular}[c]{@{}c@{}}M-AoI: 7.5\\  P-AoI: 8.9\end{tabular} &
  \begin{tabular}[c]{@{}c@{}}M-AoI: 6\\  P-AoI: 7.7\end{tabular} &
  \begin{tabular}[c]{@{}c@{}}M-AoI: 6.2\\  P-AoI: 7.9\end{tabular} \\ \hline
\multicolumn{1}{|c|}{Data Type 2} &
  \begin{tabular}[c]{@{}c@{}}M-AoI: 6.9\\ P-AoI: 8.9\end{tabular} &
  \begin{tabular}[c]{@{}c@{}}M-AoI: 6.3\\  P-AoI: 8.1\end{tabular} &
  \begin{tabular}[c]{@{}c@{}}M-AoI: 6.6\\  P-AoI: 8.3\end{tabular} \\ \hline
\end{tabular}}
\end{table}
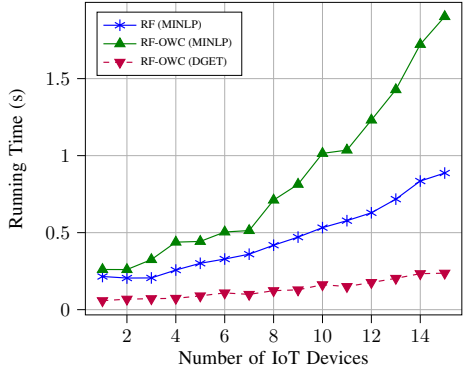
\begin{figure}[t]
\centering%
           \resizebox{0.35\textwidth}{!}{\input{figures/runningtime}}  

       \caption{\textcolor{black}{Running time (s) of the RF, RF-OWC, and DGET model vs. number of IoT Devices in the network.\vspace{-0.3cm}}}%
        \label{runningtime}
\end{figure}
In Fig.\ref{snapshotmasking}, we evaluate DGET's robustness against missing or outdated link information for RF-OWC network. We replace edge weights in the input graph \(\mathcal{G}^{I}_k(\mathcal{V}, \mathcal{E})\) with deprecated values from previous snapshots to simulate outdated channel information of the allowed communication technology. The snapshots used for replacement are non-subsequent snapshots, i.e., not directly adjacent in time, preventing the carryover of stale information across consecutive updates.  We evaluate three scenarios in which edge weight replacement occurs in $1$, $2$, or $3$ randomly selected non-subsequent snapshots among $|\mathcal{T}|=10$, with the overall replacement rate uniformly varied between $10\%$ and $50\%$. Across all settings, DGET consistently outperforms the MILP baseline. For example, when $20\%$ of all edges are outdated across $2$ snapshots, DGET maintains nearly $90\%$ accuracy, whereas MILP experiences a significantly sharper decline in performance. 
\textcolor{black}{This behavior highlights a fundamental limitation of the optimization-based approach. Although the MILP is solved optimally with respect to the provided input parameters, it implicitly assumes perfectly accurate channel state information. As a result, when the optimization is supplied with stale channel realizations, it effectively solves an optimal solution to a mismatched problem, without accounting for the stochastic temporal evolution of links or the uncertainty in their availability across snapshots. In contrast, DGET leverages historical information and learned temporal patterns to adapt its decisions under uncertainty, enabling more robust link allocation in the presence of outdated or imperfect channel knowledge.}
 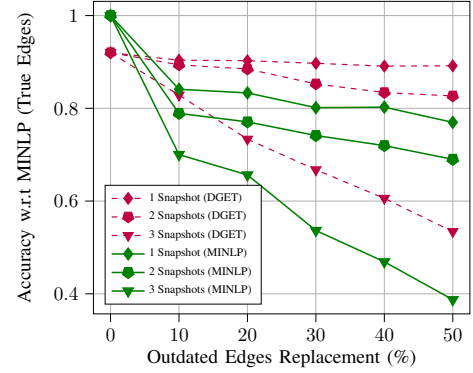
\begin{figure}[t]
\centering%
           \resizebox{0.35\textwidth}{!}{\input{figures/snapshots_2}}  
       \caption{Impact of edge staleness on DGET prediction (link class) accuracy when using outdated allowed technologies for RF-OWC network with $|\mathcal{T}|=10$. \vspace{-0.5cm}}%
        \label{snapshotmasking}
\end{figure}

\section{Conclusion and Future Work}
\label{sec7}

This paper investigated the resource allocation in hybrid RF-OWC IoT networks through deterministic optimization and a designed DGET framework that combines GNNs with Transformer-based learning to address the computational challenges of traditional MILP optimization. Experimental results showed that hybrid RF-OWC achieves significant improvements in AoI while maintaining comparable energy efficiency to the RF system, making it highly effective in environments where timely data delivery is critical. Moreover, DGET achieves near-optimal scheduling with notably reduced computational overhead and robustness to outdated channel information. \textcolor{black}{While DGET is demonstrated in the context of RF and optical communication, the proposed framework can, in principle, be applied to any dual-band communication system, a direction that also aligns with current research trends in 6G networks, with dual-band and multi-band operation being increasingly explored.} 
Future research will focus on extending DGET to mobile and outdoor environments, integrating real-time feedback mechanisms, and validating the approach on hardware testbeds and physical IoT infrastructures. Additionally, incorporating energy-harvesting-aware scheduling policies will be a key direction to support sustainable and self-powered IoT operations in resource-constrained networks.

\bibliographystyle{IEEEtran}
\bibliography{references}

\vfill\pagebreak

\end{document}

%% file: figures/EnergyPerbit_TTexchanged1.tex
\begin{tikzpicture}

\begin{axis}[
legend style={},
width=7cm,
y axis line style={black},
y tick label style={color=black},
ytick style={black},
ylabel=\textcolor{black}{Energy Per Bit (nJ)},
log basis y={10},
tick align=inside,
tick pos=left,
xlabel={Avg. Number of Packets/Device},
xmajorgrids,
grid style={gray},
y grid style={dashed, gray},
xmin=-38.75, xmax=835.75,
xtick style={color=dimgray85},
ymajorgrids,
ymin=0.010051294575221, ymax=15.6224910240376,
ymode=log,
ylabel style={yshift=-4pt},
label style={font=\footnotesize},
tick label style={font=\footnotesize},
ytick={1e-05,0.0001,0.001,0.01,0.1,1,10},
yticklabels={
  \(\displaystyle {10^{-5}}\),
  \(\displaystyle {10^{-4}}\),
  \(\displaystyle {10^{-3}}\),
  \(\displaystyle {10^{-2}}\),
  \(\displaystyle {10^{-1}}\),
  \(\displaystyle {10^{0}}\),
  \(\displaystyle {10^{1}}\)
}
]
\addplot [semithick, blue, mark=asterisk, mark size=3, mark repeat=20, mark options={solid}]
table [x=x, y=y, header=true]{%
x  y
1 3.47223015472699
6 0.578705285732974
11 0.315657428581622
16 0.217014482149865
21 0.165344367237974
26 0.133547374869794
31 0.112007475759859
36 0.0964508818504065
41 0.0846885791857228
46 0.0754832988394486
51 0.0680829754238163
56 0.0620041383324042
61 0.0569218319435347
66 0.0526095719184035
71 0.0489046725148678
76 0.0456872598238768
81 0.0428670586001807
86 0.040374787751333
91 0.038156392819941
96 0.0361690806939024
101 0.0343785321446994
106 0.0327569032699494
111 0.0312813670866183
116 0.029933032298402
121 0.028696130137311
126 0.0275573948144019
131 0.0265055858520201
136 0.0255311157839311
141 0.0246257570681889
146 0.0237824092366823
151 0.022994912229236
156 0.0222578958116323
161 0.0215666568233268
166 0.0209170587262386
171 0.0203054488106119
176 0.0198413130009115
181 0.0198413156622465
186 0.0198413158965002
191 0.019841312100345
196 0.0198413213136897
201 0.0198413149302609
206 0.0198413145199481
211 0.019841316493514
216 0.0198413189059043
221 0.0198413156473231
226 0.0198413142774519
231 0.0198413125771101
236 0.0198413144657239
241 0.0198413199555808
246 0.0198413165108519
251 0.0198413167208804
256 0.0198413232006379
261 0.0198413103979919
266 0.019841321263126
271 0.0198413135238778
276 0.0198413122910147
281 0.0198413249306793
286 0.0198413144639973
291 0.0198413144953448
296 0.0198413144476597
301 0.0198413130473824
306 0.0198413243979039
311 0.0198413141540338
316 0.0198413227153845
321 0.0198413222080474
326 0.0198413215618625
331 0.019841317616604
336 0.019841320209899
341 0.0198413301622987
346 0.0198413258448902
351 0.0198413138530679
356 0.0198413164386351
361 0.0198413153900032
366 0.0198413173541194
371 0.0198413166707496
376 0.0198413184351603
381 0.0198413162809403
386 0.0198413179147941
391 0.019841318515982
396 0.0198413237236784
401 0.0198413193273401
406 0.0198413189141565
411 0.0198413183327888
416 0.0198413229078234
421 0.019841322607031
426 0.0198413135119454
431 0.0198413323575819
436 0.0198413209039925
441 0.0198413179287128
446 0.0198413176408742
451 0.0198413196348067
456 0.0198413172868686
461 0.0198413202561414
466 0.0198413167554997
471 0.0198413194252797
476 0.0198413192758385
481 0.0198413191264044
486 0.0198413201491452
491 0.0198413122454959
496 0.0198413121356941
501 0.0198413170821854
506 0.0198413172592915
511 0.0198413161485753
516 0.0198413167523302
521 0.0198413109116899
526 0.0198413172166042
531 0.0198413164044892
536 0.0198413205768492
541 0.0198413189776191
546 0.0198413197765752
551 0.0198413191070925
556 0.0198413193758343
561 0.0198413227182516
566 0.0198413125833662
571 0.0198413171913668
576 0.0198413184577418
581 0.019841323397302
586 0.0198413171998045
591 0.0198413189799603
596 0.0198413187997346
601 0.0198413170285007
606 0.0198413236749948
611 0.0198413177498693
616 0.0198413184710385
621 0.0198413176250974
626 0.0198413183471464
631 0.0198413225681873
636 0.0198413237080917
641 0.0198413177916936
646 0.0198413240526003
651 0.0198413178153591
656 0.0198413233795003
661 0.0198413179533008
666 0.0198413202795538
671 0.0198413186918717
676 0.0198413193925549
681 0.019841320683123
686 0.0198413233795003
691 0.0198413202955881
696 0.0198413207151426
701 0.0198413157630897
706 0.0198413178085289
711 0.0198413172168601
716 0.0198413179253868
721 0.0198413196438124
726 0.0198413185182082
731 0.0198413185988703
736 0.0198413122000302
741 0.019841317927623
746 0.0198413171413462
751 0.0198413211383481
756 0.0198413182904232
761 0.0198413179529381
766 0.0198413171933188
771 0.0198413203339556
776 0.0198413171998978
781 0.0198413206385873
786 0.0198413180580561
791 0.0198413202862939
796 0.019841318555677
};
\label{plotyyref:leg4}
\addplot [semithick, green, mark=triangle*, mark size=3, mark repeat=20, mark options={solid}]
table [x=x, y=y, header=true]{%
x  y
1 4.32371852938987
6 0.840304935492974
11 0.42304768427443
16 0.290826822124672
21 0.21993205062125
26 0.181376454721911
31 0.153095323900992
36 0.128910085918633
41 0.112648123488933
46 0.0982113625318041
51 0.0924664278766446
56 0.0842797484731287
61 0.0778025416546024
66 0.0703145923192542
71 0.0653628604657856
76 0.060247384861379
81 0.0570194205314351
86 0.0541107503141713
91 0.0510653447297454
96 0.0476400953735031
101 0.0466909685422741
106 0.0444885643557441
111 0.0425195127432
116 0.0410699787918891
121 0.0393728722390079
126 0.0376663098486567
131 0.0354256724661892
136 0.0341232580372851
141 0.0329132134431695
146 0.0317806771275516
151 0.0303231870825484
156 0.0293512896260425
161 0.0285599377447529
166 0.0278227295024744
171 0.0270091995956048
176 0.0614095043662895
181 0.0597306927791006
186 0.0580967505134304
191 0.0566444336891625
196 0.0552797947464683
201 0.054000946620367
206 0.0526611643670671
211 0.0514480717944858
216 0.0503317464925973
221 0.04920210211824
226 0.0482426004369376
231 0.0473123371828184
236 0.0462846339072472
241 0.045326790898605
246 0.0443589506787894
251 0.0435771743080886
256 0.0428252031513258
261 0.0421151092554651
266 0.0412549137176954
271 0.0403864390200638
276 0.0397554211657012
281 0.0390736715423073
286 0.0384345907216754
291 0.0378632503739816
296 0.0372305581415305
301 0.0365629262213003
306 0.0360115411256552
311 0.0353423433907153
316 0.0348960891428439
321 0.034385409324843
326 0.033701242653277
331 0.033401067560667
336 0.0329078755396984
341 0.0324515078934095
346 0.031917026163486
351 0.0314589755515997
356 0.0310266982965408
361 0.0306957672971865
366 0.0302593098199169
371 0.0298303832020619
376 0.0294688689115782
381 0.0291186793101326
386 0.028746450729928
391 0.0283805229227621
396 0.0280207508716597
401 0.0276588706775307
406 0.0273467504719101
411 0.0270093160909953
416 0.0266751436723162
421 0.0263401914163505
426 0.0260302834864539
431 0.0257596672267771
436 0.0255223813563422
441 0.0252553547217576
446 0.0249024111549783
451 0.0246356508520132
456 0.0243522060141153
461 0.0240347040888562
466 0.0239204929017048
471 0.0235316958040117
476 0.02338068579355
481 0.0231300382314882
486 0.0228805715099018
491 0.0226713068470356
496 0.0224399304845381
501 0.0222219926960659
506 0.0219843719213426
511 0.0217877696098477
516 0.0215400261210929
521 0.0213885220775411
526 0.0211312591828917
531 0.0208974403434416
536 0.0207569923437475
541 0.0205504428389196
546 0.0203542660410073
551 0.0201972554970364
556 0.0200428064696355
561 0.0198201693563842
566 0.0197137583586252
571 0.0195051303093879
576 0.0193697464222597
581 0.0191670155164014
586 0.0191144728840494
591 0.0189590441314782
596 0.0187755280286931
601 0.0185739512308527
606 0.01839475270491
611 0.0182745052872482
616 0.0181435844439479
621 0.0179470739211602
626 0.0178334864534345
631 0.0177177562622007
636 0.0175822313130339
641 0.0174745556827921
646 0.0173411398371106
651 0.0172086577479177
656 0.0170982058141826
661 0.0169294938159096
666 0.0168311139131627
671 0.0167067526707415
676 0.0165945337396589
681 0.0164763024004021
686 0.0163781351146692
691 0.0162368337729726
696 0.01610955109529
701 0.0159812225396326
706 0.0158679481739469
711 0.0157466658424637
716 0.0156186274409262
721 0.015537118334947
726 0.0154535041898221
731 0.0153601228008156
736 0.0152572749928524
741 0.0151309814203481
746 0.0150500994633892
751 0.0149380062515189
756 0.0148104349028812
761 0.0146948329544117
766 0.0146081093693329
771 0.0145205940620158
776 0.0144411795028453
781 0.0143426196448687
786 0.0142727688686464
791 0.0141690155465288
796 0.0140918304882348
};
\label{plotyyref:leg2}
\addplot [semithick, orange, mark=triangle*, mark size=3, mark repeat=20, mark options={solid,rotate=90}]
table [x=x, y=y, header=true]{%
x  y
1 11.1861545781196
6 1.86221062019689
11 1.01575124738012
16 0.698481585240899
21 0.532060177199112
26 0.429834821686707
31 0.360506624640464
36 0.310368436699482
41 0.272518627345886
46 0.242897037416986
51 0.219131477722635
56 0.199566167211685
61 0.183208284653351
66 0.169291874563354
71 0.157369911565934
76 0.147016627910281
81 0.137941527421992
86 0.129921671176527
91 0.12278311781518
96 0.116388163762306
101 0.110650548156974
106 0.105431182677872
111 0.100682030304994
116 0.0963422876194343
121 0.0923612013541685
126 0.0886960743163046
131 0.0852920894746668
136 0.0821563508910393
141 0.0792430051147613
146 0.0765292035697352
151 0.0739951239813334
156 0.0716234853921881
161 0.0693991535477102
166 0.067308817597478
171 0.0653407235156804
176 0.0634844529612577
181 0.0617307387910572
186 0.060071310328932
191 0.0584987629381222
196 0.0570189049176244
201 0.0556005241982805
206 0.0542509969119145
211 0.0529654282647127
216 0.051739376684511
221 0.0505688025513773
226 0.049450023733869
231 0.0483796768998025
236 0.0473546837451457
241 0.0463722214267817
246 0.0454296966010341
251 0.0445247225651569
256 0.0436550990775562
261 0.0428094395447561
266 0.0420047508315088
271 0.0412297554287135
276 0.0404828395694976
281 0.0397625043458411
286 0.0390673556684662
291 0.0383960952617916
296 0.0377475125715586
301 0.0371204774790078
306 0.0365139337293508
311 0.0359268929941522
316 0.0353539320638478
321 0.0348032477637878
326 0.0342694556201714
331 0.0337394418588217
336 0.0332373668311607
341 0.0327500154113489
346 0.0322767492926878
351 0.0318169665392307
356 0.0313700990316573
361 0.0309356101254016
366 0.0305129925007377
371 0.0301017661867116
376 0.0297014767427393
381 0.0292934862792923
386 0.028942844247609
391 0.028575465782873
396 0.0282203326767418
401 0.0278730853768851
406 0.02750623742138
411 0.027180684169701
416 0.0268539932541997
421 0.0265350622179266
426 0.0262236178256974
431 0.025898931578715
436 0.0256019254826288
441 0.025320339499118
446 0.0249980443001607
451 0.024720210799794
456 0.0244528066783077
461 0.0241886251689561
466 0.023929090564139
471 0.0236913070634748
476 0.0234353908368479
481 0.0231838166936689
486 0.0229530069404502
491 0.0227116429906949
496 0.0224832016454181
501 0.0222553556885832
506 0.0220201858869276
511 0.02180472418549
516 0.0216006828025321
521 0.0213933825836978
526 0.0211919365491162
531 0.0210001028831777
536 0.0208042729772769
541 0.020593159482209
546 0.0204068698653149
551 0.0202344153637211
556 0.0200446000035079
561 0.0198753706444645
566 0.0197035594433628
571 0.0195373010248181
576 0.0193700634210025
581 0.0191834719765368
586 0.0190234438432558
591 0.0188625010019423
596 0.018699446903526
601 0.0185438774617329
606 0.0183908751724447
611 0.0182366343466683
616 0.0181086103519397
621 0.0179628083362236
626 0.0178188481995777
631 0.017698463830045
636 0.0175628979696092
641 0.0174184949686175
646 0.0173118781895389
651 0.0171648222174522
656 0.0170364168308727
661 0.0169067243087375
666 0.0167923527398693
671 0.0166980430442339
676 0.0165542791513819
681 0.0164312392408217
686 0.0163088266774549
691 0.0161892506969563
696 0.0160545164798984
701 0.0159580111491929
706 0.0158345607446532
711 0.0157091614527239
716 0.015625046222765
721 0.0155167066277867
726 0.0154098422570719
731 0.0152930135036442
736 0.0151935040608238
741 0.0150863513643687
746 0.0149859100409564
751 0.0148997681743182
756 0.0148161061598773
761 0.0147200147557346
766 0.0145932472399487
771 0.0145053884566311
776 0.014390530782654
781 0.014293003311245
786 0.0142338706564852
791 0.0141176601209849
796 0.014037554924236
};\label{plotyyref:leg3}
\end{axis}
\begin{axis}[
ytick pos=right,
width=7cm,
axis y line=right,
axis x line=none,
axis line style={-},
tick align=inside,
legend columns=2,
legend style={  at={(0.5,0.7)},
        anchor=south,font=\fontsize{5}{6}\selectfont},
legend style={
	/tikz/column 2/.style={
		column sep=5pt,
	},
},scaled y ticks=false,
xmajorgrids,
ylabel style={yshift=4pt},
label style={font=\footnotesize},
tick label style={font=\footnotesize},
xmin=-38.75, xmax=835.75,
xtick pos=left,
y tick style={black},
ylabel={Avg. Exchanged Packets},
ymajorgrids,
ymin=-14930, ymax=321912,
yticklabels={
  \(\displaystyle {0}\),
    \(\displaystyle {0}\),
  \(\displaystyle {5000}\),
  \(\displaystyle {10000}\),
  \(\displaystyle {15000}\)
}
]
\addlegendimage{/pgfplots/refstyle=plotyyref:leg4}
\addplot [thick, blue, dashed]
table [x=x, y=y, header=true]{%
x  y
1 400
6 2400
11 4400
16 6400
21 8400
26 10400
31 12400
36 14400
41 16400
46 18400
51 20400
56 22400
61 24400
66 26400
71 28400
76 30400
81 32400
86 34400
91 36400
96 38400
101 40400
106 42400
111 44400
116 46400
121 48400
126 50400
131 52400
136 54400
141 56400
146 58400
151 60400
156 62400
161 64400
166 66400
171 68400
176 70000
181 70000
186 70000
191 70000
196 70000
201 70000
206 70000
211 70000
216 70000
221 70000
226 70000
231 70000
236 70000
241 70000
246 70000
251 70000
256 70000
261 70000
266 70000
271 70000
276 70000
281 70000
286 70000
291 70000
296 70000
301 70000
306 70000
311 70000
316 70000
321 70000
326 70000
331 70000
336 70000
341 70000
346 70000
351 70000
356 70000
361 70000
366 70000
371 70000
376 70000
381 70000
386 70000
391 70000
396 70000
401 70000
406 70000
411 70000
416 70000
421 70000
426 70000
431 70000
436 70000
441 70000
446 70000
451 70000
456 70000
461 70000
466 70000
471 70000
476 70000
481 70000
486 70000
491 70000
496 70000
501 70000
506 70000
511 70000
516 70000
521 70000
526 70000
531 70000
536 70000
541 70000
546 70000
551 70000
556 70000
561 70000
566 70000
571 70000
576 70000
581 70000
586 70000
591 70000
596 70000
601 70000
606 70000
611 70000
616 70000
621 70000
626 70000
631 70000
636 70000
641 70000
646 70000
651 70000
656 70000
661 70000
666 70000
671 70000
676 70000
681 70000
686 70000
691 70000
696 70000
701 70000
706 70000
711 70000
716 70000
721 70000
726 70000
731 70000
736 70000
741 70000
746 70000
751 70000
756 70000
761 70000
766 70000
771 70000
776 70000
781 70000
786 70000
791 70000
796 70000
}; \addlegendentry{RF (Energy)}
\addlegendimage{/pgfplots/refstyle=plotyyref:leg2}
\addlegendentry{RF (Packets)}
\addplot [thick, green, dashed]
table [x=x, y=y, header=true]{%
x  y
1 400
6 2400
11 4400
16 6400
21 8400
26 10400
31 12400
36 14400
41 16400
46 18400
51 20400
56 22400
61 24400
66 26400
71 28400
76 30400
81 32400
86 34400
91 36400
96 38400
101 40400
106 42400
111 44400
116 46400
121 48400
126 50400
131 52400
136 54400
141 56400
146 58400
151 60400
156 62400
161 64400
166 66400
171 68400
176 70381
181 72286
186 74191
191 76096
196 78001
201 79906
206 81811
211 83716
216 85621
221 87526
226 89431
231 91336
236 93241
241 95146
246 97051
251 98956
256 100861
261 102766
266 104671
271 106576
276 108481
281 110386
286 112291
291 114196
296 116101
301 118006
306 119911
311 121816
316 123721
321 125626
326 127531
331 129436
336 131341
341 133246
346 135151
351 137056
356 138961
361 140866
366 142771
371 144676
376 146581
381 148486
386 150391
391 152296
396 154201
401 156106
406 158011
411 159916
416 161821
421 163726
426 165631
431 167536
436 169441
441 171346
446 173251
451 175156
456 177061
461 178966
466 180871
471 182776
476 184681
481 186586
486 188491
491 190396
496 192301
501 194206
506 196111
511 198016
516 199921
521 201826
526 203731
531 205636
536 207541
541 209446
546 211351
551 213256
556 215161
561 217066
566 218971
571 220876
576 222781
581 224686
586 226591
591 228496
596 230401
601 232306
606 234211
611 236116
616 238021
621 239926
626 241831
631 243736
636 245641
641 247546
646 249451
651 251356
656 253261
661 255166
666 257071
671 258976
676 260881
681 262786
686 264691
691 266596
696 268501
701 270406
706 272311
711 274216
716 276121
721 278026
726 279931
731 281836
736 283741
741 285646
746 287551
751 289456
756 291361
761 293266
766 295171
771 297076
776 298981
781 300886
786 302791
791 304696
796 306601
};\addlegendentry{RF-OWC (Energy)}
\addlegendimage{/pgfplots/refstyle=plotyyref:leg3}
\addlegendentry{RF-OWC (Packets)}
\addplot [thick, orange, dashed]
table [x=x, y=y, header=true]{%
x  y
1 381
6 2286
11 4191
16 6096
21 8001
26 9906
31 11811
36 13716
41 15621
46 17526
51 19431
56 21336
61 23241
66 25146
71 27051
76 28956
81 30861
86 32766
91 34671
96 36576
101 38481
106 40386
111 42291
116 44196
121 46101
126 48006
131 49911
136 51816
141 53721
146 55626
151 57531
156 59436
161 61341
166 63246
171 65151
176 67056
181 68961
186 70866
191 72771
196 74676
201 76581
206 78486
211 80391
216 82296
221 84201
226 86106
231 88011
236 89916
241 91821
246 93726
251 95631
256 97536
261 99441
266 101346
271 103251
276 105156
281 107061
286 108966
291 110871
296 112776
301 114681
306 116586
311 118491
316 120396
321 122301
326 124206
331 126111
336 128016
341 129921
346 131826
351 133731
356 135636
361 137541
366 139446
371 141351
376 143256
381 145161
386 147066
391 148971
396 150876
401 152781
406 154686
411 156591
416 158496
421 160401
426 162306
431 164211
436 166116
441 168021
446 169926
451 171831
456 173736
461 175641
466 177546
471 179451
476 181356
481 183261
486 185166
491 187071
496 188976
501 190881
506 192786
511 194691
516 196596
521 198501
526 200406
531 202311
536 204216
541 206121
546 208026
551 209931
556 211836
561 213741
566 215646
571 217551
576 219456
581 221361
586 223266
591 225171
596 227076
601 228981
606 230886
611 232791
616 234696
621 236601
626 238506
631 240411
636 242316
641 244221
646 246126
651 248031
656 249936
661 251841
666 253746
671 255651
676 257556
681 259461
686 261366
691 263271
696 265176
701 267081
706 268986
711 270891
716 272796
721 274701
726 276606
731 278511
736 280416
741 282321
746 284226
751 286131
756 288036
761 289941
766 291846
771 293751
776 295656
781 297561
786 299466
791 301371
796 303276
};
\addlegendentry{OWC (Energy)}
\addlegendimage{/pgfplots/refstyle=plotyyref:leg3}
\addlegendentry{OWC (Packets)}
\end{axis}
\end{tikzpicture}

%% file: figures/Switching_TTexchanged.tex
\begin{tikzpicture}

\begin{axis}[
y axis line style={black},
width=7cm,
y tick label style={color=black},
ytick style={black},
xlabel=\textcolor{black}{Avg. Number of Packets/Device},
xmajorgrids,
xmin=-8.17, xmax=167.17,
xtick style={color=dimgray85},
xtick={0,50,100,150,200},
xticklabels={0,200,400,600,800},
ylabel=\textcolor{black}{Avg. Technology Switching/Node},
ylabel style={yshift=-5pt},
ymajorgrids,
label style={font=\footnotesize},
tick label style={font=\footnotesize},
ymin=-0.1, ymax=3.675,
y grid style={dashed, gray},
ytick style={color=black}
]
\addlegendimage{/pgfplots/refstyle=plotyyref:leg1}
\draw[draw=none,fill=gray128,very thin] (axis cs:-0.2,0) rectangle (axis cs:0.2,0);
\addlegendimage{ybar,ybar legend,draw=none,fill=gray128,very thin}

\draw[draw=none,fill=gray,very thin] (axis cs:0.8,0) rectangle (axis cs:0,0);
\draw[draw=none,fill=gray,very thin] (axis cs:1.8,0) rectangle (axis cs:0,0);
\draw[draw=none,fill=gray,very thin] (axis cs:2.8,0) rectangle (axis cs:0,0);
\draw[draw=none,fill=gray,very thin] (axis cs:3.8,0) rectangle (axis cs:4,0);
\draw[draw=none,fill=gray,very thin] (axis cs:4.8,0) rectangle (axis cs:5,0);
\draw[draw=none,fill=gray,very thin] (axis cs:5.8,0) rectangle (axis cs:6.2,0);
\draw[draw=none,fill=gray,very thin] (axis cs:6.8,0) rectangle (axis cs:7.2,0);
\draw[draw=none,fill=gray,very thin] (axis cs:7.8,0) rectangle (axis cs:8.2,0);
\draw[draw=none,fill=gray,very thin] (axis cs:8.8,0) rectangle (axis cs:9.2,2.01);
\draw[draw=none,fill=gray,very thin] (axis cs:9.8,0) rectangle (axis cs:10.2,1.79);
\draw[draw=none,fill=gray,very thin] (axis cs:10.8,0) rectangle (axis cs:11.2,1.83);
\draw[draw=none,fill=gray,very thin] (axis cs:11.8,0) rectangle (axis cs:12.2,1.69);
\draw[draw=none,fill=gray,very thin] (axis cs:12.8,0) rectangle (axis cs:13.2,1.83);
\draw[draw=none,fill=gray,very thin] (axis cs:13.8,0) rectangle (axis cs:14.2,1.83);
\draw[draw=none,fill=gray,very thin] (axis cs:14.8,0) rectangle (axis cs:15.2,1.93);
\draw[draw=none,fill=gray,very thin] (axis cs:15.8,0) rectangle (axis cs:16.2,1.89);
\draw[draw=none,fill=gray,very thin] (axis cs:16.8,0) rectangle (axis cs:17.2,1.83);
\draw[draw=none,fill=gray,very thin] (axis cs:17.8,0) rectangle (axis cs:18.2,2.05);
\draw[draw=none,fill=gray, thin] (axis cs:18.8,0) rectangle (axis cs:19.2,2.19);
\draw[draw=none,fill=gray,very thin] (axis cs:19.8,0) rectangle (axis cs:20.2,1.79);
\draw[draw=none,fill=gray,very thin] (axis cs:20.8,0) rectangle (axis cs:21.2,1.83);
\draw[draw=none,fill=gray,very thin] (axis cs:21.8,0) rectangle (axis cs:22.2,1.83);
\draw[draw=none,fill=gray,very thin] (axis cs:22.8,0) rectangle (axis cs:23.2,1.69);
\draw[draw=none,fill=gray,very thin] (axis cs:23.8,0) rectangle (axis cs:24.2,1.73);
\draw[draw=none,fill=gray,very thin] (axis cs:24.8,0) rectangle (axis cs:25.2,1.73);
\draw[draw=none,fill=gray,very thin] (axis cs:25.8,0) rectangle (axis cs:26.2,1.83);
\draw[draw=none,fill=gray,very thin] (axis cs:26.8,0) rectangle (axis cs:27.2,1.83);
\draw[draw=none,fill=gray,very thin] (axis cs:27.8,0) rectangle (axis cs:28.2,2.81);
\draw[draw=none,fill=gray,very thin] (axis cs:28.8,0) rectangle (axis cs:29.2,2.81);
\draw[draw=none,fill=gray,very thin] (axis cs:29.8,0) rectangle (axis cs:30.2,2.93);
\draw[draw=none,fill=gray,very thin] (axis cs:30.8,0) rectangle (axis cs:31.2,2.89);
\draw[draw=none,fill=gray,very thin] (axis cs:31.8,0) rectangle (axis cs:32.2,2.88);
\draw[draw=none,fill=gray,very thin] (axis cs:32.8,0) rectangle (axis cs:33.2,2.89);
\draw[draw=none,fill=gray,very thin] (axis cs:33.8,0) rectangle (axis cs:34.2,2.89);
\draw[draw=none,fill=gray,very thin] (axis cs:34.8,0) rectangle (axis cs:35.2,1.58);
\draw[draw=none,fill=gray,very thin] (axis cs:35.8,0) rectangle (axis cs:36.2,1.64);
\draw[draw=none,fill=gray,very thin] (axis cs:36.8,0) rectangle (axis cs:37.2,1.62);
\draw[draw=none,fill=gray,very thin] (axis cs:37.8,0) rectangle (axis cs:38.2,1.6);
\draw[draw=none,fill=gray,very thin] (axis cs:38.8,0) rectangle (axis cs:39.2,1.61);
\draw[draw=none,fill=gray,very thin] (axis cs:39.8,0) rectangle (axis cs:40.2,1.61);
\draw[draw=none,fill=gray,very thin] (axis cs:40.8,0) rectangle (axis cs:41.2,1.59);
\draw[draw=none,fill=gray,very thin] (axis cs:41.8,0) rectangle (axis cs:42.2,1.59);
\draw[draw=none,fill=gray,very thin] (axis cs:42.8,0) rectangle (axis cs:43.2,1.61);
\draw[draw=none,fill=gray,very thin] (axis cs:43.8,0) rectangle (axis cs:44.2,1.58);
\draw[draw=none,fill=gray,very thin] (axis cs:44.8,0) rectangle (axis cs:45.2,1.59);
\draw[draw=none,fill=gray,very thin] (axis cs:45.8,0) rectangle (axis cs:46.2,1.59);
\draw[draw=none,fill=gray,very thin] (axis cs:46.8,0) rectangle (axis cs:47.2,1.61);
\draw[draw=none,fill=gray,very thin] (axis cs:47.8,0) rectangle (axis cs:48.2,1.55);
\draw[draw=none,fill=gray,very thin] (axis cs:48.8,0) rectangle (axis cs:49.2,1.59);
\draw[draw=none,fill=gray,very thin] (axis cs:49.8,0) rectangle (axis cs:50.2,1.59);
\draw[draw=none,fill=gray,very thin] (axis cs:50.8,0) rectangle (axis cs:51.2,1.59);
\draw[draw=none,fill=gray,very thin] (axis cs:51.8,0) rectangle (axis cs:52.2,1.57);
\draw[draw=none,fill=gray,very thin] (axis cs:52.8,0) rectangle (axis cs:53.2,1.59);
\draw[draw=none,fill=gray,very thin] (axis cs:53.8,0) rectangle (axis cs:54.2,1.62);
\draw[draw=none,fill=gray,very thin] (axis cs:54.8,0) rectangle (axis cs:55.2,1.59);
\draw[draw=none,fill=gray,very thin] (axis cs:55.8,0) rectangle (axis cs:56.2,1.57);
\draw[draw=none,fill=gray,very thin] (axis cs:56.8,0) rectangle (axis cs:57.2,1.58);
\draw[draw=none,fill=gray,very thin] (axis cs:57.8,0) rectangle (axis cs:58.2,1.57);
\draw[draw=none,fill=gray,very thin] (axis cs:58.8,0) rectangle (axis cs:59.2,1.59);
\draw[draw=none,fill=gray,very thin] (axis cs:59.8,0) rectangle (axis cs:60.2,1.59);
\draw[draw=none,fill=gray,very thin] (axis cs:60.8,0) rectangle (axis cs:61.2,1.61);
\draw[draw=none,fill=gray,very thin] (axis cs:61.8,0) rectangle (axis cs:62.2,1.61);
\draw[draw=none,fill=gray,very thin] (axis cs:62.8,0) rectangle (axis cs:63.2,1.57);
\draw[draw=none,fill=gray,very thin] (axis cs:63.8,0) rectangle (axis cs:64.2,1.61);
\draw[draw=none,fill=gray,very thin] (axis cs:64.8,0) rectangle (axis cs:65.2,1.59);
\draw[draw=none,fill=gray,very thin] (axis cs:65.8,0) rectangle (axis cs:66.2,1.57);
\draw[draw=none,fill=gray,very thin] (axis cs:66.8,0) rectangle (axis cs:67.2,1.55);
\draw[draw=none,fill=gray,very thin] (axis cs:67.8,0) rectangle (axis cs:68.2,1.6);
\draw[draw=none,fill=gray,very thin] (axis cs:68.8,0) rectangle (axis cs:69.2,1.62);
\draw[draw=none,fill=gray,very thin] (axis cs:69.8,0) rectangle (axis cs:70.2,1.6);
\draw[draw=none,fill=gray,very thin] (axis cs:70.8,0) rectangle (axis cs:71.2,1.61);
\draw[draw=none,fill=gray,very thin] (axis cs:71.8,0) rectangle (axis cs:72.2,1.65);
\draw[draw=none,fill=gray,very thin] (axis cs:72.8,0) rectangle (axis cs:73.2,1.64);
\draw[draw=none,fill=gray,very thin] (axis cs:73.8,0) rectangle (axis cs:74.2,1.62);
\draw[draw=none,fill=gray,very thin] (axis cs:74.8,0) rectangle (axis cs:75.2,1.61);
\draw[draw=none,fill=gray,very thin] (axis cs:75.8,0) rectangle (axis cs:76.2,1.62);
\draw[draw=none,fill=gray,very thin] (axis cs:76.8,0) rectangle (axis cs:77.2,1.62);
\draw[draw=none,fill=gray,very thin] (axis cs:77.8,0) rectangle (axis cs:78.2,1.6);
\draw[draw=none,fill=gray,very thin] (axis cs:78.8,0) rectangle (axis cs:79.2,1.6);
\draw[draw=none,fill=gray,very thin] (axis cs:79.8,0) rectangle (axis cs:80.2,1.61);
\draw[draw=none,fill=gray,very thin] (axis cs:80.8,0) rectangle (axis cs:81.2,1.6);
\draw[draw=none,fill=gray,very thin] (axis cs:81.8,0) rectangle (axis cs:82.2,1.6);
\draw[draw=none,fill=gray,very thin] (axis cs:82.8,0) rectangle (axis cs:83.2,1.64);
\draw[draw=none,fill=gray,very thin] (axis cs:83.8,0) rectangle (axis cs:84.2,1.6);
\draw[draw=none,fill=gray,very thin] (axis cs:84.8,0) rectangle (axis cs:85.2,1.58);
\draw[draw=none,fill=gray,very thin] (axis cs:85.8,0) rectangle (axis cs:86.2,1.65);
\draw[draw=none,fill=gray,very thin] (axis cs:86.8,0) rectangle (axis cs:87.2,1.64);
\draw[draw=none,fill=gray,very thin] (axis cs:87.8,0) rectangle (axis cs:88.2,1.64);
\draw[draw=none,fill=gray,very thin] (axis cs:88.8,0) rectangle (axis cs:89.2,1.64);
\draw[draw=none,fill=gray,very thin] (axis cs:89.8,0) rectangle (axis cs:90.2,1.64);
\draw[draw=none,fill=gray,very thin] (axis cs:90.8,0) rectangle (axis cs:91.2,1.64);
\draw[draw=none,fill=gray,very thin] (axis cs:91.8,0) rectangle (axis cs:92.2,1.64);
\draw[draw=none,fill=gray,very thin] (axis cs:92.8,0) rectangle (axis cs:93.2,1.64);
\draw[draw=none,fill=gray,very thin] (axis cs:93.8,0) rectangle (axis cs:94.2,1.64);
\draw[draw=none,fill=gray,very thin] (axis cs:94.8,0) rectangle (axis cs:95.2,1.64);
\draw[draw=none,fill=gray,very thin] (axis cs:95.8,0) rectangle (axis cs:96.2,1.64);
\draw[draw=none,fill=gray,very thin] (axis cs:96.8,0) rectangle (axis cs:97.2,1.62);
\draw[draw=none,fill=gray,very thin] (axis cs:97.8,0) rectangle (axis cs:98.2,1.64);
\draw[draw=none,fill=gray,very thin] (axis cs:98.8,0) rectangle (axis cs:99.2,1.61);
\draw[draw=none,fill=gray,very thin] (axis cs:99.8,0) rectangle (axis cs:100.2,1.63);
\draw[draw=none,fill=gray,very thin] (axis cs:100.8,0) rectangle (axis cs:101.2,1.63);
\draw[draw=none,fill=gray,very thin] (axis cs:101.8,0) rectangle (axis cs:102.2,1.61);
\draw[draw=none,fill=gray,very thin] (axis cs:102.8,0) rectangle (axis cs:103.2,1.61);
\draw[draw=none,fill=gray,very thin] (axis cs:103.8,0) rectangle (axis cs:104.2,1.62);
\draw[draw=none,fill=gray,very thin] (axis cs:104.8,0) rectangle (axis cs:105.2,1.62);
\draw[draw=none,fill=gray,very thin] (axis cs:105.8,0) rectangle (axis cs:106.2,1.63);
\draw[draw=none,fill=gray,very thin] (axis cs:106.8,0) rectangle (axis cs:107.2,1.63);
\draw[draw=none,fill=gray,very thin] (axis cs:107.8,0) rectangle (axis cs:108.2,1.65);
\draw[draw=none,fill=gray,very thin] (axis cs:108.8,0) rectangle (axis cs:109.2,1.63);
\draw[draw=none,fill=gray,very thin] (axis cs:109.8,0) rectangle (axis cs:110.2,1.63);
\draw[draw=none,fill=gray,very thin] (axis cs:110.8,0) rectangle (axis cs:111.2,1.63);
\draw[draw=none,fill=gray,very thin] (axis cs:111.8,0) rectangle (axis cs:112.2,1.64);
\draw[draw=none,fill=gray,very thin] (axis cs:112.8,0) rectangle (axis cs:113.2,1.63);
\draw[draw=none,fill=gray,very thin] (axis cs:113.8,0) rectangle (axis cs:114.2,1.62);
\draw[draw=none,fill=gray,very thin] (axis cs:114.8,0) rectangle (axis cs:115.2,1.63);
\draw[draw=none,fill=gray,very thin] (axis cs:115.8,0) rectangle (axis cs:116.2,1.63);
\draw[draw=none,fill=gray,very thin] (axis cs:116.8,0) rectangle (axis cs:117.2,1.64);
\draw[draw=none,fill=gray,very thin] (axis cs:117.8,0) rectangle (axis cs:118.2,1.63);
\draw[draw=none,fill=gray,very thin] (axis cs:118.8,0) rectangle (axis cs:119.2,1.64);
\draw[draw=none,fill=gray,very thin] (axis cs:119.8,0) rectangle (axis cs:120.2,1.6);
\draw[draw=none,fill=gray,very thin] (axis cs:120.8,0) rectangle (axis cs:121.2,1.62);
\draw[draw=none,fill=gray,very thin] (axis cs:121.8,0) rectangle (axis cs:122.2,1.63);
\draw[draw=none,fill=gray,very thin] (axis cs:122.8,0) rectangle (axis cs:123.2,1.63);
\draw[draw=none,fill=gray,very thin] (axis cs:123.8,0) rectangle (axis cs:124.2,1.61);
\draw[draw=none,fill=gray,very thin] (axis cs:124.8,0) rectangle (axis cs:125.2,1.61);
\draw[draw=none,fill=gray,very thin] (axis cs:125.8,0) rectangle (axis cs:126.2,1.61);
\draw[draw=none,fill=gray,very thin] (axis cs:126.8,0) rectangle (axis cs:127.2,1.58);
\draw[draw=none,fill=gray,very thin] (axis cs:127.8,0) rectangle (axis cs:128.2,1.6);
\draw[draw=none,fill=gray,very thin] (axis cs:128.8,0) rectangle (axis cs:129.2,1.6);
\draw[draw=none,fill=gray,very thin] (axis cs:129.8,0) rectangle (axis cs:130.2,1.58);
\draw[draw=none,fill=gray,very thin] (axis cs:130.8,0) rectangle (axis cs:131.2,1.58);
\draw[draw=none,fill=gray,very thin] (axis cs:131.8,0) rectangle (axis cs:132.2,1.58);
\draw[draw=none,fill=gray,very thin] (axis cs:132.8,0) rectangle (axis cs:133.2,1.59);
\draw[draw=none,fill=gray,very thin] (axis cs:133.8,0) rectangle (axis cs:134.2,1.58);
\draw[draw=none,fill=gray,very thin] (axis cs:134.8,0) rectangle (axis cs:135.2,1.61);
\draw[draw=none,fill=gray,very thin] (axis cs:135.8,0) rectangle (axis cs:136.2,1.59);
\draw[draw=none,fill=gray,very thin] (axis cs:136.8,0) rectangle (axis cs:137.2,1.58);
\draw[draw=none,fill=gray,very thin] (axis cs:137.8,0) rectangle (axis cs:138.2,1.61);
\draw[draw=none,fill=gray,very thin] (axis cs:138.8,0) rectangle (axis cs:139.2,1.62);
\draw[draw=none,fill=gray,very thin] (axis cs:139.8,0) rectangle (axis cs:140.2,1.63);
\draw[draw=none,fill=gray,very thin] (axis cs:140.8,0) rectangle (axis cs:141.2,1.63);
\draw[draw=none,fill=gray,very thin] (axis cs:141.8,0) rectangle (axis cs:142.2,1.64);
\draw[draw=none,fill=gray,very thin] (axis cs:142.8,0) rectangle (axis cs:143.2,1.65);
\draw[draw=none,fill=gray,very thin] (axis cs:143.8,0) rectangle (axis cs:144.2,1.63);
\draw[draw=none,fill=gray,very thin] (axis cs:144.8,0) rectangle (axis cs:145.2,1.6);
\draw[draw=none,fill=gray,very thin] (axis cs:145.8,0) rectangle (axis cs:146.2,1.63);
\draw[draw=none,fill=gray,very thin] (axis cs:146.8,0) rectangle (axis cs:147.2,1.64);
\draw[draw=none,fill=gray,very thin] (axis cs:147.8,0) rectangle (axis cs:148.2,1.64);
\draw[draw=none,fill=gray,very thin] (axis cs:148.8,0) rectangle (axis cs:149.2,1.62);
\draw[draw=none,fill=gray,very thin] (axis cs:149.8,0) rectangle (axis cs:150.2,1.64);
\draw[draw=none,fill=gray,very thin] (axis cs:150.8,0) rectangle (axis cs:151.2,1.64);
\draw[draw=none,fill=gray,very thin] (axis cs:151.8,0) rectangle (axis cs:152.2,1.64);
\draw[draw=none,fill=gray,very thin] (axis cs:152.8,0) rectangle (axis cs:153.2,1.65);
\draw[draw=none,fill=gray,very thin] (axis cs:153.8,0) rectangle (axis cs:154.2,1.63);
\draw[draw=none,fill=gray,very thin] (axis cs:154.8,0) rectangle (axis cs:155.2,1.61);
\draw[draw=none,fill=gray,very thin] (axis cs:155.8,0) rectangle (axis cs:156.2,1.62);
\draw[draw=none,fill=gray,very thin] (axis cs:156.8,0) rectangle (axis cs:157.2,1.64);
\draw[draw=none,fill=gray,very thin] (axis cs:157.8,0) rectangle (axis cs:158.2,1.64);
\draw[draw=none,fill=gray,very thin] (axis cs:158.8,0) rectangle (axis cs:159.2,1.61);
\label{plotyyref:leg1}
\end{axis}

\begin{axis}[
ytick pos=right,
axis y line=right,
width=7cm,
axis x line=none,
axis line style={-},
tick align=inside,
x grid style={light gray},
legend columns=1,
legend style={  at={(0.6,0.1)},
        anchor=south,font=\fontsize{5}{6}\selectfont},
ylabel={Avg. Energy Consumption/Device (nJ)},
ymajorgrids,
xmin=-8.17, xmax=167.17,
y tick label style={color=black},
ylabel style={yshift=5pt},
label style={font=\footnotesize},
tick label style={font=\footnotesize},ymin=510.970783064303, ymax=1436.25305582325,
ytick style={color=black},
yticklabel style={anchor=west}
]
\addplot [thick, red, mark=*, mark size=2, mark repeat=10, mark options={solid}]
table [x=x, y=y, header=true]{%
x  y
0 553.02906818971
1 553.287081119965
2 553.982401567611
3 553.926151968281
4 553.303444836323
5 553.844754190245
6 553.451404082024
7 553.876746170645
8 553.303444836323
9 553.723439932587
10 580.329343858356
11 580.855863457555
12 597.451404082024
13 610.876746263301
14 610.876746170645
15 617.709512709988
16 617.303444836323
17 617.982401567611
18 632.026621520537
19 643.253747773239
20 643.329343858356
21 643.855863457555
22 643.855863457555
23 643.451404082024
24 643.847919672386
25 643.786984501767
26 643.876746170645
27 715.876746263301
28 715.613547961143
29 715.613547961143
30 715.709512709988
31 880.303446166579
32 880.629876343063
33 880.303453511942
34 880.303444836323
35 1390.50003455644
36 1387.16786750007
37 1383.64631337448
38 1387.49960603602
39 1385.34268885247
40 1385.2173632575
41 1387.63663961429
42 1385.12763194167
43 1386.92492303579
44 1385.64222426461
45 1382.55226278689
46 1388.20508936398
47 1386.06976696621
48 1385.77872821643
49 1384.80966004004
50 1385.77793633695
51 1387.88992477044
52 1386.03629123476
53 1386.90522013647
54 1386.50317323273
55 1386.38572934981
56 1385.93387170516
57 1385.97994920401
58 1383.79839299678
59 1384.00417888733
60 1385.31253304585
61 1387.12140260842
62 1389.55198743993
63 1385.16095070729
64 1386.62118321762
65 1388.32595943213
66 1388.91597343049
67 1390.21244836702
68 1387.08538315711
69 1384.92180089844
70 1386.48194093764
71 1390.06802008621
72 1385.00599988359
73 1387.09890205215
74 1385.84263013662
75 1385.82824579238
76 1390.19542517514
77 1388.22439342596
78 1388.66606964368
79 1388.57666094977
80 1388.21708820262
81 1386.14625211458
82 1387.26462841112
83 1387.08061886021
84 1385.42624101872
85 1383.60328819846
86 1389.29063041444
87 1383.64654923765
88 1384.87331155535
89 1382.83058584965
90 1384.04125867421
91 1382.3662293214
92 1385.52433607057
93 1384.55374214717
94 1382.24857620058
95 1384.18534340549
96 1383.8578313371
97 1381.34686771006
98 1381.81581969466
99 1381.35317781059
100 1381.48698621651
101 1380.50361481086
102 1381.03075924956
103 1382.78629054018
104 1385.46219799451
105 1385.44051691939
106 1384.60287240806
107 1384.17527547392
108 1389.26888181136
109 1387.27273725571
110 1390.46131592171
111 1388.83210192341
112 1387.60062117745
113 1391.66780133965
114 1388.9557425926
115 1388.35176600595
116 1390.8035078462
117 1388.05640051899
118 1387.59245771249
119 1389.18103159948
120 1388.9398349986
121 1383.83704683289
122 1386.17995964078
123 1386.94925440442
124 1384.2605728017
125 1386.49355541596
126 1387.91025473615
127 1390.36204079764
128 1390.91648066626
129 1387.59930672331
130 1388.2723793193
131 1387.36204013303
132 1385.58949527136
133 1388.68777933274
134 1384.38371124035
135 1391.44280693417
136 1391.56602048888
137 1387.54016338556
138 1388.96341301774
139 1387.19655396571
140 1387.08534346663
141 1388.85848311638
142 1390.42001168163
143 1388.86021057809
144 1391.54003688977
145 1392.68996760858
146 1390.51585833347
147 1394.19477069784
148 1391.44671811625
149 1393.61669307043
150 1390.0612992901
151 1390.89446692242
152 1391.91604464447
153 1390.67130879888
154 1391.4783127705
155 1392.02412922222
156 1389.18271616291
157 1389.40486848462
158 1391.41910495764
159 1390.26108539237
};
\addlegendentry{\scriptsize{Energy Consumption}}
\addlegendimage{/pgfplots/refstyle=plotyyref:leg1}
\addlegendentry{\scriptsize{Tech. Switching}}
\end{axis}

\end{tikzpicture}

%% file: figures/AoIvs.Nodes.tex
\begin{tikzpicture}

\begin{axis}[
legend cell align={left},
legend style={
   at={(0.97,0.03)},
  anchor=south east},
tick align=outside,
tick pos=left,
x grid style={darkgray176},
xlabel={Number of Nodes},
xmajorgrids,
xmin=1.65, xmax=9.35,
xminorgrids,
xtick style={color=black},
xtick={1,2,3,4,5,6,7,8,9,10},
xticklabels={
  \(\displaystyle {1}\),
  \(\displaystyle {2}\),
  \(\displaystyle {3}\),
  \(\displaystyle {4}\),
  \(\displaystyle {5}\),
  \(\displaystyle {6}\),
  \(\displaystyle {7}\),
  \(\displaystyle {8}\),
  \(\displaystyle {9}\),
  \(\displaystyle {10}\)
},
y grid style={darkgray176},
ylabel={Age Of Information},
ymajorgrids,
ymin=4.61441270232938, ymax=9.89475749350678,
yminorgrids,
ytick style={color=black},
ytick={4,5,6,7,8,9,10},
yticklabels={
  \(\displaystyle {4}\),
  \(\displaystyle {5}\),
  \(\displaystyle {6}\),
  \(\displaystyle {7}\),
  \(\displaystyle {8}\),
  \(\displaystyle {9}\),
  \(\displaystyle {10}\)
}
]
\addplot [semithick, blue, mark=asterisk, mark size=5, mark repeat=1, mark options={solid}]
table [x=x, y=y, header=true]{%
x  y
2 5.93196848408324
3 6.89641193176912
4 7.44318182776175
5 7.37567421997204
6 7.35066842684523
7 7.64965594178517
8 7.73552731841855
9 7.84186365012079
};
\addlegendentry{M-AoI RF}
\addplot [semithick, green, mark=triangle*, mark size=5, mark repeat=1, mark options={solid}]
table [x=x, y=y, header=true]{%
x  y
2 4.85442837465563
3 5.83479395744546
4 6.17405348040824
5 6.60062944353644
6 6.47670126353215
7 6.6796122517013
8 6.87521474116161
9 6.93818145757066
};
\addlegendentry{M-AoI RF-OWC}
\addplot [thick, blue, dashed]
table [x=x, y=y, header=true]{%
x  y
2 8.98273846801341
3 9.31254512432009
4 9.49063907734508
5 9.53190185545511
6 9.4601119079144
7 9.56067568526934
8 9.65474182118053
9 9.65139958913329
};
\addlegendentry{P-AoI RF}
\addplot [thick, green, dash pattern=on 1pt off 3pt on 3pt off 3pt]
table [x=x, y=y, header=true]{%
x  y
2 8.33227272727267
3 8.66472891306221
4 8.98547056068509
5 9
6 8.97269671049146
7 9.03843060705029
8 9.14279374999998
9 9.15540150889086
};
\addlegendentry{P-AoI RF-OWC}
\end{axis}

\end{tikzpicture}

%% file: figures/AoIvs.Aps.tex
\begin{tikzpicture}

\begin{axis}[
legend cell align={left},
legend style={
  at={(0.09,0.55)},
  anchor=west,
},
tick align=outside,
tick pos=left,
x grid style={darkgray176},
xlabel={Number of APs},
xmajorgrids,
xmin=1.65, xmax=9.35,
xminorgrids,
xtick style={color=black},
xtick={1,2,3,4,5,6,7,8,9,10},
xticklabels={
  \(\displaystyle {1}\),
  \(\displaystyle {2}\),
  \(\displaystyle {3}\),
  \(\displaystyle {4}\),
  \(\displaystyle {5}\),
  \(\displaystyle {6}\),
  \(\displaystyle {7}\),
  \(\displaystyle {8}\),
  \(\displaystyle {9}\),
  \(\displaystyle {10}\)
},
y grid style={darkgray176},
ylabel={Age Of Information},
ymajorgrids,
ymin=6.23493225800434, ymax=9.84874020460524,
yminorgrids,
ytick style={color=black},
ytick={6,6.5,7,7.5,8,8.5,9,9.5,10},
yticklabels={
  \(\displaystyle {6.0}\),
  \(\displaystyle {6.5}\),
  \(\displaystyle {7.0}\),
  \(\displaystyle {7.5}\),
  \(\displaystyle {8.0}\),
  \(\displaystyle {8.5}\),
  \(\displaystyle {9.0}\),
  \(\displaystyle {9.5}\),
  \(\displaystyle {10.0}\)
}
]
\addplot [semithick, blue, mark=asterisk, mark size=5, mark repeat=1, mark options={solid}]
table [x=x, y=y, header=true]{%
x  y
2 7.5
3 7.3
4 7.39494034766563
5 7.53631501515298
6 7.5
7 7.63360505633515
8 7.72118551586073
9 7.88643966024665
};
\addlegendentry{M-AoI RF}
\addplot [semithick, green, mark=triangle*, mark size=5, mark repeat=1, mark options={solid}]
table [x=x, y=y, header=true]{%
x  y
2 6.55
3 6.43687359251479
4 6.48803941787571
5 6.39919625557711
6 6.5
7 6.5
8 6.6
9 6.45
};
\addlegendentry{M-AoI RF-OWC}
\addplot [thick, blue, dashed]
table [x=x, y=y, header=true]{%
x  y
2 9.65
3 9.51562644773563
4 9.51271160052172
5 9.58983776157848
6 9.56779599426847
7 9.57114031044655
8 9.65840605579268
9 9.68447620703247
};
\addlegendentry{P-AoI RF}
\addplot [thick, green, dash pattern=on 1pt off 3pt on 3pt off 3pt]
table [x=x, y=y, header=true]{%
x  y
2 9.15
3 9.02855000971189
4 9.0153366611049
5 9.00400614755299
6 9.00290567216389
7 9.04855617672061
8 9
9 8.89
};
\addlegendentry{P-AoI RF-OWC}
\end{axis}

\end{tikzpicture}

%% file: figures/PDRvs.Nodes_NOWOC.tex
\begin{tikzpicture}

\begin{axis}[
legend cell align={left},
legend style={},
tick align=outside,
tick pos=left,
x grid style={darkgray176},
xlabel={Number of Nodes},
xmajorgrids,
xtick={1,2,3,4,5,6,7,8,9},
xmin=1.65, xmax=9.35,
xminorgrids,
xtick style={color=black},
y grid style={darkgray176},
ylabel={Successful Transmission Rate},
ymajorgrids,
ymin=0.71735, ymax=0.88565,
yminorgrids,
ytick style={color=black}
]
\addplot [semithick, blue, mark=asterisk, mark size=5, mark repeat=1, mark options={solid}]
table [x=x, y=y, header=true]{%
x  y
2 0.835
3 0.845
4 0.815
5 0.81
6 0.76
7 0.755
8 0.725
9 0.7255
};
\addlegendentry{RF}
\addplot [semithick, green, mark=triangle*, mark size=5, mark repeat=1, mark options={solid}]
table [x=x, y=y, header=true]{%
x  y
2 0.87
3 0.878
4 0.853
5 0.854
6 0.82
7 0.815
8 0.741
9 0.741
};
\addlegendentry{RF-OWC}
\end{axis}

\end{tikzpicture}

%% file: figures/Energyvs.Nodes_NOWOC.tex
\begin{tikzpicture}

\definecolor{darkgray176}{RGB}{176,176,176}
\definecolor{green}{RGB}{0,128,0}
\definecolor{lightgray204}{RGB}{204,204,204}

\begin{axis}[
legend cell align={left},
legend style={},
tick align=outside,
tick pos=left,
x grid style={darkgray176},
xlabel={Number of Nodes},
xmajorgrids,
xmin=1.7, xmax=8.3,
xminorgrids,
xtick style={color=black},
xtick={1,2,3,4,5,6,7,8,9},
xticklabels={
  \(\displaystyle {1}\),
  \(\displaystyle {2}\),
  \(\displaystyle {3}\),
  \(\displaystyle {4}\),
  \(\displaystyle {5}\),
  \(\displaystyle {6}\),
  \(\displaystyle {7}\),
  \(\displaystyle {8}\),
  \(\displaystyle {9}\)
},
y grid style={darkgray176},
ylabel={Energy Consumption Per Node (J)},
ymajorgrids,
ymin=1.780815e-08, ymax=7.610815e-08,
yminorgrids,
ytick style={color=black},
ytick={1e-08,2e-08,3e-08,4e-08,5e-08,6e-08,7e-08,8e-08},
yticklabels={
  \(\displaystyle {1}\),
  \(\displaystyle {2}\),
  \(\displaystyle {3}\),
  \(\displaystyle {4}\),
  \(\displaystyle {5}\),
  \(\displaystyle {6}\),
  \(\displaystyle {7}\),
  \(\displaystyle {8}\)
}
]
\addplot [semithick, blue, mark=asterisk, mark size=5, mark repeat=1, mark options={solid}]
table [x=x, y=y, header=true]{%
x  y
2 2.145815e-08
3 2.045815e-08
4 2.545815e-08
5 2.245815e-08
6 2.265815e-08
7 2.365815e-08
8 2.465815e-08
};
\addlegendentry{RF}
\addplot [semithick, green, mark=triangle*, mark size=5, mark repeat=1, mark options={solid}]
table [x=x, y=y, header=true]{%
x  y
2 5.245815e-08
3 6.945815e-08
4 7.345815e-08
5 7.045815e-08
6 4.945815e-08
7 3.945815e-08
8 3.545815e-08
};
\addlegendentry{RF-OWC}
\end{axis}

\end{tikzpicture}

%% file: figures/PDRvs.APs_NOWOC.tex
\begin{tikzpicture}

\definecolor{darkgray176}{RGB}{176,176,176}
\definecolor{green}{RGB}{0,128,0}
\definecolor{lightgray204}{RGB}{204,204,204}

\begin{axis}[
legend cell align={left},
legend style={
  at={(0.03,0.97)},
  anchor=north west,
},
tick align=outside,
tick pos=left,
x grid style={darkgray176},
xlabel={Number of APs},
xmajorgrids,
xmin=1.65, xmax=9.35,
xminorgrids,
xtick={1,2,3,4,5,6,7,8,9},
xtick style={color=black},
y grid style={darkgray176},
ylabel={Successful Transmission Rate},
ymajorgrids,
ymin=0.42675, ymax=0.93825,
yminorgrids,
ytick style={color=black}
]
\addplot [semithick, blue, mark=asterisk, mark size=5, mark repeat=1, mark options={solid}]
table [x=x, y=y, header=true]{%
x  y
2 0.45
3 0.58
4 0.65
5 0.668
6 0.72
7 0.73
8 0.78
9 0.798
};
\addlegendentry{RF}
\addplot [semithick, green, mark=triangle*, mark size=5, mark repeat=1, mark options={solid}]
table [x=x, y=y, header=true]{%
x  y
2 0.5
3 0.72
4 0.78
5 0.79
6 0.88
7 0.89
8 0.915
9 0.913
};
\addlegendentry{RF-OWC}
\end{axis}

\end{tikzpicture}

%% file: figures/Energyvs.APs_NOWOC.tex
\begin{tikzpicture}

\begin{axis}[
legend cell align={left},
legend style={
  at={(0.03,0.97)},
  anchor=north west,
},
tick align=outside,
tick pos=left,
x grid style={darkgray176},
xlabel={Number of APs},
xmajorgrids,
xmin=1.7, xmax=8.3,
xminorgrids,
xtick style={color=black},
xtick={1,2,3,4,5,6,7,8,9},
xticklabels={
  \(\displaystyle {1}\),
  \(\displaystyle {2}\),
  \(\displaystyle {3}\),
  \(\displaystyle {4}\),
  \(\displaystyle {5}\),
  \(\displaystyle {6}\),
  \(\displaystyle {7}\),
  \(\displaystyle {8}\),
  \(\displaystyle {9}\)
},
y grid style={darkgray176},
ylabel={Energy Consumption Per Node (J)},
ymajorgrids,
ymin=5.3671622719744e-08, ymax=8.6121622719744e-08,
yminorgrids,
ytick style={color=black},
ytick={5e-08,5.5e-08,6e-08,6.5e-08,7e-08,7.5e-08,8e-08,8.5e-08,9e-08},
yticklabels={
  \(\displaystyle {5.0}\),
  \(\displaystyle {5.5}\),
  \(\displaystyle {6.0}\),
  \(\displaystyle {6.5}\),
  \(\displaystyle {7.0}\),
  \(\displaystyle {7.5}\),
  \(\displaystyle {8.0}\),
  \(\displaystyle {8.5}\),
  \(\displaystyle {9.0}\)
}
]
\addplot [semithick, blue, mark=asterisk, mark size=5, mark repeat=1, mark options={solid}]
table [x=x, y=y, header=true]{%
x  y
2 5.6646622719744e-08
3 5.5146622719744e-08
4 6.0146622719744e-08
5 6.0946622719744e-08
6 6.1346622719744e-08
7 6.1946622719744e-08
8 6.21662271974402e-08
};
\addlegendentry{RF}
\addplot [semithick, green, mark=triangle*, mark size=5, mark repeat=1, mark options={solid}]
table [x=x, y=y, header=true]{%
x  y
2 6.0646622719744e-08
3 7.2646622719744e-08
4 7.6646622719744e-08
5 8.1646622719744e-08
6 8.4646622719744e-08
7 8.2146622719744e-08
8 8.4646622719744e-08
};
\addlegendentry{RF-OWC}
\end{axis}

\end{tikzpicture}

%% file: figures/LabelOccurencesfigure.tex
\begin{tikzpicture}

\definecolor{darkgray176}{RGB}{176,176,176}
\definecolor{green}{RGB}{0,128,0}
\definecolor{lightgray204}{RGB}{204,204,204}

\begin{axis}[
legend cell align={left},
legend style={},
tick align=outside,
tick pos=left,
ylabel style={yshift=-2pt},
x grid style={darkgray176},
xlabel={Labels},
xmajorgrids,
xmin=-1.09, xmax=7.49,
xtick style={color=black},
y grid style={darkgray176},
ylabel={\textcolor{black}{Count}},
ymajorgrids,
xtick={0,1,...,7},
ymin=0, ymax=9753945.6,
ytick style={color=black}
]
\draw[draw=none,fill=red,fill opacity=0.7] (axis cs:-0.4,0) rectangle (axis cs:0.4,9289472);
\addlegendimage{ybar,ybar legend,draw=none,fill=red,fill opacity=0.7}
\addlegendentry{Original Classes}

\draw[draw=none,fill=red,fill opacity=0.7] (axis cs:0.6,0) rectangle (axis cs:1.4,1099852);
\draw[draw=none,fill=red,fill opacity=0.7] (axis cs:1.6,0) rectangle (axis cs:2.4,411036);
\draw[draw=none,fill=darkgray176,fill opacity=0.9] (axis cs:-0.7,0) rectangle (axis cs:0.1,4155687);
\addlegendimage{ybar,ybar legend,draw=none,fill=darkgray176,fill opacity=0.9}
\addlegendentry{Augmented Classes}

\draw[draw=none,fill=darkgray176,fill opacity=0.9] (axis cs:0.3,0) rectangle (axis cs:1.1,1501655);
\draw[draw=none,fill=darkgray176,fill opacity=0.9] (axis cs:1.3,0) rectangle (axis cs:2.1,395672);
\draw[draw=none,fill=darkgray176,fill opacity=0.9] (axis cs:2.3,0) rectangle (axis cs:3.1,837525);
\draw[draw=none,fill=darkgray176,fill opacity=0.9] (axis cs:3.3,0) rectangle (axis cs:4.1,112289);
\draw[draw=none,fill=darkgray176,fill opacity=0.9] (axis cs:4.3,0) rectangle (axis cs:5.1,2794605);
\draw[draw=none,fill=darkgray176,fill opacity=0.9] (axis cs:5.3,0) rectangle (axis cs:6.1,704180);
\draw[draw=none,fill=darkgray176,fill opacity=0.9] (axis cs:6.3,0) rectangle (axis cs:7.1,298747);
\end{axis}

\end{tikzpicture}

%% file: figures/accuracy.tex
\begin{tikzpicture}

\begin{axis}[
legend cell align={left},
legend style={
  at={(0.97,0.03)},
  anchor=south east,
},
tick align=outside,
tick pos=left,
x grid style={darkgray176},
xlabel={Epochs},
xmajorgrids,
xmin=-20.35, xmax=510,
ylabel style={yshift=-5pt},
xtick style={color=black},
y grid style={darkgray176},
ylabel={Accuracy},
ymajorgrids,
ymin=0.44047619047619, ymax=1.01666666666667,
ytick style={color=black}
]
\addplot [semithick, red]
table [x=x, y=y]{%
x  y
0 0.528282828282828
1 0.53989898989899
2 0.547474747474747
3 0.546969696969697
4 0.54040404040404
5 0.544949494949495
6 0.551010101010101
7 0.546464646464647
8 0.561111111111111
9 0.549494949494949
10 0.576767676767677
11 0.57020202020202
12 0.568181818181818
13 0.574242424242424
14 0.590909090909091
15 0.578282828282828
16 0.580808080808081
17 0.604040404040404
18 0.583838383838384
19 0.612121212121212
20 0.608585858585859
21 0.608585858585859
22 0.587373737373737
23 0.597979797979798
24 0.603030303030303
25 0.60959595959596
26 0.626262626262626
27 0.638888888888889
28 0.619191919191919
29 0.625757575757576
30 0.641919191919192
31 0.611616161616162
32 0.618181818181818
33 0.622222222222222
34 0.624747474747475
35 0.625252525252525
36 0.646969696969697
37 0.660606060606061
38 0.669191919191919
39 0.663636363636364
40 0.67020202020202
41 0.667171717171717
42 0.667676767676768
43 0.686868686868687
44 0.668181818181818
45 0.688888888888889
46 0.685858585858586
47 0.692424242424242
48 0.697474747474747
49 0.675252525252525
50 0.693434343434344
51 0.716666666666667
52 0.705555555555556
53 0.714141414141414
54 0.715151515151515
55 0.736868686868687
56 0.722727272727273
57 0.74040404040404
58 0.742929292929293
59 0.758585858585859
60 0.737373737373737
61 0.74040404040404
62 0.733838383838384
63 0.754545454545455
64 0.758585858585859
65 0.768181818181818
66 0.753030303030303
67 0.773232323232323
68 0.774747474747475
69 0.786363636363636
70 0.785353535353535
71 0.783333333333333
72 0.794444444444445
73 0.801010101010101
74 0.808585858585859
75 0.800505050505051
76 0.800505050505051
77 0.80959595959596
78 0.809090909090909
79 0.822727272727273
80 0.823737373737374
81 0.824242424242424
82 0.825757575757576
83 0.826767676767677
84 0.83030303030303
85 0.831818181818182
86 0.826262626262626
87 0.86010101010101
88 0.847474747474747
89 0.843939393939394
90 0.841919191919192
91 0.835858585858586
92 0.853535353535354
93 0.846464646464646
94 0.842424242424243
95 0.851010101010101
96 0.845454545454545
97 0.855050505050505
98 0.854545454545454
99 0.850505050505051
100 0.865151515151515
101 0.872727272727273
102 0.858585858585859
103 0.871212121212121
104 0.865656565656566
105 0.867676767676768
106 0.875757575757576
107 0.871212121212121
108 0.876767676767677
109 0.883333333333333
110 0.877272727272727
111 0.879292929292929
112 0.873039215686275
113 0.87843137254902
114 0.881862745098039
115 0.888235294117647
116 0.889705882352941
117 0.882843137254902
118 0.884803921568627
119 0.882352941176471
120 0.880392156862745
121 0.882843137254902
122 0.882352941176471
123 0.882352941176471
124 0.879411764705882
125 0.895588235294118
126 0.880882352941176
127 0.881862745098039
128 0.893137254901961
129 0.882843137254902
130 0.877450980392157
131 0.887373737373737
132 0.903030303030303
133 0.894444444444444
134 0.884343434343434
135 0.897474747474748
136 0.892424242424242
137 0.9
138 0.887878787878788
139 0.896464646464646
140 0.893939393939394
141 0.892929292929293
142 0.888383838383838
143 0.88989898989899
144 0.894949494949495
145 0.895454545454546
146 0.903030303030303
147 0.891414141414141
148 0.895959595959596
149 0.904545454545455
150 0.906060606060606
151 0.902525252525253
152 0.906060606060606
153 0.904545454545455
154 0.9
155 0.902525252525253
156 0.904545454545455
157 0.904545454545455
158 0.897474747474748
159 0.9
160 0.90050505050505
161 0.904545454545455
162 0.898989898989899
163 0.906565656565657
164 0.906060606060606
165 0.905555555555556
166 0.905050505050505
167 0.901010101010101
168 0.904040404040404
169 0.9
170 0.901515151515151
171 0.906565656565657
172 0.911111111111111
173 0.907575757575758
174 0.91969696969697
175 0.902525252525253
176 0.907070707070707
177 0.904545454545455
178 0.906060606060606
179 0.907575757575758
180 0.898484848484848
181 0.907070707070707
182 0.914141414141414
183 0.898484848484848
184 0.901010101010101
185 0.90959595959596
186 0.898484848484848
187 0.896969696969697
188 0.91010101010101
189 0.905555555555556
190 0.904545454545455
191 0.904040404040404
192 0.914141414141414
193 0.912626262626263
194 0.90959595959596
195 0.914141414141414
196 0.916161616161616
197 0.894444444444444
198 0.904545454545455
199 0.903030303030303
200 0.903535353535354
201 0.902941176470588
202 0.90343137254902
203 0.900980392156863
204 0.905882352941176
205 0.900490196078431
206 0.906372549019608
207 0.902941176470588
208 0.900490196078431
209 0.90343137254902
210 0.897058823529412
211 0.906862745098039
212 0.895588235294118
213 0.90343137254902
214 0.901960784313726
215 0.901470588235294
216 0.896078431372549
217 0.903921568627451
218 0.893137254901961
219 0.906862745098039
220 0.904411764705882
221 0.904411764705882
222 0.901010101010101
223 0.906060606060606
224 0.908585858585859
225 0.904545454545455
226 0.902020202020202
227 0.904545454545455
228 0.906565656565657
229 0.88989898989899
230 0.891919191919192
231 0.906060606060606
232 0.893939393939394
233 0.903030303030303
234 0.901010101010101
235 0.903030303030303
236 0.904545454545455
237 0.895454545454546
238 0.896969696969697
239 0.897474747474748
240 0.902020202020202
241 0.904545454545455
242 0.895959595959596
243 0.894949494949495
244 0.90050505050505
245 0.91010101010101
246 0.894444444444444
247 0.896969696969697
248 0.902525252525253
249 0.89949494949495
250 0.907575757575758
251 0.894949494949495
252 0.902020202020202
253 0.903030303030303
254 0.895959595959596
255 0.891414141414141
256 0.905555555555556
257 0.897474747474748
258 0.88989898989899
259 0.902020202020202
260 0.893434343434343
261 0.905050505050505
262 0.906565656565657
263 0.9
264 0.903030303030303
265 0.890909090909091
266 0.894444444444444
267 0.903535353535354
268 0.895098039215686
269 0.898529411764706
270 0.890196078431372
271 0.896078431372549
272 0.893627450980392
273 0.905392156862745
274 0.899019607843137
275 0.905882352941176
276 0.886764705882353
277 0.9
278 0.892647058823529
279 0.907843137254902
280 0.900490196078431
281 0.889705882352941
282 0.900980392156863
283 0.902450980392157
284 0.897549019607843
285 0.894607843137255
286 0.888235294117647
287 0.895588235294118
288 0.890686274509804
289 0.909090909090909
290 0.894949494949495
291 0.890909090909091
292 0.90050505050505
293 0.88989898989899
294 0.897979797979798
295 0.89040404040404
296 0.895959595959596
297 0.894949494949495
298 0.901010101010101
299 0.892929292929293
300 0.895959595959596
301 0.882323232323232
302 0.89949494949495
303 0.908585858585859
304 0.898484848484848
305 0.902020202020202
306 0.903535353535354
307 0.90050505050505
308 0.890909090909091
309 0.896464646464646
310 0.892424242424242
311 0.893939393939394
312 0.89040404040404
313 0.895959595959596
314 0.903535353535354
315 0.892424242424242
316 0.891414141414141
317 0.892929292929293
318 0.903030303030303
319 0.898484848484848
320 0.90959595959596
321 0.894949494949495
322 0.891919191919192
323 0.899019607843137
324 0.895098039215686
325 0.905882352941176
326 0.900490196078431
327 0.891666666666667
328 0.902450980392157
329 0.898039215686275
330 0.904901960784314
331 0.897058823529412
332 0.897549019607843
333 0.890196078431372
334 0.88921568627451
335 0.912745098039216
336 0.89656862745098
337 0.898039215686275
338 0.904901960784314
339 0.897058823529412
340 0.905882352941176
341 0.893137254901961
342 0.89656862745098
343 0.905392156862745
344 0.899019607843137
345 0.901470588235294
346 0.905882352941176
347 0.905555555555556
348 0.897979797979798
349 0.901515151515151
350 0.907070707070707
351 0.914141414141414
352 0.897979797979798
353 0.891919191919192
354 0.912121212121212
355 0.901515151515151
356 0.903535353535354
357 0.898484848484848
358 0.902525252525253
359 0.893434343434343
360 0.901010101010101
361 0.913636363636364
362 0.893939393939394
363 0.912626262626263
364 0.897474747474748
365 0.908080808080808
366 0.894949494949495
367 0.907575757575758
368 0.895959595959596
369 0.897979797979798
370 0.886363636363636
371 0.906060606060606
372 0.905050505050505
373 0.911111111111111
374 0.901010101010101
375 0.914141414141414
376 0.893434343434343
377 0.903030303030303
378 0.885858585858586
379 0.897979797979798
380 0.89949494949495
381 0.893939393939394
382 0.902525252525253
383 0.883838383838384
384 0.897979797979798
385 0.914646464646465
386 0.896969696969697
387 0.892424242424242
388 0.901010101010101
389 0.901010101010101
390 0.894444444444444
391 0.903030303030303
392 0.892929292929293
393 0.897979797979798
394 0.898989898989899
395 0.908333333333333
396 0.910294117647059
397 0.902941176470588
398 0.91078431372549
399 0.909313725490196
400 0.893137254901961
401 0.906372549019608
402 0.907352941176471
403 0.904901960784314
404 0.909313725490196
405 0.908823529411765
406 0.899019607843137
407 0.899509803921569
408 0.890909090909091
409 0.896464646464646
410 0.892424242424242
411 0.893939393939394
412 0.89040404040404
413 0.895959595959596
414 0.903535353535354
415 0.892424242424242
416 0.891414141414141
417 0.892929292929293
418 0.903030303030303
419 0.898484848484848
420 0.90959595959596
421 0.894949494949495
422 0.891919191919192
423 0.899019607843137
424 0.895098039215686
425 0.905882352941176
426 0.900490196078431
427 0.891666666666667
428 0.902450980392157
429 0.898039215686275
430 0.904901960784314
431 0.897058823529412
432 0.897549019607843
433 0.890196078431372
434 0.88921568627451
435 0.912745098039216
436 0.89656862745098
437 0.898039215686275
438 0.904901960784314
439 0.897058823529412
440 0.905882352941176
441 0.893137254901961
442 0.89656862745098
443 0.905392156862745
444 0.899019607843137
445 0.901470588235294
446 0.905882352941176
447 0.905555555555556
448 0.897979797979798
449 0.901515151515151
450 0.907070707070707
451 0.914141414141414
452 0.897979797979798
453 0.891919191919192
454 0.912121212121212
455 0.901515151515151
456 0.903535353535354
457 0.898484848484848
458 0.902525252525253
459 0.893434343434343
460 0.901010101010101
461 0.913636363636364
462 0.893939393939394
463 0.912626262626263
464 0.897474747474748
465 0.908080808080808
466 0.894949494949495
467 0.907575757575758
468 0.895959595959596
469 0.897979797979798
470 0.886363636363636
471 0.906060606060606
472 0.905050505050505
473 0.911111111111111
474 0.901010101010101
475 0.914141414141414
476 0.893434343434343
477 0.903030303030303
478 0.885858585858586
479 0.897979797979798
480 0.89949494949495
481 0.893939393939394
482 0.902525252525253
483 0.883838383838384
484 0.897979797979798
485 0.914646464646465
486 0.896969696969697
487 0.892424242424242
488 0.901010101010101
489 0.901010101010101
490 0.894444444444444
491 0.903030303030303
492 0.892929292929293
493 0.897979797979798
494 0.898989898989899
495 0.908333333333333
496 0.910294117647059
497 0.902941176470588
498 0.91078431372549
499 0.909313725490196
400 0.893137254901961

};
\addlegendentry{Train}

\addplot [semithick, black]
table [x=x, y=y]{%
x  y
0 0.466666666666667
1 0.466666666666667
2 0.466666666666667
3 0.466666666666667
4 0.466666666666667
5 0.466666666666667
6 0.466666666666667
7 0.466666666666667
8 0.466666666666667
9 0.466666666666667
10 0.466666666666667
11 0.466666666666667
12 0.466666666666667
13 0.466666666666667
14 0.466666666666667
15 0.466666666666667
16 0.466666666666667
17 0.466666666666667
18 0.466666666666667
19 0.466666666666667
20 0.466666666666667
21 0.466666666666667
22 0.466666666666667
23 0.471428571428571
24 0.471428571428571
25 0.471428571428571
26 0.471428571428571
27 0.471428571428571
28 0.471428571428571
29 0.471428571428571
30 0.471428571428571
31 0.471428571428571
32 0.471428571428571
33 0.471428571428571
34 0.471428571428571
35 0.471428571428571
36 0.471428571428571
37 0.471428571428571
38 0.471428571428571
39 0.471428571428571
40 0.471428571428571
41 0.471428571428571
42 0.471428571428571
43 0.471428571428571
44 0.471428571428571
45 0.471428571428571
46 0.473809523809524
47 0.490476190476191
48 0.497619047619048
49 0.497619047619048
50 0.497619047619048
51 0.497619047619048
52 0.497619047619048
53 0.502380952380952
54 0.519047619047619
55 0.554761904761905
56 0.573809523809524
57 0.60952380952381
58 0.645238095238095
59 0.664285714285714
60 0.683333333333333
61 0.707142857142857
62 0.735714285714286
63 0.75
64 0.723809523809524
65 0.728571428571429
66 0.728571428571429
67 0.735714285714286
68 0.771428571428572
69 0.807142857142857
70 0.838095238095238
71 0.842857142857143
72 0.842857142857143
73 0.845238095238095
74 0.885714285714286
75 0.95
76 0.954761904761905
77 0.954761904761905
78 0.954761904761905
79 0.954761904761905
80 0.954761904761905
81 0.954761904761905
82 0.954761904761905
83 0.954761904761905
84 0.954761904761905
85 0.954761904761905
86 0.954761904761905
87 0.954761904761905
88 0.954761904761905
89 0.954761904761905
90 0.980952380952381
91 0.980952380952381
92 0.980952380952381
93 0.980952380952381
94 0.980952380952381
95 0.980952380952381
96 0.980952380952381
97 0.980952380952381
98 0.980952380952381
99 0.980952380952381
100 0.980952380952381
101 0.980952380952381
102 0.980952380952381
103 0.980952380952381
104 0.980952380952381
105 0.980952380952381
106 0.980952380952381
107 0.980952380952381
108 0.980952380952381
109 0.980952380952381
110 0.980952380952381
111 0.980952380952381
112 0.972222222222222
113 0.972222222222222
114 0.972222222222222
115 0.972222222222222
116 0.972222222222222
117 0.972222222222222
118 0.972222222222222
119 0.972222222222222
120 0.972222222222222
121 0.972222222222222
122 0.988888888888889
123 0.988888888888889
124 0.988888888888889
125 0.988888888888889
126 0.988888888888889
127 0.988888888888889
128 0.988888888888889
129 0.988888888888889
130 0.988888888888889
131 0.978571428571429
132 0.978571428571429
133 0.978571428571429
134 0.978571428571429
135 0.978571428571429
136 0.978571428571429
137 0.978571428571429
138 0.978571428571429
139 0.978571428571429
140 0.978571428571429
141 0.978571428571429
142 0.978571428571429
143 0.978571428571429
144 0.978571428571429
145 0.978571428571429
146 0.964285714285714
147 0.964285714285714
148 0.964285714285714
149 0.964285714285714
150 0.964285714285714
151 0.964285714285714
152 0.964285714285714
153 0.964285714285714
154 0.964285714285714
155 0.978571428571429
156 0.978571428571429
157 0.978571428571429
158 0.978571428571429
159 0.978571428571429
160 0.978571428571429
161 0.978571428571429
162 0.978571428571429
163 0.978571428571429
164 0.978571428571429
165 0.978571428571429
166 0.978571428571429
167 0.978571428571429
168 0.978571428571429
169 0.978571428571429
170 0.978571428571429
171 0.978571428571429
172 0.978571428571429
173 0.978571428571429
174 0.978571428571429
175 0.978571428571429
176 0.978571428571429
177 0.978571428571429
178 0.978571428571429
179 0.978571428571429
180 0.978571428571429
181 0.978571428571429
182 0.978571428571429
183 0.978571428571429
184 0.978571428571429
185 0.978571428571429
186 0.978571428571429
187 0.978571428571429
188 0.978571428571429
189 0.978571428571429
190 0.978571428571429
191 0.978571428571429
192 0.978571428571429
193 0.978571428571429
194 0.978571428571429
195 0.973809523809524
196 0.973809523809524
197 0.973809523809524
198 0.973809523809524
199 0.973809523809524
200 0.973809523809524
201 0.977777777777778
202 0.977777777777778
203 0.977777777777778
204 0.977777777777778
205 0.977777777777778
206 0.977777777777778
207 0.977777777777778
208 0.977777777777778
209 0.977777777777778
210 0.975
211 0.975
212 0.975
213 0.975
214 0.975
215 0.975
216 0.975
217 0.975
218 0.975
219 0.975
220 0.975
221 0.975
222 0.971428571428572
223 0.971428571428572
224 0.971428571428572
225 0.971428571428572
226 0.971428571428572
227 0.971428571428572
228 0.971428571428572
229 0.971428571428572
230 0.971428571428572
231 0.971428571428572
232 0.971428571428572
233 0.971428571428572
234 0.971428571428572
235 0.971428571428572
236 0.971428571428572
237 0.971428571428572
238 0.971428571428572
239 0.971428571428572
240 0.966666666666667
241 0.966666666666667
242 0.966666666666667
243 0.966666666666667
244 0.966666666666667
245 0.966666666666667
246 0.966666666666667
247 0.966666666666667
248 0.966666666666667
249 0.973809523809524
250 0.973809523809524
251 0.973809523809524
252 0.973809523809524
253 0.973809523809524
254 0.973809523809524
255 0.973809523809524
256 0.973809523809524
257 0.973809523809524
258 0.973809523809524
259 0.973809523809524
260 0.973809523809524
261 0.973809523809524
262 0.973809523809524
263 0.973809523809524
264 0.973809523809524
265 0.973809523809524
266 0.973809523809524
267 0.973809523809524
268 0.980555555555556
269 0.980555555555556
270 0.980555555555556
271 0.980555555555556
272 0.980555555555556
273 0.980555555555556
274 0.980555555555556
275 0.980555555555556
276 0.980555555555556
277 0.980555555555556
278 0.980555555555556
279 0.983333333333333
280 0.983333333333333
281 0.983333333333333
282 0.983333333333333
283 0.983333333333333
284 0.983333333333333
285 0.983333333333333
286 0.983333333333333
287 0.983333333333333
288 0.983333333333333
289 0.978571428571429
290 0.978571428571429
291 0.978571428571429
292 0.978571428571429
293 0.978571428571429
294 0.978571428571429
295 0.978571428571429
296 0.973809523809524
297 0.973809523809524
298 0.973809523809524
299 0.973809523809524
300 0.973809523809524
301 0.973809523809524
302 0.961904761904762
303 0.961904761904762
304 0.961904761904762
305 0.961904761904762
306 0.961904761904762
307 0.961904761904762
308 0.961904761904762
309 0.961904761904762
310 0.980952380952381
311 0.980952380952381
312 0.980952380952381
313 0.980952380952381
314 0.980952380952381
315 0.980952380952381
316 0.980952380952381
317 0.980952380952381
318 0.980952380952381
319 0.980952380952381
320 0.980952380952381
321 0.980952380952381
322 0.980952380952381
323 0.980555555555556
324 0.980555555555556
325 0.980555555555556
326 0.980555555555556
327 0.980555555555556
328 0.980555555555556
329 0.980555555555556
330 0.980555555555556
331 0.980555555555556
332 0.980555555555556
333 0.980555555555556
334 0.980555555555556
335 0.980555555555556
336 0.980555555555556
337 0.980555555555556
338 0.980555555555556
339 0.980555555555556
340 0.980555555555556
341 0.972222222222222
342 0.972222222222222
343 0.972222222222222
344 0.972222222222222
345 0.972222222222222
346 0.972222222222222
347 0.976190476190476
348 0.976190476190476
349 0.976190476190476
350 0.976190476190476
351 0.976190476190476
352 0.976190476190476
353 0.976190476190476
354 0.976190476190476
355 0.976190476190476
356 0.976190476190476
357 0.976190476190476
358 0.976190476190476
359 0.976190476190476
360 0.976190476190476
361 0.976190476190476
362 0.976190476190476
363 0.990476190476191
364 0.990476190476191
365 0.990476190476191
366 0.990476190476191
367 0.990476190476191
368 0.990476190476191
369 0.990476190476191
370 0.990476190476191
371 0.95952380952381
372 0.95952380952381
373 0.95952380952381
374 0.95952380952381
375 0.95952380952381
376 0.95952380952381
377 0.95952380952381
378 0.95952380952381
379 0.95952380952381
380 0.95952380952381
381 0.983333333333333
382 0.983333333333333
383 0.983333333333333
384 0.983333333333333
385 0.983333333333333
386 0.983333333333333
387 0.983333333333333
388 0.983333333333333
389 0.983333333333333
390 0.983333333333333
391 0.983333333333333
392 0.983333333333333
393 0.983333333333333
394 0.983333333333333
395 0.963888888888889
396 0.963888888888889
397 0.963888888888889
398 0.963888888888889
399 0.963888888888889
400 0.963888888888889
401 0.963888888888889
402 0.972222222222222
403 0.972222222222222
404 0.972222222222222
405 0.972222222222222
406 0.972222222222222
407 0.972222222222222
408 0.961904761904762
409 0.961904761904762
410 0.980952380952381
411 0.980952380952381
412 0.980952380952381
413 0.980952380952381
414 0.980952380952381
415 0.980952380952381
416 0.980952380952381
417 0.980952380952381
418 0.980952380952381
419 0.980952380952381
420 0.980952380952381
421 0.980952380952381
422 0.980952380952381
423 0.980555555555556
424 0.980555555555556
425 0.980555555555556
426 0.980555555555556
427 0.980555555555556
428 0.980555555555556
429 0.980555555555556
430 0.980555555555556
431 0.980555555555556
432 0.980555555555556
433 0.980555555555556
434 0.980555555555556
435 0.980555555555556
436 0.980555555555556
437 0.980555555555556
438 0.980555555555556
439 0.980555555555556
440 0.980555555555556
441 0.972222222222222
442 0.972222222222222
443 0.972222222222222
444 0.972222222222222
445 0.972222222222222
446 0.972222222222222
447 0.976190476190476
448 0.976190476190476
449 0.976190476190476
450 0.976190476190476
451 0.976190476190476
452 0.976190476190476
453 0.976190476190476
454 0.976190476190476
455 0.976190476190476
456 0.976190476190476
457 0.976190476190476
458 0.986190476190476
459 0.976190476190476
460 0.976190476190476
461 0.976190476190476
462 0.976190476190476
463 0.990476190476191
464 0.980476190476191
465 0.990476190476191
466 0.990476190476191
467 0.990476190476191
468 0.990476190476191
469 0.990476190476191
470 0.990476190476191
471 0.95952380952381
472 0.95952380952381
473 0.95952380952381
474 0.95952380952381
475 0.93952380952381
476 0.95952380952381
477 0.96952380952381
478 0.95952380952381
479 0.95952380952381
480 0.95952380952381
481 0.983333333333333
482 0.983333333333333
483 0.973333333333333
484 0.983333333333333
485 0.983333333333333
486 0.983333333333333
487 0.983333333333333
488 0.983333333333333
489 0.983333333333333
490 0.983333333333333
491 0.983333333333333
492 0.983333333333333
493 0.983333333333333
494 0.983333333333333
495 0.983888888888889
496 0.983888888888889
497 0.983888888888889
498 0.983888888888889
499 0.983888888888889
500 0.983888888888889

};
\addlegendentry{Validation}
\end{axis}

\end{tikzpicture}

%% file: figures/ConfusionMatrix.tex
\begin{tikzpicture}

\definecolor{darkgray176}{RGB}{176,176,176}
\definecolor{green}{RGB}{0,128,0}

\begin{axis}[
tick align=outside,
tick pos=left,
x grid style={darkgray176},
xlabel={Predicted Class},
xmin=0, xmax=8,
xtick style={color=black},
xtick={0.5,1.5,2.5,3.5,4.5,5.5,6.5,7.5},
xticklabels={0,1,2,3,4,5,6,7},
y dir=reverse,
y grid style={darkgray176},
ylabel={True Class},
ymin=0, ymax=8,
ytick style={color=black},
ytick={0.5,1.5,2.5,3.5,4.5,5.5,6.5,7.5},
yticklabel style={rotate=90.0},
yticklabels={0,1,2,3,4,5,6,7}
]
\addplot graphics [includegraphics cmd=\pgfimage,xmin=0, xmax=8, ymin=8, ymax=0] {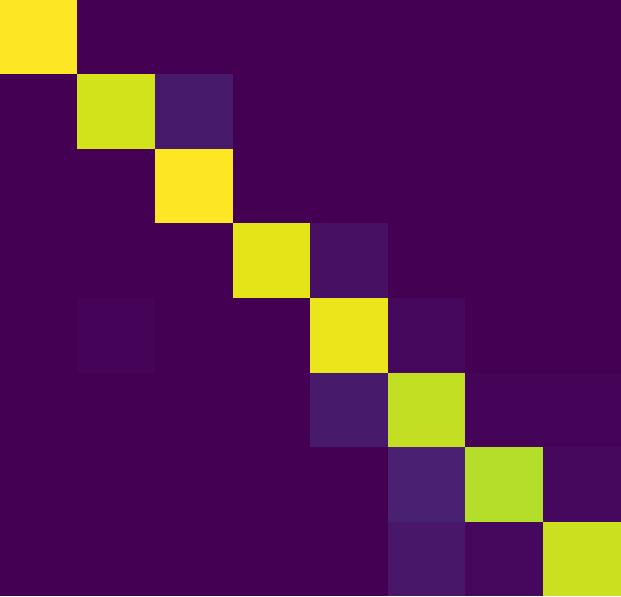};
\draw (axis cs:0.5,0.5) node[
  scale=0.8,
  text=black,
  rotate=0.0
]{\bfseries 1.00};
\draw (axis cs:1.5,1.5) node[
  scale=0.8,
  text=green,
  rotate=0.0,
  align=center
]{\bfseries 0.94 \\(+0.02)};
\draw (axis cs:2.5,1.5) node[
  scale=0.8,
  text=green,
  rotate=0.0,
  align=center
]{\bfseries 0.06 \\(-0.02)};
\draw (axis cs:1.5,2.5) node[
  scale=0.8,
  text=red,
  rotate=0.0,
  align=center
]{\bfseries 0.04 \\(+0.03)};
\draw (axis cs:2.5,2.5) node[
  scale=0.8,
  text=red,
  rotate=0.0,
  align=center
]{\bfseries 0.96 \\(-0.02)};
\draw (axis cs:5.5,2.5) node[
  scale=0.8,
  text=green,
  rotate=0.0,
  align=center
]{\bfseries 0.00 \\(-0.01)};
\draw (axis cs:3.5,3.5) node[
  scale=0.8,
  text=green,
  rotate=0.0,
  align=center
]{\bfseries 0.93 \\(+0.05)};
\draw (axis cs:4.5,3.5) node[
  scale=0.8,
  text=green,
  rotate=0.0,
  align=center
]{\bfseries 0.07 \\(-0.05)};
\draw (axis cs:3.5,4.5) node[
  scale=0.8,
  text=green,
  rotate=0.0,
  align=center
]{\bfseries 0.10 \\(-0.01)};
\draw (axis cs:4.5,4.5) node[
  scale=0.8,
  text=green,
  rotate=0.0,
  align=center
]{\bfseries 0.90 \\(+0.09)};
\draw (axis cs:5.5,4.5) node[
  scale=0.8,
  text=green,
  rotate=0.0,
  align=center
]{\bfseries 0.00 \\(-0.02)};
\draw (axis cs:6.5,4.5) node[
  scale=0.8,
  text=green,
  rotate=0.0,
  align=center
]{\bfseries 0.00 \\(-0.06)};
\draw (axis cs:4.5,5.5) node[
  scale=0.8,
  text=green,
  rotate=0.0,
  align=center
]{\bfseries 0.00 \\(-0.01)};
\draw (axis cs:5.5,5.5) node[
  scale=0.8,
  text=green,
  rotate=0.0,
  align=center
]{\bfseries 0.92 \\(+0.04)};
\draw (axis cs:6.5,5.5) node[
  scale=0.8,
  text=green,
  rotate=0.0,
  align=center
]{\bfseries 0.08 \\(-0.02)};
\draw (axis cs:7.5,5.5) node[
  scale=0.8,
  text=green,
  rotate=0.0,
  align=center
]{\bfseries 0.00 \\(-0.01)};
\draw (axis cs:5.5,6.5) node[
  scale=0.8,
  text=green,
  rotate=0.0,
  align=center
]{\bfseries 0.03 \\(-0.01)};
\draw (axis cs:6.5,6.5) node[
  scale=0.8,
  text=green,
  rotate=0.0,
  align=center
]{\bfseries 0.90 \\(+0.02)};
\draw (axis cs:7.5,6.5) node[
  scale=0.8,
  text=green,
  rotate=0.0,
  align=center
]{\bfseries 0.07 \\(-0.01)};
\draw (axis cs:4.5,7.5) node[
  scale=0.8,
  text=green,
  rotate=0.0,
  align=center
]{\bfseries 0.00 \\(-0.03)};
\draw (axis cs:5.5,7.5) node[
  scale=0.8,
  text=green,
  rotate=0.0,
  align=center
]{\bfseries 0.03 \\(-0.03)};
\draw (axis cs:6.5,7.5) node[
  scale=0.8,
  text=red,
  rotate=0.0,
  align=center
]{\bfseries 0.05 \\(+0.02)};
\draw (axis cs:7.5,7.5) node[
  scale=0.8,
  text=green,
  rotate=0.0,
  align=center
]{\bfseries 0.92 \\(+0.04)};
\end{axis}

\end{tikzpicture}

%% file: figures/Trans_Embeddings.tex
\begin{tikzpicture}

\definecolor{darkgray176}{RGB}{176,176,176}

\begin{groupplot}[group style={group size=3 by 1, horizontal sep=1.3cm},label style={font=\fontsize{16}{16}\selectfont},      
  ticklabel style={font=\fontsize{15}{15}\selectfont}  
  ] 
\nextgroupplot[
tick align=outside,
tick pos=left,
x grid style={darkgray176},
xlabel={t-SNE Dimension 1},
xlabel style={yshift=-1.5ex},
xmajorgrids,
xmin=-0.401639840006828, xmax=2.13482685983181,
xtick style={color=black},
xtick={-0.5,0,0.5,1,1.5,2,2.5},
xticklabels={\ensuremath{-}1,0,1,2,3,,},
y grid style={darkgray176},
ylabel={t-SNE Dimension 2},
ylabel style={yshift=1.5ex},
ymajorgrids,
ymin=-11.6160777568817, ymax=-9.36682000160217,
ytick style={color=black}
]
\addplot [
  colormap={mymap}{[1pt]
 rgb(0pt)=(0.894117647058824,0.101960784313725,0.109803921568627);
  rgb(1pt)=(0.215686274509804,0.494117647058824,0.72156862745098);
  rgb(2pt)=(0.301960784313725,0.686274509803922,0.290196078431373);
  rgb(3pt)=(0.596078431372549,0.305882352941176,0.63921568627451);
  rgb(4pt)=(1,0.498039215686275,0);
  rgb(5pt)=(1,1,0.2);
  rgb(6pt)=(0.650980392156863,0.337254901960784,0.156862745098039);
  rgb(7pt)=(0.968627450980392,0.505882352941176,0.749019607843137);
  rgb(8pt)=(0.6,0.6,0.6)
},
  only marks,
  scatter,
  scatter src=explicit
]
table [x=x, y=y, meta=colordata]{%
x  y  colordata
0.843450427055359 -9.5073881149292 0.0
-0.139912083745003 -10.1987972259521 5.0
-0.238732397556305 -10.3510522842407 5.0
-0.286345899105072 -10.4386835098267 5.0
0.881970763206482 -9.46905899047852 0.0
-0.114110760390759 -10.2459774017334 5.0
-0.220774218440056 -10.4212207794189 7.0
-0.261887222528458 -10.4853820800781 5.0
1.87329292297363 -10.1982727050781 5.0
1.85452377796173 -10.2425012588501 5.0
0.748403787612915 -11.4163618087769 1.0
0.711647868156433 -11.4472246170044 1.0
1.97253608703613 -10.3420829772949 5.0
1.95223641395569 -10.4217805862427 5.0
0.984982907772064 -11.42103099823 1.0
0.815788924694061 -11.5121812820435 1.0
2.01953291893005 -10.4371252059937 5.0
1.99609100818634 -10.4760255813599 5.0
1.02349650859833 -11.4457874298096 2.0
0.917373836040497 -11.5138387680054 1.0
};

\nextgroupplot[
tick align=outside,
tick pos=left,
x grid style={darkgray176},
xlabel={t-SNE Dimension 1},
xlabel style={yshift=-1.5ex},
xmajorgrids,
xmin=-5.59057717323303, xmax=-3.50990672111511,
xtick style={color=black},
y grid style={darkgray176},
ymajorgrids,
ymin=0.0440159447491169, ymax=2.38661302551627,
ytick style={color=black}
]
\addplot [
  colormap={mymap}{[1pt]
 rgb(0pt)=(0.894117647058824,0.101960784313725,0.109803921568627);
  rgb(1pt)=(0.215686274509804,0.494117647058824,0.72156862745098);
  rgb(2pt)=(0.301960784313725,0.686274509803922,0.290196078431373);
  rgb(3pt)=(0.596078431372549,0.305882352941176,0.63921568627451);
  rgb(4pt)=(1,0.498039215686275,0);
  rgb(5pt)=(1,1,0.2);
  rgb(6pt)=(0.650980392156863,0.337254901960784,0.156862745098039);
  rgb(7pt)=(0.968627450980392,0.505882352941176,0.749019607843137);
  rgb(8pt)=(0.6,0.6,0.6)
},
  only marks,
  scatter,
  scatter src=explicit
]
table [x=x, y=y, meta=colordata]{%
x  y  colordata
-5.49246978759766 1.32637131214142 0.0
-4.76631689071655 0.164412334561348 5.0
-4.74696588516235 0.150497630238533 5.0
-4.78611326217651 0.178616464138031 7.0
-5.49600124359131 1.2949070930481 0.0
-4.83264827728271 0.166820555925369 5.0
-4.8131103515625 0.152782276272774 5.0
-4.84935998916626 0.178955838084221 5.0
-4.52043151855469 2.25792789459229 5.0
-4.58584976196289 2.27095150947571 5.0
-3.60671997070312 1.07970702648163 1.0
-3.61303639411926 1.10009932518005 1.0
-4.49873638153076 2.26699042320251 5.0
-4.56388330459595 2.28013134002686 5.0
-3.60448265075684 1.09929871559143 2.0
-3.60760545730591 1.10934150218964 1.0
-4.54348230361938 2.24867844581604 5.0
-4.60461664199829 2.26295161247253 5.0
-3.61531186103821 1.0809360742569 1.0
-3.61227345466614 1.07065153121948 1.0
};

\nextgroupplot[
colorbar,
colorbar style={ylabel={}},
colormap={mymap}{[1pt]
 rgb(0pt)=(0.894117647058824,0.101960784313725,0.109803921568627);
  rgb(1pt)=(0.215686274509804,0.494117647058824,0.72156862745098);
  rgb(2pt)=(0.301960784313725,0.686274509803922,0.290196078431373);
  rgb(3pt)=(0.596078431372549,0.305882352941176,0.63921568627451);
  rgb(4pt)=(1,0.498039215686275,0);
  rgb(5pt)=(1,1,0.2);
  rgb(6pt)=(0.650980392156863,0.337254901960784,0.156862745098039);
  rgb(7pt)=(0.968627450980392,0.505882352941176,0.749019607843137);
  rgb(8pt)=(0.6,0.6,0.6)
},
point meta max=7,
point meta min=0,
tick align=outside,
tick pos=left,
x grid style={darkgray176},
xlabel={t-SNE Dimension 1},
xlabel style={yshift=-1.5ex},
xmajorgrids,
xmin=-5.03339666128159, xmax=-3.02251082658768,
xtick style={color=black},
y grid style={darkgray176},
ymajorgrids,
ymin=1.84164279103279, ymax=4.10199612975121,
ytick style={color=black}
]
\addplot [
  colormap={mymap}{[1pt]
 rgb(0pt)=(0.894117647058824,0.101960784313725,0.109803921568627);
  rgb(1pt)=(0.215686274509804,0.494117647058824,0.72156862745098);
  rgb(2pt)=(0.301960784313725,0.686274509803922,0.290196078431373);
  rgb(3pt)=(0.596078431372549,0.305882352941176,0.63921568627451);
  rgb(4pt)=(1,0.498039215686275,0);
  rgb(5pt)=(1,1,0.2);
  rgb(6pt)=(0.650980392156863,0.337254901960784,0.156862745098039);
  rgb(7pt)=(0.968627450980392,0.505882352941176,0.749019607843137);
  rgb(8pt)=(0.6,0.6,0.6)
},
  only marks,
  scatter,
  scatter src=explicit
]
table [x=x, y=y, meta=colordata]{%
x  y  colordata
-3.11391472816467 2.75132203102112 0.0
-4.20363521575928 1.97164189815521 5.0
-4.24149990081787 1.96341395378113 5.0
-4.24218797683716 1.96327412128448 5.0
-3.11977458000183 2.72709918022156 0.0
-4.15689325332642 1.95221710205078 5.0
-4.19085741043091 1.94465732574463 5.0
-4.19140386581421 1.9443861246109 6.0
-3.61841320991516 3.97151875495911 5.0
-3.56868839263916 3.96263980865479 5.0
-4.93963956832886 3.25675272941589 1.0
-4.93989706039429 3.25653100013733 1.0
-3.64595365524292 3.99863839149475 7.0
-3.59303641319275 3.98728036880493 5.0
-4.93111991882324 3.28576469421387 1.0
-4.94199275970459 3.2729754447937 1.0
-3.64640474319458 3.9992527961731 5.0
-3.59351086616516 3.98787879943848 5.0
-4.93123149871826 3.28602719306946 1.0
-4.94188404083252 3.27335143089294 1.0
};
\end{groupplot}

\end{tikzpicture}

%% file: figures/Indu_Embeddings.tex
\begin{tikzpicture}

\definecolor{darkgray176}{RGB}{176,176,176}

\begin{groupplot}[group style={group size=3 by 1, horizontal sep=1.3cm},label style={font=\fontsize{16}{16}\selectfont},      
  ticklabel style={font=\fontsize{15}{15}\selectfont},  
] 
\nextgroupplot[
tick align=outside,
tick pos=left,
x grid style={darkgray176},
xlabel={t-SNE Dimension 1},
xlabel style={yshift=-1.5ex},
xmajorgrids,
xmin=-65.9490507125854, xmax=57.5926473617554,
xtick style={color=black},
y grid style={darkgray176},
ylabel={t-SNE Dimension 2},
ylabel style={yshift=1.5ex},
ymajorgrids,
ymin=-45.7058378219605, ymax=56.7875677108765,
ytick style={color=black}
]
\addplot [
  colormap={mymap}{[1pt]
 rgb(0pt)=(0.894117647058824,0.101960784313725,0.109803921568627);
  rgb(1pt)=(0.215686274509804,0.494117647058824,0.72156862745098);
  rgb(2pt)=(0.301960784313725,0.686274509803922,0.290196078431373);
  rgb(3pt)=(0.596078431372549,0.305882352941176,0.63921568627451);
  rgb(4pt)=(1,0.498039215686275,0);
  rgb(5pt)=(1,1,0.2);
  rgb(6pt)=(0.650980392156863,0.337254901960784,0.156862745098039);
  rgb(7pt)=(0.968627450980392,0.505882352941176,0.749019607843137);
  rgb(8pt)=(0.6,0.6,0.6)
},
  only marks,
  scatter,
  scatter src=explicit
]
table [x=x, y=y, meta=colordata]{%
x  y  colordata
-27.5071468353271 -38.7696800231934 0.0
-22.0588645935059 5.34617948532104 5.0
-60.3335189819336 4.40356540679932 5.0
-55.4123458862305 29.6053142547607 5.0
-13.4049587249756 -26.3022956848145 0.0
-41.0210838317871 -8.96914672851562 5.0
-31.847448348999 32.9709777832031 7.0
-40.506534576416 13.6002731323242 5.0
15.2404680252075 -39.5500907897949 5.0
11.9888477325439 -14.6747035980225 5.0
-4.41874170303345 33.8662872314453 1.0
6.22063398361206 14.2239093780518 1.0
33.3332862854004 -2.93809413909912 5.0
30.5040702819824 -23.5117340087891 5.0
7.16435194015503 52.128776550293 1.0
36.5827484130859 25.2264366149902 1.0
51.9771156311035 -16.4935417175293 5.0
41.277271270752 -41.047046661377 5.0
16.9663372039795 30.9429206848145 2.0
30.0708885192871 47.967845916748 1.0
};

\nextgroupplot[
tick align=outside,
tick pos=left,
x grid style={darkgray176},
xlabel={t-SNE Dimension 1},
xlabel style={yshift=-1.5ex},
xmajorgrids,
xmin=-51.5791721343994, xmax=55.3653049468994,
xtick style={color=black},
xtick={-60,-40,-20,0,20,40,60},
xticklabels={\ensuremath{-}75,\ensuremath{-}50,\ensuremath{-}25,0,25,50,75},
y grid style={darkgray176},
ymajorgrids,
ymin=-46.506099319458, ymax=41.3135410308838,
ytick style={color=black}
]
\addplot [
  colormap={mymap}{[1pt]
 rgb(0pt)=(0.894117647058824,0.101960784313725,0.109803921568627);
  rgb(1pt)=(0.215686274509804,0.494117647058824,0.72156862745098);
  rgb(2pt)=(0.301960784313725,0.686274509803922,0.290196078431373);
  rgb(3pt)=(0.596078431372549,0.305882352941176,0.63921568627451);
  rgb(4pt)=(1,0.498039215686275,0);
  rgb(5pt)=(1,1,0.2);
  rgb(6pt)=(0.650980392156863,0.337254901960784,0.156862745098039);
  rgb(7pt)=(0.968627450980392,0.505882352941176,0.749019607843137);
  rgb(8pt)=(0.6,0.6,0.6)
},
  only marks,
  scatter,
  scatter src=explicit
]
table [x=x, y=y, meta=colordata]{%
x  y  colordata
-3.94766426086426 -42.5142974853516 0.0
20.480884552002 -32.3006782531738 5.0
50.5041923522949 -17.3035259246826 5.0
18.0951919555664 -13.0725831985474 7.0
-5.52342844009399 -27.2860221862793 0.0
33.1511154174805 -19.4319305419922 5.0
39.4497604370117 -35.8371658325195 5.0
37.7099266052246 -2.50907588005066 5.0
-21.3883743286133 14.1762542724609 5.0
-15.4098100662231 -5.91421222686768 5.0
4.6695351600647 5.59593820571899 1.0
19.1712055206299 14.1024761199951 1.0
-40.2907791137695 14.8128671646118 5.0
-30.5291481018066 -18.7493324279785 5.0
31.0143013000488 26.1102313995361 2.0
2.50103449821472 21.7134227752686 1.0
-30.8496227264404 -0.32833731174469 5.0
-46.7180595397949 -6.28073263168335 5.0
-3.2829532623291 37.3217391967773 1.0
15.3490047454834 34.6963729858398 1.0
};

\nextgroupplot[
colorbar,
colorbar style={ylabel={}},
colormap={mymap}{[1pt]
 rgb(0pt)=(0.894117647058824,0.101960784313725,0.109803921568627);
  rgb(1pt)=(0.215686274509804,0.494117647058824,0.72156862745098);
  rgb(2pt)=(0.301960784313725,0.686274509803922,0.290196078431373);
  rgb(3pt)=(0.596078431372549,0.305882352941176,0.63921568627451);
  rgb(4pt)=(1,0.498039215686275,0);
  rgb(5pt)=(1,1,0.2);
  rgb(6pt)=(0.650980392156863,0.337254901960784,0.156862745098039);
  rgb(7pt)=(0.968627450980392,0.505882352941176,0.749019607843137);
  rgb(8pt)=(0.6,0.6,0.6)
},
point meta max=7,
point meta min=0,
tick align=outside,
tick pos=left,
x grid style={darkgray176},
xlabel={t-SNE Dimension 1},
xlabel style={yshift=-1.5ex},
xmajorgrids,
xmin=-28.2503950595856, xmax=12.3829834461212,
xtick style={color=black},
y grid style={darkgray176},
ymajorgrids,
ymin=-35.0493062496185, ymax=11.6869586467743,
ytick style={color=black}
]
\addplot [
  colormap={mymap}{[1pt]
 rgb(0pt)=(0.894117647058824,0.101960784313725,0.109803921568627);
  rgb(1pt)=(0.215686274509804,0.494117647058824,0.72156862745098);
  rgb(2pt)=(0.301960784313725,0.686274509803922,0.290196078431373);
  rgb(3pt)=(0.596078431372549,0.305882352941176,0.63921568627451);
  rgb(4pt)=(1,0.498039215686275,0);
  rgb(5pt)=(1,1,0.2);
  rgb(6pt)=(0.650980392156863,0.337254901960784,0.156862745098039);
  rgb(7pt)=(0.968627450980392,0.505882352941176,0.749019607843137);
  rgb(8pt)=(0.6,0.6,0.6)
},
  only marks,
  scatter,
  scatter src=explicit
]
table [x=x, y=y, meta=colordata]{%
x  y  colordata
-24.6337928771973 -4.25798463821411 0.0
-18.496509552002 -24.7399597167969 5.0
-11.489818572998 -28.8182544708252 5.0
-14.247184753418 -18.4951419830322 5.0
-18.3927459716797 -7.6269006729126 0.0
-23.557991027832 -18.0980701446533 5.0
-26.4034233093262 -26.5389003753662 5.0
-19.2102279663086 -32.9249305725098 6.0
-5.6990532875061 9.56258296966553 5.0
-12.3401002883911 3.18854570388794 5.0
2.49654054641724 -27.5588188171387 1.0
7.88211917877197 -12.8146495819092 1.0
-7.64324474334717 -4.78529977798462 7.0
2.15077066421509 -2.99458646774292 5.0
-1.73477387428284 -13.6204853057861 1.0
-3.79429960250854 -21.6019554138184 1.0
-4.05251026153564 1.96765446662903 5.0
3.41836714744568 5.5376443862915 5.0
10.5360116958618 -21.9837074279785 1.0
3.45138239860535 -19.2927379608154 1.0
};
\end{groupplot}

\end{tikzpicture}

%% file: figures/runningtime.tex
\begin{tikzpicture}

\begin{axis}[
legend cell align={left},
legend style={
  at={(0.03,0.97)},
  anchor=north west,
},
legend style={font=\fontsize{6}{5}\selectfont},
tick align=outside,
tick pos=left,
x grid style={darkgray176},
xlabel={Number of IoT Devices},
xmajorgrids,
xmin=0.3, xmax=15.7,
xtick style={color=black},
y grid style={darkgray176},
ylabel={Running Time (s)},
ymajorgrids,
ymin=-0.0339343935120293, ymax=1.99500699942209,
ytick style={color=black}
]
\addplot [semithick, blue, mark=asterisk, mark size=3, mark options={solid}]
table [x=x, y=y, header=true]{%
x  y
1 0.215292210880307
2 0.20528557718136
3 0.206017109914375
4 0.257177531096242
5 0.301035409573149
6 0.328677036126517
7 0.360136897294846
8 0.418421338253196
9 0.47093432242097
10 0.53257151318642
11 0.57773095377887
12 0.628926420541561
13 0.717491028931511
14 0.834996119831005
15 0.887023727887261
};
\addlegendentry{RF (MILP)}
\addplot [semithick, green, mark=triangle*, mark size=3, mark options={solid}]
table [x=x, y=y, header=true]{%
x  y
1 0.260800513989302
2 0.260304820898833
3 0.325099678244775
4 0.438346353208467
5 0.44295495672353
6 0.504242449639197
7 0.513537416563637
8 0.71133606284867
9 0.813759315196449
10 1.01450055023289
11 1.03600705368832
12 1.23088700926579
13 1.4282067003034
14 1.7221066465017
15 1.90278239065236
};
\addlegendentry{RF-OWC (MILP)}
\addplot [semithick, purple, dashed, mark=triangle*, mark size=3, mark options={solid,rotate=180}]
table [x=x, y=y, header=true]{%
x  y
1 0.0582902152577034
2 0.0680789650971654
3 0.0713660912315636
4 0.0735356638544947
5 0.0894408999289831
6 0.108736317498598
7 0.0992348470058894
8 0.123329696741469
9 0.128498791850608
10 0.161950466289248
11 0.150268130952162
12 0.176410781527284
13 0.203272118750645
14 0.233655240292343
15 0.23684249738927
};
\addlegendentry{RF-OWC (DGET)}
\end{axis}

\end{tikzpicture}

%% file: figures/snapshots_2.tex
\begin{tikzpicture}

\definecolor{darkgray176}{RGB}{176,176,176}
\definecolor{darkorange25512714}{RGB}{255,127,14}
\definecolor{forestgreen4416044}{RGB}{44,160,44}
\definecolor{steelblue31119180}{RGB}{31,119,180}

\begin{axis}[
legend cell align={left},
legend style={
  at={(0.03,0.03)},
  anchor=south west, font=\fontsize{6}{5}\selectfont,
},
tick align=outside,
tick pos=left,
x grid style={darkgray176},
xlabel={Outdated Edges Replacement (\%)},
xmajorgrids,
xmin=-2.5, xmax=52.5,
xtick style={color=black},
y grid style={darkgray176},
ylabel={Accuracy w.r.t MILP (True Edges) },
ymajorgrids,
ymin=0.356767282275928, ymax=1.03063012941543,
ytick style={color=black}
]
\addplot [semithick, purple, dashed, mark=diamond*, mark size=3, mark options={solid,rotate=180}]
table [x=x, y=y, header=true]{%
x  y
0 0.92
10 0.903769677800086
20 0.902729810068996
30 0.897061352309238
40 0.890961295920426
50 0.891615726522979
};
\addlegendentry{1 Snapshot (DGET)}
\addplot [semithick,purple, dashed, mark=triangle*, mark size=3, mark options={solid,rotate=180}]
table [x=x, y=y, header=true]{%
x  y
0 0.92
10 0.828618439325606
20 0.733496273088443
30 0.667514622928803
40 0.605873289151355
50 0.534209634663842
};
\addlegendentry{3 Snapshots (DGET)}

\addplot [thick,green, mark=diamond*, mark size=3, mark options={rotate=180}]
table [x=x, y=y, header=true]{%
x  y
0 1
10 0.841043624130064
20 0.833242291991912
30 0.801425676844251
40 0.802468001448853
50 0.769707091682615
};
\addlegendentry{1 Snapshot (MILP)}
\addplot [thick,green, mark=triangle*, mark size=3, mark options={rotate=180}]
table [x=x, y=y, header=true]{%
x  y
0 1
10 0.700189065249821
20 0.656298853001967
30 0.536524587956529
40 0.469049513968413
50 0.38739741169136
};
\addlegendentry{3 Snapshots (MILP)}
\end{axis}

\end{tikzpicture}